\PassOptionsToPackage{unicode}{hyperref}
\PassOptionsToPackage{hyphens}{url}
\PassOptionsToPackage{dvipsnames,svgnames,x11names}{xcolor}
\documentclass[
  12pt]{article}

\usepackage{amsmath,amssymb}
\usepackage{iftex}
\ifPDFTeX
  \usepackage[T1]{fontenc}
  \usepackage[utf8]{inputenc}
  \usepackage{textcomp} 
\else 
  \usepackage{unicode-math}
  \defaultfontfeatures{Scale=MatchLowercase}
  \defaultfontfeatures[\rmfamily]{Ligatures=TeX,Scale=1}
\fi
\usepackage{lmodern}
\ifPDFTeX\else  
\fi
\IfFileExists{upquote.sty}{\usepackage{upquote}}{}
\IfFileExists{microtype.sty}{
  \usepackage[]{microtype}
  \UseMicrotypeSet[protrusion]{basicmath} 
}{}
\makeatletter
\@ifundefined{KOMAClassName}{
  \IfFileExists{parskip.sty}{%
    \usepackage{parskip}
  }{
    \setlength{\parindent}{0pt}
    \setlength{\parskip}{6pt plus 2pt minus 1pt}}
}{
  \KOMAoptions{parskip=half}}
\makeatother
\usepackage{xcolor}
\makeatletter
\ifx\paragraph\undefined\else
  \let\oldparagraph\paragraph
  \renewcommand{\paragraph}{
    \@ifstar
      \xxxParagraphStar
      \xxxParagraphNoStar
  }
  \newcommand{\xxxParagraphStar}[1]{\oldparagraph*{#1}\mbox{}}
  \newcommand{\xxxParagraphNoStar}[1]{\oldparagraph{#1}\mbox{}}
\fi
\ifx\subparagraph\undefined\else
  \let\oldsubparagraph\subparagraph
  \renewcommand{\subparagraph}{
    \@ifstar
      \xxxSubParagraphStar
      \xxxSubParagraphNoStar
  }
  \newcommand{\xxxSubParagraphStar}[1]{\oldsubparagraph*{#1}\mbox{}}
  \newcommand{\xxxSubParagraphNoStar}[1]{\oldsubparagraph{#1}\mbox{}}

\fi
\makeatother

\usepackage{longtable,booktabs,array}
\usepackage{calc} 
\usepackage{etoolbox}
\makeatletter
\patchcmd\longtable{\par}{\if@noskipsec\mbox{}\fi\par}{}{}
\makeatother
\IfFileExists{footnotehyper.sty}{\usepackage{footnotehyper}}{\usepackage{footnote}}
\makesavenoteenv{longtable}
\usepackage{graphicx}
\makeatletter
\def\maxwidth{\ifdim\Gin@nat@width>\linewidth\linewidth\else\Gin@nat@width\fi}
\def\maxheight{\ifdim\Gin@nat@height>\textheight\textheight\else\Gin@nat@height\fi}
\makeatother
\setkeys{Gin}{width=\maxwidth,height=\maxheight,keepaspectratio}
\makeatletter
\def\fps@figure{htbp}
\makeatother

\makeatletter
\@ifpackageloaded{caption}{}{\usepackage{caption}}
\AtBeginDocument{%
\ifdefined\contentsname
  \renewcommand*\contentsname{Table of contents}
\else
  \newcommand\contentsname{Table of contents}
\fi
\ifdefined\listfigurename
  \renewcommand*\listfigurename{List of Figures}
\else
  \newcommand\listfigurename{List of Figures}
\fi
\ifdefined\listtablename
  \renewcommand*\listtablename{List of Tables}
\else
  \newcommand\listtablename{List of Tables}
\fi
\ifdefined\figurename
  \renewcommand*\figurename{Figure}
\else
  \newcommand\figurename{Figure}
\fi
\ifdefined\tablename
  \renewcommand*\tablename{Table}
\else
  \newcommand\tablename{Table}
\fi
}
\@ifpackageloaded{float}{}{\usepackage{float}}
\floatstyle{ruled}
\@ifundefined{c@chapter}{\newfloat{codelisting}{h}{lop}}{\newfloat{codelisting}{h}{lop}[chapter]}
\floatname{codelisting}{Listing}

\makeatother
\makeatletter
\@ifpackageloaded{caption}{}{\usepackage{caption}}
\@ifpackageloaded{subcaption}{}{\usepackage{subcaption}}
\makeatother

\ifLuaTeX
  \usepackage{selnolig}  
\fi
\usepackage[]{natbib}
\usepackage{bookmark}

\IfFileExists{xurl.sty}{\usepackage{xurl}}{} 
\hypersetup{
  pdftitle={Model-Based Clustering of Longitudinal Functional Data with Hidden Semi-Markov Models},
  pdfkeywords={contaminated Gaussian distribution, explicit-duration model, functional principal subspace, regime switching, sojourn distribution},
  colorlinks=true,
  linkcolor={blue},
  filecolor={Maroon},
  citecolor={Blue},
  urlcolor={Blue} 
}

\newcommand{\anon}{1}
\if1\anon\hypersetup{pdfauthor={Bagas Caesar Suherlan; Emilio Porcu; Antonello Maruotti}}\fi

\usepackage{bm}
\usepackage{amsthm}

\newtheorem{proposition}{Proposition}

\newtheorem{lemma}{Lemma}
\newtheorem{remark}{Remark}

\newtheorem{assumption}{Assumption}

\begin{document}

\def\spacingset#1{\renewcommand{\baselinestretch}%
{#1}\small\normalsize} \spacingset{1}


\if1\anon
{
  \title{\bf Model-Based Clustering of\\ Longitudinal Functional Data with\\ Hidden Semi-Markov Models}
  \author{Bagas Caesar Suherlan\thanks{Department of Mathematics, Khalifa University (100066890@ku.ac.ae). 
  The authors report there is no funding associated with the work featured in this article.}
    \and Emilio Porcu\thanks{Department of Mathematics and Center for Biotechnology, Khalifa University (emilio.porcu@ku.ac.ae).}
    \and Antonello Maruotti\thanks{Department of Public Health and Epidemiology and Center for Biotechnology, Khalifa University, and Department GEPLI, LUMSA University (antonello.maruotti@ku.ac.ae).}}
  \date{}
  \maketitle
} \fi

\if0\anon
{
  \bigskip
  \bigskip
  \bigskip
  \begin{center}
    {\LARGE\bf Model-Based Clustering of\\ Longitudinal Functional Data with\\ Hidden Semi-Markov Models\par}
\end{center}
  \medskip
} \fi

\bigskip
\begin{abstract}
Model-based clustering of longitudinal functional data must account both for heterogeneity
among curves and for temporal dependence in their latent group memberships. Functional
mixture models typically treat curves as exchangeable, while temporal approaches may
restrict regime persistence or ignore state-specific functional variation. We propose hidden
semi-Markov models for ordered sequences of multivariate functional observations in
which each latent regime has a parsimonious, state-specific functional principal subspace,
extending functional high-dimensional data clustering (funHDDC) to longitudinal settings.
Gaussian, contaminated-Gaussian and multivariate-$t$ emissions accommodate atypical
observations, and explicit state-specific sojourn distributions govern regime persistence,
with hidden Markov models as the geometric special case. Estimation uses
expectation--maximization (EM) and expectation/conditional-maximization (ECM) algorithms
with segmental forward--backward recursions, and identifiability is established up to label
permutations and spectral indeterminacies. In simulations, temporal dependence yields
the largest gains when single curves are weakly informative; with robust emissions,
state-specific subspaces outperform a global functional principal component representation;
and explicit durations add smaller gains when regimes overlap and sojourns are non-geometric.
Applications to air-quality profiles and smartphone activity recordings, evaluated on
held-out years and subjects, show improved predictive fit and illustrate how apparent
duration effects can reflect time-varying transitions.
\end{abstract}

\noindent%
{\it Keywords:} Contaminated Gaussian distribution; Explicit-duration model; Functional principal subspace;
Regime switching; Sojourn distribution.
\vfill

\newpage
\spacingset{1.8} 

\section{Introduction}\label{sec-intro}

Many contemporary datasets consist of longitudinal sequences of
multivariate functional observations. Examples include daily
pollution profiles, repeated physiological measurements, and
successive windows of sensor recordings. Such data exhibit two
distinct forms of structure: variation within each functional
observation and dependence across observations collected over
time. When the underlying system evolves through persistent
regimes, clustering must account for both structures. In
particular, curves with similar shapes need not belong to the
same regime, while atypical observations should not
automatically be interpreted as regime changes. These
difficulties are compounded when the functional representation
is high-dimensional relative to the information available
within individual sequences.

Clustering longitudinal data has been extensively investigated
through finite mixtures, latent class models, mixed-effects
models, and functional approaches
\citep{lu2025clustering,zeldow2021functional,zhou2023clustermld}.
A fundamental distinction concerns the object being clustered.
Many longitudinal clustering methods assign an entire
individual trajectory to a latent group, thereby describing
heterogeneity between subjects while allowing variation
within each group. Functional clustering similarly seeks
homogeneous groups of curves, commonly through basis
expansions or functional principal component analysis
\citep{Jacques2014,Bouveyron2011}.
Extensions to multivariate and longitudinal trajectories
have addressed additional sources of heterogeneity,
including phase variation and subject-specific functional
patterns \citep{park2017clustering,zeldow2021functional}.
These approaches are valuable when group membership is
stable over the observation period. They are less directly
suited to settings in which the same observational unit
may successively occupy different latent regimes.

A related challenge is the dimensionality of the functional
representation. Functional high-dimensional data clustering
(funHDDC), which extends high-dimensional data clustering
\citep{Bouveyron2007} to curves, addresses this problem by associating each cluster
with a low-dimensional functional principal subspace and
parsimonious residual variation
\citep{Bouveyron2011,Schmutz2020}.
Unlike a global functional principal component representation,
state-specific subspaces allow different regimes to exhibit
distinct directions of functional variability. However,
conventional functional mixture models generally treat curves
as exchangeable. Consequently, they cannot exploit the
ordering of repeated observations to distinguish persistent
regimes from isolated atypical curves or to resolve
classification uncertainty when regime-specific distributions
overlap.

Latent-state models provide a natural probabilistic connection
between functional clustering and longitudinal dependence.
Hidden Markov models (HMMs) represent the evolution of
unobserved states through transition probabilities
\citep{Zucchini2016,bartolucci2012latent}, and extensions to
functional observations have demonstrated their usefulness
for modeling sequences of multivariate curves
\citep{Martino2020,Sidrow2022}.
Related factor and subspace hidden-state models also show
how temporal dependence can be combined with parsimonious
representations of high-dimensional observations \citep{golini2026dynamic,maruotti2017dynamic}.
The distinctive feature of these approaches is that cluster
membership may change over time, rather than being fixed
for an entire longitudinal trajectory. Nevertheless, the
ordinary HMM imposes geometric state durations:
conditional on the current state, the probability of leaving
it is independent of the time already spent there.

This restriction may be consequential for longitudinal
functional observations. The duration of a physiological,
environmental, or behavioral regime may itself carry
information about the underlying process, beyond that
contained in individual curves and successive transition
probabilities. Hidden semi-Markov models (HSMMs) accommodate
this feature by specifying the distribution of state
durations separately from the transition probabilities
between successive regimes \citep{Guedon2003,Yu2010}.
The HMM is recovered as the special case of geometric
sojourn distributions. The statistical challenge is therefore
to integrate explicit-duration dependence with parsimonious
functional emissions, while retaining sufficient flexibility
to accommodate atypical curves without introducing
spurious regimes.

We develop a family of hidden semi-Markov models for
clustering longitudinal sequences of multivariate functional
observations. Conditional on a latent regime, each curve
is represented through a state-specific functional principal
subspace, with an unrestricted mean function, leading
functional scale eigenvalues, and parsimonious residual
variation. Gaussian, contaminated-Gaussian, and
multivariate-$t$ emissions accommodate different patterns
of within-regime variability and atypical observations,
building on robust functional mixture models
\citep{AmovinAssagba2022,Anton2023}.
An embedded transition matrix describes changes between
regimes, while shifted Poisson, shifted negative-binomial,
and nonparametric sojourn distributions allow different
forms of persistence. The resulting framework combines
functional clustering and longitudinal regime segmentation
within a likelihood-based model, with the HMM retained
as a nested benchmark.

Estimation is based on segmental forward--backward
recursions and expectation--maximization (EM) \citep{Dempster1977} or
expectation/conditional-maximization (ECM) \citep{MengRubin1993}
algorithms.
The parsimonious subspace structure permits low-rank
evaluation of emission densities, while the likelihood
accounts for terminal censoring and, when appropriate,
initial residual durations. We establish sufficient
conditions for identifiability of the functional emissions,
state-specific subspaces, and semi-Markov dynamics,
accounting for eigenvalue multiplicities, contaminated
component pairing, and the observation horizon.

The numerical investigation is designed to assess the
contribution of each modeling component rather than
presume the superiority of the most flexible specification.
Simulation studies compare state-specific functional
subspaces with global functional principal component
representations under matched emission and temporal
structures, distinguishing the effects of robustness,
Markov persistence, and explicit durations. Applications
to air-quality profiles and smartphone activity recordings
are evaluated on held-out years and subjects. The results
show that temporal dependence and state-specific subspaces
can improve clustering and predictive fit, whereas the
additional benefit of explicit durations depends on regime
overlap and the underlying persistence structure. In
particular, apparent non-geometric durations may reflect
time-varying transition dynamics, highlighting the
importance of distinguishing duration dependence from
nonstationarity.
Section~\ref{sec:framework} introduces the model and a concise
identifiability statement, Section~\ref{sec:inference} develops likelihood
inference and computation, Sections~\ref{sec:simulations}
and~\ref{sec:applications} report the simulation studies and applications,
and Section~\ref{sec:conclusion} concludes.

\section{Functional representation and hidden semi-Markov clustering}\label{sec:framework}

\subsection{Longitudinal multivariate functional observations}
\label{subsec:functional-representation}

We consider $N$ independent sequences $\mathcal D_i=\{\bm X_{i\ell}:\ell=1,\ldots,L_i\}$, $i=1,\ldots,N$, of multivariate functional observations collected in a prescribed order, where $L_i$ is the number of functional observations in sequence $i$ and each observation $\bm X_{i\ell}(t)=(X_{i\ell}^{(1)}(t),\ldots,X_{i\ell}^{(p)}(t))^\top$, $t\in[0,T]$, belongs to the Hilbert space $\mathcal H=L^2([0,T];\mathbb R^p)$. The data therefore involve two distinct indices. The argument $t$ describes variation within an individual functional observation, whereas $\ell$ indexes successive observations within the same longitudinal sequence. For example, $t$ may represent the hour of the day and $\ell$ the calendar day, or $t$ may represent time within a sensor window and $\ell$ the position of that window in a recording. The observational units indexed by $i$ may correspond to subjects, monitoring locations, or independent recording sessions.

Each functional component is observed at finitely many time points and represented through a basis expansion,
\begin{equation}
\label{eq:basis-component}
X_{i\ell}^{(s)}(t)=\sum_{r=1}^{R_s}c_{i\ell r}^{(s)}\xi_r^{(s)}(t),\qquad s=1,\ldots,p,
\end{equation}
where $\{\xi_r^{(s)}\}_{r=1}^{R_s}$ is a set of linearly independent basis functions in $L^2[0,T]$ and $c_{i\ell r}^{(s)}$ are the corresponding coefficients. Equivalently,
\begin{equation}
\label{eq:basis-multivariate}
\bm X_{i\ell}(t)=\xi(t)\bm c_{i\ell},\qquad\bm c_{i\ell}\in\mathbb R^R,\quad R=\sum_{s=1}^pR_s,
\end{equation}
where $\xi(t)\in\mathbb R^{p\times R}$ is the block-diagonal basis matrix and $\bm c_{i\ell}$ stacks the component-specific coefficients. The geometry of the functional representation is determined by the Gram matrix $W=\int_0^T\xi(t)^\top\xi(t)\,\mathrm dt$, which is symmetric positive definite because the selected basis functions are linearly independent. Since $\xi(t)$ is block diagonal, $W=\operatorname{diag}(W_1,\ldots,W_p)$ with $W_s=\int_0^T\xi^{(s)}(t)\,\xi^{(s)}(t)^\top\,\mathrm dt$, where $\xi^{(s)}(t)=(\xi_1^{(s)}(t),\ldots,\xi_{R_s}^{(s)}(t))^\top$. For two coefficient vectors $\bm c$ and $\bm c'$ representing functions $\bm X$ and $\bm X'$, respectively, $\langle\bm X,\bm X'\rangle_{\mathcal H}=\bm c^\top W\bm c'$, so the transformed coefficients $W^{1/2}\bm c_{i\ell}$ preserve the functional inner-product geometry in Euclidean coordinates.

We assume that the longitudinal observations arise from $K$ latent states. Let $S_{i\ell}\in\{1,\ldots,K\}$ denote the unobserved state associated with the $\ell$-th functional observation of sequence $i$, and define $Z_{i\ell k}=\mathbf 1\{S_{i\ell}=k\}$. Throughout, a \emph{state} denotes a latent category of the stochastic process, a \emph{regime} refers to its interpretation as a persistent condition of the underlying system, and a \emph{cluster} is the collection of functional observations assigned to that state. Conditional on the latent states, the functional observations are assumed independent, with state-dependent coefficient densities $\bm c_{i\ell}\mid S_{i\ell}=k\sim f_k(\cdot;\bm\theta_k)$. This conditional independence does not imply marginal independence across successive observations: longitudinal dependence is introduced through the joint distribution of the latent states, specified by the semi-Markov structure below. No transition is defined between the end of one sequence and the beginning of another.

\subsection{State-specific functional subspaces}
\label{subsec:functional-subspaces}

The functional representation in \eqref{eq:basis-multivariate} may be high-dimensional, particularly when several functional components are observed, so estimating an unrestricted covariance matrix for each latent state can be unstable or computationally demanding. We address this problem through the parsimonious covariance structure of functional high-dimensional data clustering (funHDDC), allowing each state to be associated with its own low-dimensional functional principal subspace. For state $k$, let $\mathcal E_k\subset\mathcal H$ denote a functional subspace of dimension $d_k$, with $1\le d_k\le R-1$, whose orthonormal basis functions are represented as $\zeta_{kr}(t)=\xi(t)W^{-1/2}\bm u_{kr}$, $r=1,\ldots,d_k$, where $U_k=[\bm u_{k1},\ldots,\bm u_{kd_k}]$ and $U_k^\top U_k=I_{d_k}$. Indeed, $\langle\zeta_{kj},\zeta_{kh}\rangle_{\mathcal H}=\bm u_{kj}^\top\bm u_{kh}=\delta_{jh}$, so the matrix $U_k$ identifies the state-specific functional principal directions in the whitened coefficient space. Conditional on $S_{i\ell}=k$, we assume the latent score representation
\begin{equation}
\label{eq:score-relation}
\bm c_{i\ell}=\bm\mu_k+W^{-1/2}U_k\bm\delta_{i\ell}+\bm\varepsilon_{i\ell},
\end{equation}
where $\bm\mu_k\in\mathbb R^R$ is unrestricted, and the mutually independent terms satisfy $\bm\delta_{i\ell}\sim N_{d_k}(\bm0,\Gamma_k)$ and $\bm\varepsilon_{i\ell}\sim N_R(\bm0,b_kW^{-1})$, with $\Gamma_k=\operatorname{diag}(a_{k1}-b_k,\ldots,a_{kd_k}-b_k)$ and $a_{k1}\ge\cdots\ge a_{kd_k}>b_k>0$. It follows that the conditional Gaussian distribution of the coefficient vector has mean $\bm\mu_k$ and covariance matrix
\begin{equation}
\label{eq:covariance-funhddc}
\Sigma_k=W^{-1/2}\left[U_kA_kU_k^\top+b_k(I_R-U_kU_k^\top)\right]W^{-1/2},\qquad A_k=\operatorname{diag}(a_{k1},\ldots,a_{kd_k}).
\end{equation}
The leading eigenvalues $a_{kj}$ describe variation along the state-specific functional principal directions, while $b_k$ represents residual variation in their orthogonal complement. Consequently, different states may exhibit distinct mean functions, intrinsic dimensions, principal directions, and patterns of functional variability, which is more flexible than imposing a common global functional principal subspace across all states. We denote the general covariance structure by $[a_{kj},b_k,U_k,d_k]$. Parsimonious variants are obtained by imposing equality constraints on selected parameters within or across states, as described in Appendix~\ref{subsec:emission-updates}.

\subsection{Functional emission families}
\label{subsec:emission-families}

We consider three emission families, sharing the location parameter $\bm\mu_k$ and the structured scale matrix $\Sigma_k$ in \eqref{eq:covariance-funhddc}, but differing in their treatment of atypical observations and tail behavior. Under the Gaussian specification, $\bm c_{i\ell}\mid S_{i\ell}=k\sim N_R(\bm\mu_k,\Sigma_k)$, with emission density $f_k(\bm c;\bm\theta_k)=\phi_R(\bm c;\bm\mu_k,\Sigma_k)=(2\pi)^{-R/2}|\Sigma_k|^{-1/2}\exp\{-\Delta_k(\bm c)/2\}$, where $\Delta_k(\bm c)=(\bm c-\bm\mu_k)^\top\Sigma_k^{-1}(\bm c-\bm\mu_k)$ is the squared Mahalanobis distance. This specification corresponds to the Gaussian funHDDC model. To accommodate atypical functional observations, we consider the contaminated-Gaussian emission \citep{punzo2016clustering,Punzo2016} $f_k(\bm c;\bm\theta_k)=\alpha_k\phi_R(\bm c;\bm\mu_k,\Sigma_k)+(1-\alpha_k)\phi_R(\bm c;\bm\mu_k,\eta_k\Sigma_k)$, where $0<\alpha_k<1$ is the probability that an observation is ordinary and $\eta_k>1$ is the covariance inflation factor for atypical observations. Equivalently, introducing the latent indicator $\Upsilon_{i\ell}\in\{0,1\}$ with $\Pr(\Upsilon_{i\ell}=1\mid S_{i\ell}=k)=\alpha_k$, the conditional distribution is Gaussian with covariance $\Sigma_k$ when $\Upsilon_{i\ell}=1$ and $\eta_k\Sigma_k$ otherwise. Both components share the same location and principal directions, so atypical curves can be accommodated without necessarily introducing additional states.

As a heavy-tailed alternative \citep{maruotti2019hidden}, we consider multivariate-$t$ emissions \citep{peel2000robust,bernardi2017multiple}, $\bm c_{i\ell}\mid S_{i\ell}=k\sim t_R(\bm\mu_k,\Sigma_k,\nu_k)$, with density
\[
f_k(\bm c;\bm\theta_k)=\frac{\Gamma((\nu_k+R)/2)}{\Gamma(\nu_k/2)(\pi\nu_k)^{R/2}|\Sigma_k|^{1/2}}\left(1+\frac{\Delta_k(\bm c)}{\nu_k}\right)^{-(\nu_k+R)/2},
\]
where $\nu_k>0$ denotes the degrees of freedom. Here $\Sigma_k$ is a scale matrix rather than the covariance matrix. When $\nu_k>2$, $\operatorname{Cov}(\bm c_{i\ell}\mid S_{i\ell}=k)=\nu_k\Sigma_k/(\nu_k-2)$, so its principal directions coincide with those of the structured scale matrix. The degrees of freedom may be state-specific or constrained to a common value $\nu_1=\cdots=\nu_K=\nu$. The multivariate-$t$ distribution also admits the latent-scale representation $H_{i\ell}\mid S_{i\ell}=k\sim\operatorname{Gamma}(\nu_k/2,\nu_k/2)$, using the shape--rate parameterization, and $\bm c_{i\ell}\mid H_{i\ell}=h,S_{i\ell}=k\sim N_R(\bm\mu_k,\Sigma_k/h)$. The auxiliary variables $\Upsilon_{i\ell}$ and $H_{i\ell}$ facilitate the EM/ECM estimation described in Section~\ref{sec:inference}, with detailed emission updates in Appendix~\ref{app:emission-updates}.

We refer to the three specifications as funHDDC, C-funHDDC, and T-funHDDC emissions, respectively. In each case, $\bm\theta_k$ contains the state-specific location, subspace, and scale parameters $(\bm\mu_k,U_k,A_k,b_k)$, together with $(\alpha_k,\eta_k)$ for C-funHDDC or $\nu_k$ for T-funHDDC. The intrinsic dimension $d_k$ and any covariance-sharing restrictions define the selected emission specification.

\subsection{Explicit-duration hidden semi-Markov model}
\label{subsec:hsm}

We model the evolution of the latent states through a hidden semi-Markov process, in which persistence is represented by explicit state-specific sojourn distributions. This separates the duration of a regime from the selection of the regime entered next. For sequence $i$, let $M_i$ denote the number of latent segments intersecting the observation window. Segment $j$ has state $B_{ij}\in\{1,\ldots,K\}$ and full duration $V_{ij}\in\{1,2,\ldots\}$, measured in numbers of consecutive functional observations. Writing $\kappa_{i0}=0$ and $\kappa_{ij}=\sum_{j'=1}^{j}V_{ij'}$, the observation window contains the first $L_i$ time points of the process, with $\kappa_{i,M_i-1}<L_i\le\kappa_{iM_i}$. Thus, the final segment may continue beyond the observation window. The state associated with observation $\ell$ is $S_{i\ell}=B_{ij}$ if $\kappa_{i,j-1}<\ell\le\kappa_{ij}$.

The segment states follow an embedded Markov chain with initial distribution $\rho_k=\Pr(B_{i1}=k)$, $\sum_{k=1}^K\rho_k=1$, and transition matrix $P^\ast=(p^\ast_{hk})$, where $p^\ast_{hk}=\Pr(B_{ij}=k\mid B_{i,j-1}=h)$ for $j\ge2$. Because persistence is represented through segment durations, self-transitions are excluded: $p^\ast_{hh}=0$, $p^\ast_{hk}\ge0$ and $\sum_{k\ne h}p^\ast_{hk}=1$. Conditional on $B_{ij}=k$, the full duration has probability mass function $\varphi_k(v;\bm\beta_k)=\Pr(V_{ij}=v\mid B_{ij}=k)$, $v=1,2,\ldots$, where $\bm\beta_k$ denotes its parameters. We assume that, conditional on the current segment state, its duration is independent of the state selected for the next segment and of the preceding segment history $\mathcal H_{ij}$. Consequently, $\Pr(V_{ij}=v,B_{i,j+1}=h\mid B_{ij}=k,\mathcal H_{ij})=\varphi_k(v;\bm\beta_k)p^\ast_{kh}$ for $h\ne k$.

We consider three duration specifications. The shifted Poisson (P) model assumes $V_{ij}=1+\widetilde V_{ij}$ with $\widetilde V_{ij}\sim\operatorname{Pois}(\lambda_k)$, giving $\mathbb E(V_{ij}\mid B_{ij}=k)=1+\lambda_k$ and $\operatorname{Var}(V_{ij}\mid B_{ij}=k)=\lambda_k$. The shifted negative-binomial (NB) model assumes $V_{ij}=1+\widetilde V_{ij}$ with $\widetilde V_{ij}\sim\operatorname{NB}(m_k,r_k)$, where the negative binomial is parameterized by mean $m_k>0$ and size $r_k>0$, with
\[
\Pr(\widetilde V_{ij}=v)=\frac{\Gamma(v+r_k)}{\Gamma(r_k)v!}\left(\frac{r_k}{r_k+m_k}\right)^{r_k}\left(\frac{m_k}{r_k+m_k}\right)^v.
\]
Its mean and variance are $\mathbb E(V_{ij}\mid B_{ij}=k)=1+m_k$ and $\operatorname{Var}(V_{ij}\mid B_{ij}=k)=m_k+m_k^2/r_k$. The shifted Poisson is obtained as $r_k\to\infty$, while $r_k=1$ gives a geometric distribution on $\{1,2,\ldots\}$. Finally, the nonparametric (NP) specification assigns probabilities $\varphi_k(1),\ldots,\varphi_k(V_{\max})$ to a finite duration support, subject to $\varphi_k(v)\ge0$ and $\sum_{v=1}^{V_{\max}}\varphi_k(v)=1$.

Combining these duration distributions and the embedded transition process with the state-dependent emissions of Section~\ref{subsec:emission-families} defines the proposed functional HSMM family. We denote its Gaussian, contaminated-Gaussian, and multivariate-$t$ versions by HSMM-funHDDC, HSMM-C-funHDDC, and HSMM-T-funHDDC, respectively, with the same parsimonious covariance variants as in the underlying funHDDC family. The ordinary hidden Markov model is recovered when all sojourn distributions are geometric. Specifically, for an HMM transition matrix $P=(p_{hk})$ with $p_{kk}<1$, $\varphi_k(v)=(1-p_{kk})p_{kk}^{v-1}$ and $p^\ast_{kh}=p_{kh}/(1-p_{kk})$ for $h\ne k$. Conversely, a semi-Markov model with geometric sojourns induces an HMM with $p_{kk}=1-q_k$ and $p_{kh}=q_kp^\ast_{kh}$ for $h\ne k$, where $q_k$ is the geometric exit probability. The HMM formulation and its inference are provided in Appendix~\ref{app:hmm-model}.
For model selection, the semi-Markov latent structure contributes $(K-1)+K(K-2)+\tau_{\mathrm{dur}}$ free parameters, where the terms correspond to the initial distribution, embedded transition matrix, and duration distributions, and the duration contribution is $\tau_{\mathrm{dur}}=K$ for the shifted Poisson, $2K$ for the shifted negative binomial and $K(V_{\max}-1)$ for the nonparametric sojourn. The remaining parameters arise from the selected emission specification; the complete counts and model-selection criteria are given in Appendix~\ref{subsec:hyperparameter-selection}.

Under the fixed-window observation convention, the final segment is right-censored. If its observed length is $u$, its duration contribution is the survivor probability
\begin{equation}
\bar\varphi_k(u;\bm\beta_k)=\Pr(V_{ij}\ge u\mid B_{ij}=k)=\sum_{v\ge u}\varphi_k(v;\bm\beta_k),
\label{eq:hsm-survivor}
\end{equation}
rather than the probability of an exact duration $u$ \citep{Guedon2003}. The initial segment is assumed to begin at a regime-entry time and is therefore not left-censored, although it may also be right-censored when it spans the entire observation window. Windows beginning within an ongoing regime require an initial residual-duration distribution, as described in Appendix~\ref{subsubsec:hsm-leftcens} and used in Section~\ref{sec:milano}.

\subsection{Identifiability of the functional HSMM}
\label{subsec:identifiability-summary}
For a known number of states $K$ and a fixed functional basis, the
identification problem concerns the state-dependent emission densities,
their functional spectral subspaces, the initial distribution $\bm\rho$,
the embedded transition matrix $P^\ast$, and the duration parameters
$\{\bm\beta_k\}$. Proposition~\ref{prop:hsmm-id} in
Appendix~\ref{app:hsmm-proof} gives sufficient conditions for their
identification from the population law of the observed coefficient
sequences, up to a common permutation of state labels and the standard
spectral indeterminacies. Crucially, the proposition assumes a known $K$,
identifiable state emissions, and sequences that begin at state-entry
times (Assumption~\ref{id:A}\textup{(A5)}); its finite-horizon condition
ensures that observable duration probabilities determine the duration
parameters within the chosen sojourn family. The conclusion does not automatically extend to windows
starting inside an ongoing regime (Remark~\ref{id:rem-boundary}), for which the initial observed
sojourn follows a residual-duration law. The proofs also treat tied leading
eigenvalues and the duration-family conditions.

\section{Likelihood inference and computation}\label{sec:inference}
The HSMM parameter vector is $\tilde{\bm\Theta}=\{\bm\rho,P^\ast,\bm\beta_1,\ldots,\bm\beta_K,\bm\theta_1,\ldots,\bm\theta_K\}$. The E-step yields the posterior state probabilities $\gamma_{i\ell k}=\Pr(S_{i\ell}=k\mid\mathcal{D}_i)$, posterior segments and embedded-transition counts, and all emission updates use these posterior state probabilities. A superscript $(m)$ denotes a value at iteration $m$ of the algorithm.

\subsection{Efficient evaluation of functional emissions}\label{subsec:emission-evaluation}
For a coefficient vector $\bm{c}$ and regime $k$, we define $\bm{z}_k=W^{1/2}(\bm{c}-\bm{\mu}_k)$ and $\bm{y}_k=U_k^\top\bm{z}_k$, with components $y_{kj}$.
The funHDDC structure gives
\[
  \log(|\Sigma_k|) = -\log(|W|) + \sum_{j=1}^{d_k}\log(a_{kj}) + (R-d_k)\log(b_k),
  \qquad
  \Delta_k(\bm{c}) = \sum_{j=1}^{d_k}\frac{y_{kj}^{2}}{a_{kj}} + \frac{\|\bm{z}_k-U_k\bm{y}_k\|^{2}}{b_k},
\]
for the squared Mahalanobis distance of Section~\ref{subsec:emission-families}. Thus, emission evaluation requires neither a full covariance
inverse nor a basis for the residual subspace. 

Since $W$ is fixed, $W^{1/2}\bm{c}_{i\ell}$ and $\log(|W|)$ can be precomputed, and $W^{1/2}\bm{\mu}_k$ and log-determinants are updated once per regime per iteration. Evaluation then costs $O(Rd_k)$ per observation and regime for all three emission families. For the inflated Gaussian component, the squared Mahalanobis distance is $\Delta_k(\bm{c})/\eta_k$ and the log-determinant is $\log(|\Sigma_k|) + R\log(\eta_k)$.

\subsection{Segmental likelihood and HSMM E-step}
\label{subsec:inference-hsm}
We use the notation of Section~\ref{subsec:hsm}: $B_{ij}$ is the state occupied by the $j$-th segment of sequence $i$, $V_{ij}$ its duration, $\kappa_{ij}=V_{i1}+\cdots+V_{ij}$ the cumulative segment endpoints, $M_i$ the number of segments, itself latent, $P^\ast$ the embedded transition matrix and $\varphi_k(v;\bm{\beta}_k)$ the sojourn probability mass function. For a block of $u$ consecutive functional observations ending at time $r$, we define $\Psi_i(r,u;k)=\prod_{\ell=r-u+1}^{r}f_k(\bm{c}_{i\ell};\bm{\theta}_k)$ and write $\Psi_i^{(m)}(r,u;k)$ for the same product evaluated at $\bm{\theta}_k^{(m)}$. Each sequence is observed over a fixed window, so its final segment is right-censored. Accordingly, wherever a segment ends at $L_i$ the duration mass $\varphi_k(u)$ is replaced by the survivor~\eqref{eq:hsm-survivor}, while interior segments are unchanged. It is convenient to carry the substitution in a single symbol: for a segment of duration $u$ ending at time $r$,
\begin{equation}
  \varphi_k^{\dagger}(u\mid r;\bm{\beta}_k)=
  \begin{cases}
    \varphi_k(u;\bm{\beta}_k), & r<L_i,\\[3pt]
    \bar{\varphi}_k(u;\bm{\beta}_k), & r=L_i,
  \end{cases}
  \label{eq:hsm-effective-duration}
\end{equation}
the dependence on the sequence entering only through whether $r$ is its final time. The observed length $u$ is the full duration of an interior segment but only a lower bound $V_{ij}\ge u$, for a terminal one, including a segment spanning the whole window. For geometric durations, we have $\varphi_k(u)=(1-p_{kk})p_{kk}^{u-1}$ and $\bar{\varphi}_k(u)=p_{kk}^{u-1}$. With the survivor for the terminal segment, the fixed-window HMM likelihood is recovered exactly, whereas the duration mass would introduce an additional terminal exit factor $1-p_{kk}$. 

The likelihood contribution of the $i$--th sequence can be written as a sum over all latent segmentations,
\[
\begin{aligned}
\mathcal{L}_i(\tilde{\bm{\Theta}}) = \displaystyle\sum_{q=1}^{L_i} \displaystyle\sum_{\substack{b_1,\ldots,b_q \\ b_j\neq b_{j+1}}} \displaystyle\sum_{\substack{u_1,\ldots,u_q\geq 1 \\ u_1+\cdots+u_q=L_i}} & \rho_{b_1}\varphi_{b_1}^{\dagger}(u_1\mid u_1;\bm{\beta}_{b_1})\Psi_i(u_1,u_1;b_1) \\
& \times \prod_{j=2}^{q} p^\ast_{b_{j-1},b_j}\,\varphi_{b_j}^{\dagger}\bigl(u_j\mid \textstyle\sum_{j'=1}^{j}u_{j'};\bm{\beta}_{b_j}\bigr)\,\Psi_i\Bigl(\displaystyle\sum_{j'=1}^{j}u_{j'},u_j;b_j\Bigr),
\end{aligned}
\]
where the sums run over the number $q$ of segments, their regimes $b_1,\ldots,b_q$ and their lengths $u_1,\ldots,u_q$, the final factor of each segmentation carrying the survivor because its block ends at $L_i$, with the full likelihood $\mathcal{L}(\tilde{\bm{\Theta}})=\prod_{i=1}^{N}\mathcal{L}_i(\tilde{\bm{\Theta}})$. The complete-data log-likelihood can be written using segment-level indicators $\widetilde Z_{ijk} = \bm{1}\{B_{ij}=k\}$ for $j=1,\ldots,M_i$. By ignoring constants that do not depend on the parameters, we have
\[
\begin{aligned}
\Lambda_c(\tilde{\bm{\Theta}}) ={} & \displaystyle\sum_{i=1}^{N} \displaystyle\sum_{k=1}^{K} \widetilde Z_{i1k}\log\rho_k + \displaystyle\sum_{i=1}^{N} \displaystyle\sum_{j=2}^{M_i} \displaystyle\sum_{h=1}^{K} \displaystyle\sum_{\substack{k=1\\k\neq h}}^{K} \widetilde Z_{i,j-1,h} \widetilde Z_{ijk} \log p^\ast_{hk} \\
& + \displaystyle\sum_{i=1}^{N} \left[\displaystyle\sum_{j=1}^{M_i-1} \displaystyle\sum_{k=1}^{K} \widetilde Z_{ijk}\log\varphi_k(V_{ij};\bm{\beta}_k) + \displaystyle\sum_{k=1}^{K} \widetilde Z_{iM_ik}\log\bar{\varphi}_k\bigl(L_i-\kappa_{i,M_i-1};\bm{\beta}_k\bigr)\right] \\
& + \displaystyle\sum_{i=1}^{N} \displaystyle\sum_{\ell=1}^{L_i} \displaystyle\sum_{k=1}^{K} Z_{i\ell k}\log f_k(\bm{c}_{i\ell};\bm{\theta}_k),
\end{aligned}
\]
the terminal segment entering through the survivor since only its observed length $L_i-\kappa_{i,M_i-1}$ is known. The contaminated-Gaussian and $t$ versions augment this complete-data log-likelihood with the terms involving the latent variables $\Upsilon_{i\ell}$ or $H_{i\ell}$ of Appendix~\ref{app:emission-latent}.

\subsubsection{E-step}
\label{subsubsec:hsm-estep}

At the current parameter value $\tilde{\bm\Theta}^{(m)}$, let
$F_i(r,k)$, $1\le r<L_i$, be the joint density of the observations
through time $r$ and a \emph{completed} state-$k$ segment at $r$.
With $F_i(0,k)$ undefined (initial segments are handled by $\bm\rho$),
\begin{equation}
\label{eq:hsm-forward-interior}
\begin{aligned}
F_i(r,k)={}&\sum_{v=1}^{\min(r,V_{\max})}
\varphi_k^{(m)}(v)\Psi_i^{(m)}(r,v;k)\\
&\times\Bigl[\mathbf1\{r=v\}\rho_k^{(m)}
+\mathbf1\{r>v\}\sum_{h\ne k}F_i(r-v,h)p_{hk}^{\ast(m)}\Bigr],
\qquad r<L_i.
\end{aligned}
\end{equation}
At the observation boundary, no state exit is observed. We therefore
use a separate terminal quantity,
\begin{equation}
\label{eq:hsm-forward-terminal}
\begin{aligned}
\widetilde F_i(L_i,k)={}&\sum_{v=1}^{\min(L_i,V_{\max})}
\bar\varphi_k^{(m)}(v)\Psi_i^{(m)}(L_i,v;k)\\
&\times\Bigl[\mathbf1\{L_i=v\}\rho_k^{(m)}
+\mathbf1\{L_i>v\}\sum_{h\ne k}F_i(L_i-v,h)p_{hk}^{\ast(m)}\Bigr].
\end{aligned}
\end{equation}
The sequence likelihood is $\mathcal L_i(\tilde{\bm\Theta}^{(m)})=\sum_{k=1}^K\widetilde F_i(L_i,k)$. Here $V_{\max}$ is the largest observed segment length allowed in the
recursions; for NP durations it is their finite support bound.

If the duration distribution is not finitely supported, $V_{\max} = \max_i L_i$ suffices for likelihood evaluation, since no longer \emph{completed} duration can be observed. Evaluated directly, the segmental recursions cost $O(L_iV_{\max}K^2)$ per sequence. Precomputing the entry weights $E_i(r,k)=\sum_{h\neq k}F_i(r,h)p^{\ast(m)}_{hk}$ for all $(r,k)$ at cost of $O(L_iK^2)$, and accumulating $\log\Psi_i^{(m)}(r,v;k)$ incrementally in $v$ reduces this to $O(L_iK^{2}+L_iV_{\max}K)$, against $O(L_iK^{2})$ for the hidden Markov recursions, so that $V_{\max}$ controls the additional cost of the semi-Markov layer. For an unbounded parametric duration family, a computational cutoff $V_{\max}<L_i$ omits admissible segmentations and is an approximation, not a renormalized duration distribution. Its effect can be assessed by increasing the cutoff, and likelihood comparisons with the nested HMM require compatible duration-support conventions. On a single core, the semi-Markov recursion takes $0.5$~s for a sequence of $10{,}000$ curves at $V_{\max}=640$, and a semi-Markov EM iteration costs two to three times a hidden Markov iteration at $V_{\max}=50$ (Appendix~\ref{app:computation}). 

The backward quantity $G_i(r,k)$ is the conditional likelihood of the observations from time $r+1$ to $L_i$, given that a segment in state $k$ ends at time $r$. For $r=L_i-1,\ldots,1$,
\[
    G_i(r,h) = \displaystyle\sum_{k\neq h} \displaystyle\sum_{v=1}^{\min(V_{\max},L_i-r)} p_{hk}^{\ast(m)}\varphi_k^{\dagger(m)}(v\mid r+v)\Psi_i^{(m)}(r+v,v;k)G_i(r+v,k),
\]
with $G_i(L_i,k)=1$, and the term $v=L_i-r$ for a terminal segment beginning at $r+1$ being weighted by the survivor. The terminal forward quantity $\widetilde F_i(L_i,k)$ is used only to evaluate the likelihood, while the interior quantities $F_i(r,k)$ enter the posterior recursions. The posterior probability of a segment in state $k$ of length $u$ ending at time $r$ is
\[
    \begin{aligned}
    \chi_i^{(m)}(r,u,k) ={}& \frac{\varphi_k^{\dagger(m)}(u\mid r)\Psi_i^{(m)}(r,u;k)G_i(r,k)}{\mathcal{L}_i(\tilde{\bm{\Theta}}^{(m)})}\\
    &\times\Bigl[\bm{1}\{r=u\}\rho_k^{(m)} + \bm{1}\{r>u\} \sum_{h\neq k} F_i(r-u,h)p_{hk}^{\ast(m)}\Bigr],
    \end{aligned}
\]
so that $\chi_i^{(m)}(L_i,u,k)$ is the posterior probability that the sequence ends inside a segment in state $k$ of observed length $u$, and $\chi_{i0}^{(m)}(u,k)=\chi_i^{(m)}(u,u,k)$ is the posterior probability that the first segment has state $k$ and duration $u$, the case $u=L_i$ covering a sequence whose first segment is also its last. 

For $r>u$, we define the posterior probability of an embedded transition from $h$ to $k$ followed by a segment of duration $u$ ending at $r$ as
\[
    \tilde{\omega}_i^{(m)}(r,u,h,k) = \frac{F_i(r-u,h)p_{hk}^{\ast(m)}\varphi_k^{\dagger(m)}(u\mid r)\Psi_i^{(m)}(r,u;k)G_i(r,k)}{\mathcal{L}_i(\tilde{\bm{\Theta}}^{(m)})}, \qquad h\neq k.
\]
The posterior state probability at time $\ell$ is obtained by summing over all posterior segments that cover $\ell$, which is
\[
    \gamma_{i\ell k}^{(m)} = \Pr\left(S_{i\ell}=k\mid\mathcal D_i;\tilde{\bm{\Theta}}^{(m)}\right) = \displaystyle\sum_{r=1}^{L_i} \displaystyle\sum_{\substack{v=1,\ldots,\min(r,V_{\max})\\ r-v+1\leq \ell\leq r}}\chi_i^{(m)}(r,v,k).
\]
The posterior segments tile $1,\ldots,L_i$, so $\sum_{k=1}^{K}\gamma_{i\ell k}^{(m)}=1$ for every $\ell$. For C-funHDDC and T-funHDDC emissions, the posterior expectations $\upsilon_{i\ell k}^{(m)}$, $h_{i\ell k}^{(m)}$ and $e_{i\ell k}^{(m)}$ of the latent variables are computed from the state-conditional emission distributions (Appendix~\ref{app:emission-latent}).

\subsubsection{M-step}
\label{subsubsec:hsm-mstep}

The initial probabilities are updated using the posterior distribution of the first segment, $\rho_k^{(m+1)}=N^{-1}\sum_{i=1}^{N}\sum_{v=1}^{\min(L_i,V_{\max})}\chi_{i0}^{(m)}(v,k)$. With the posterior expected numbers of embedded transitions from state $h$ to state $k$,
\[
    \widehat N_{hk}^{(m)} = \displaystyle\sum_{i=1}^{N} \displaystyle\sum_{r=2}^{L_i} \displaystyle\sum_{v=1}^{\min(r-1,V_{\max})} \tilde{\omega}_i^{(m)}(r,v,h,k),
    \qquad
    p_{hk}^{\ast(m+1)} = \frac{\widehat N_{hk}^{(m)}}  {\sum_{q\neq h} \widehat N_{hq}^{(m)}}, \quad h\neq k,
\]
with $p_{hh}^{\ast(m+1)}=0$. The duration data separate at $r=L_i$. Let
\[
    \widehat{M}_k^{(m)}(u) = \displaystyle\sum_{i=1}^{N} \displaystyle\sum_{r=u}^{L_i-1} \chi_i^{(m)}(r,u,k), \qquad \widehat{M}_k^{\mathrm{cens},(m)}(u) = \displaystyle\sum_{i=1}^{N} \chi_i^{(m)}(L_i,u,k)
\]
be the posterior expected numbers of segments in state $k$ whose duration is known to equal $u$ and to be at least $u$, respectively. Every sequence contributes exactly one terminal segment, so $\sum_{k}\sum_{u}\widehat{M}_k^{\mathrm{cens},(m)}(u)=N$. For a parametric duration family, the duration parameter is updated by maximizing the weighted duration log-likelihood of a right-censored sample,
\[
    \bm{\beta}_k^{(m+1)} = \arg\max_{\bm{\beta}_k} \displaystyle\sum_{u} \left[\widehat{M}_k^{(m)}(u) \log\varphi_k(u;\bm{\beta}_k) + \widehat{M}_k^{\mathrm{cens},(m)}(u) \log\bar{\varphi}_k(u;\bm{\beta}_k)\right].
\]

For the nonparametric sojourn model with support $\{1,\ldots,V_{\max}\}$, the duration probabilities are updated by augmenting each right-censored terminal segment with its unobserved full duration, using the expected counts $\widehat M_k^{(m)}(v)$ and $\widehat M_k^{\mathrm{cens},(m)}(u)$ defined above. Conditional on a censored segment of observed length $u$, the probability that its full duration equals $v\ge u$ is $\Pr(V=v\mid V\ge u,B=k;\bm\beta_k^{(m)})=\varphi_k^{(m)}(v)/\bar\varphi_k^{(m)}(u)$, so the augmented expected count of state-$k$ segments with full duration $v$ is
\begin{equation}
\label{eq:hsm-np-augmented-count}
\widehat C_k^{(m)}(v)
=
\widehat M_k^{(m)}(v)
+
\sum_{u=1}^{v}
\widehat M_k^{\mathrm{cens},(m)}(u)
\frac{\varphi_k^{(m)}(v)}
{\bar\varphi_k^{(m)}(u)},
\qquad v=1,\ldots,V_{\max}.
\end{equation}
Maximizing the corresponding augmented expected complete-data log-likelihood under the constraint $\sum_{v=1}^{V_{\max}}\varphi_k(v)=1$ gives
\begin{equation}
\label{eq:hsm-np-update}
\varphi_k^{(m+1)}(v)
=
\frac{\widehat C_k^{(m)}(v)}
{\displaystyle\sum_{w=1}^{V_{\max}}
\widehat C_k^{(m)}(w)}
,
\qquad v=1,\ldots,V_{\max}.
\end{equation}
This is the exact M-step for the duration probabilities under the augmented EM formulation, provided the relevant survivor probabilities are positive. When no terminal censoring is present, the update reduces to the normalized expected counts of completed segment durations.

The emission-parameter updates are common to the HSMM and HMM specifications, since the temporal dependence enters these updates through the posterior state probabilities $\gamma_{i\ell k}^{(m)}$. For Gaussian emissions, the state-specific location and scatter parameters are estimated using these probabilities as weights. For contaminated-Gaussian and multivariate-$t$ emissions, the updates additionally involve the conditional expectations of the contamination indicators or latent scales, which produce observation-specific precision weights and thereby reduce the influence of atypical functional observations on the estimated locations and principal subspaces.

\section{Simulation studies}\label{sec:simulations}
We investigate whether temporal dependence improves regime recovery, whether explicit durations add information beyond geometric sojourns, and whether state-specific functional subspaces improve on global FPCA. Two experiments generate Markov and semi-Markov latent paths, and every method is evaluated on the same 100 datasets per scenario by the adjusted Rand index (ARI) \citep{hubert1985comparing} of its pointwise maximum a posteriori (MAP) labels. We compare Gaussian, contaminated-Gaussian (C) and multivariate-$t$ (T) emissions under i.i.d.\ (exchangeable) mixtures and the appropriate temporal models. Models are named by temporal layer, emission and sojourn, so that T-funHDDC is the i.i.d.\ mixture with $t$ emission, HMM-C-funHDDC the HMM with contaminated-Gaussian emission and HSMM-T-funHDDC (P) the HSMM with $t$ emission and P sojourn. The global-FPCA competitors are FPCA-GMM, FPCA-$t$MM and FPCA-CMM for the Gaussian, $t$ and contaminated-Gaussian mixtures, and FPCA-HMM or FPCA-HSMM followed by -$t$ or -C for the temporal models with robust emissions. Full results and graphical diagnostics are in Appendix~\ref{app:simulation-results}.

\subsection{Design and comparisons}\label{sec:sim-design}
Each dataset consists of $N=25$ independent sequences of trivariate ($p=3$) curves on $[0,1]$, with each component expanded in $R_s=15$ orthonormal Fourier functions, so that $R=45$ and $W=I_R$. Conditional on regime $k$, a typical curve has coefficients generated from~\eqref{eq:score-relation},
\[
    \bm{c}_{i\ell} = \bm{\mu}_k + U_k\bm{\delta}_{i\ell} + \bm{\varepsilon}_{i\ell}, \qquad \bm{\delta}_{i\ell}\sim N\bigl(\bm{0},(a_k-b_k)I_{d_k}\bigr), \qquad \bm{\varepsilon}_{i\ell}\sim N(\bm{0},b_kI_R),
\]
so that the leading scale eigenvalues are tied within a regime, $a_{k1}=\cdots=a_{kd_k}=a_k$. The loading matrix $U_k$ consists of the first $d_k$ columns of a common orthonormal frame formed by the lowest-frequency non-constant Fourier modes, taken in turn from the three components, so that the regime subspaces are smooth, nested and of unequal dimension. With tied eigenvalues only the projectors $U_kU_k^{\top}$ are identifiable (Lemma~\ref{id:lemma}, Appendix~\ref{app:hsmm-proof}), and the fitted models use the general structure $[a_{kj},b_k,U_k,d_k]$, which contains the generating one. Each regime mean has three nonzero coefficients of magnitude $\mu/\sqrt{3}$, one on a low-frequency mode of each component, so that $\|\bm{\mu}_k\|=\mu$. Two regime means differ on two or three components and lie at distance $2\mu/\sqrt{3}$ or $\sqrt{2}\,\mu$, so that no single component separates all the regimes and the separation $\mu$ controls how informative a single curve is about its regime.

Contamination acts on whole curves. Independently of the regime path, each curve is atypical with probability $1-\alpha$, and its coefficients are then drawn from $N_R(\bm{\mu}_k,\eta\Sigma_k)$, which inflates all its standard deviations by $\sqrt{\eta}$. Scenario~A is clean ($\alpha=1$). Scenario~B has mild contamination ($\alpha=0.95$, $\eta=5$) and Scenario~C severe contamination ($\alpha=0.85$, $\eta=10$), so that $5\%$ and $15\%$ of the curves are atypical. The C-funHDDC emission is therefore correctly specified, the T-funHDDC emission is a heavy-tailed approximation and the Gaussian emission is misspecified.

The global-FPCA competitor projects the pooled whitened coefficients on their leading $q$ principal components, with $q$ the smallest number explaining at least $90\%$ of the variance ($q=30$--$34$ in Simulation~I and $32$--$35$ in Simulation~II). It models the scores with the same emission and temporal layer as the paired funHDDC model but with an unrestricted covariance in each state, so that the two models differ in the whole dimension-reduction step: state-specific low-rank-plus-isotropic scale matrices on all $R$ coefficients against a global $q$-dimensional projection, which discards the trailing $R-q$ dimensions, with unrestricted state covariances. Since the data are generated from the funHDDC structure, the comparison measures the advantage of this structure relative to a global FPCA representation rather than the effect of state-specific subspaces alone. Its likelihood is defined on the scores, so its Bayesian information criterion \citep[BIC;][]{Schwarz1978} is not compared with that of funHDDC. All methods share a multistart protocol with robust initial M-steps for the $t$ and contaminated emissions (Appendix~\ref{subsec:initialization}), refinement passes and warm starts of every HSMM from the fitted HMM, with $K$ and the $d_k$ fixed at their true values to isolate the temporal and emission effects (Appendix~\ref{app:simulation-details}). The ARI is computed against the true regimes, pooled over the $N$ sequences (Appendix~\ref{subsec:convergence-classification}). Because all models are fitted to the same datasets, a difference between two models is assessed by a paired $95\%$ $t$-interval over the replicates, and every difference discussed below has an interval that excludes zero unless it is described as negligible. Means marked $\dagger$ in Table~\ref{tab:sim1} are over different subsets of non-degenerate replicates, so no paired interval is defined for them, and they are compared only descriptively.

\begin{table}
\centering
\caption{Simulation I: mean ARI (standard deviation) over 100 replicates. $^{\dagger}$Mean over the non-degenerate fits (86, 70 and 69 replicates, in reading order).
}
\label{tab:sim1}
\begin{tabular}{@{}lccc@{}} 
\toprule
Method & Scenario A & Scenario B & Scenario C \\
\midrule
funHDDC & 0.768 (0.026) & 0.549 (0.056) & 0.349 (0.089) \\
T-funHDDC & 0.766 (0.026) & 0.686 (0.036) & 0.539 (0.071) \\
C-funHDDC & 0.767 (0.027) & 0.761 (0.031) & 0.719 (0.031) \\
HMM-funHDDC & 0.969 (0.012) & 0.676 (0.051) & 0.440 (0.103) \\
HMM-T-funHDDC & 0.969 (0.011) & 0.941 (0.025) & 0.886 (0.076) \\
HMM-C-funHDDC & 0.969 (0.011) & \textbf{0.970 (0.011)} & \textbf{0.961 (0.014)} \\
\midrule
FPCA-GMM & 0.669 (0.060) & 0.312 (0.135)$^{\dagger}$ & 0.292 (0.076)$^{\dagger}$ \\
FPCA-$t$MM & 0.646 (0.053) & 0.540 (0.071) & 0.348 (0.089) \\
FPCA-CMM & 0.648 (0.062) & 0.634 (0.074) & 0.592 (0.039) \\
FPCA-HMM & 0.954 (0.019) & 0.880 (0.095) & 0.402 (0.116)$^{\dagger}$ \\
FPCA-HMM-$t$ & 0.949 (0.022) & 0.927 (0.028) & 0.856 (0.083) \\
FPCA-HMM-C & 0.951 (0.020) & 0.948 (0.019) & 0.935 (0.021) \\
\bottomrule
\end{tabular}
\end{table}

\subsection{Simulation I: Markov persistence and robust emissions}\label{sec:sim1}
The regime path is a homogeneous Markov chain with $K=4$ regimes, sequences of length $L=40$ and separation $\mu=2$, with $\bm{d}=(2,4,6,8)$, $\bm{a}=(6,4,3,2)$ and $\bm{b}=(0.5,0.35,0.25,0.2)$. Each regime persists with probability $p_{kk}=0.92$, and the leaving mass is weighted towards the cyclic forward neighbor, with $p_{k,k+1}\approx0.053$ (indices modulo $K$) and about $0.013$ for each of the two other regimes. The transition matrix is doubly stochastic and the chain starts from its uniform stationary law, so that runs have mean length $12.5$ and a sequence contains on average $3.1$ regime changes (Figure~\ref{fig:sim1curves} in Appendix~\ref{app:detail-sec:sim1} shows the curves). The purpose of this experiment is to quantify the benefit of dependence when the true duration law is geometric, rather than to demonstrate an HSMM advantage. Table~\ref{tab:sim1} reports the main comparison. On clean data the HMM raises ARI from approximately 0.77 to 0.97. Under severe contamination the Gaussian HMM falls to 0.440, whereas the $t$ and contaminated-Gaussian HMMs attain 0.886 and 0.961, respectively. Thus temporal dependence cannot compensate for severe emission misspecification. State-specific subspaces improve on global FPCA for both robust HMMs across all scenarios, although the differences are smaller than those between exchangeable and temporally dependent models. The misspecified Gaussian comparison is not uniformly favorable, and some global-FPCA Gaussian fits are degenerate (Appendix~\ref{app:detail-sec:sim1}).

Figure~\ref{fig:sim1labels} shows the MAP labels on the first three sequences of replicate~1 in Scenario~B. The hidden Markov fits reproduce the true regime runs, whereas the i.i.d.\ fits fragment them. Over the whole replicate, T-funHDDC and C-funHDDC mislabel $119$ and $92$ of the $1000$ curves, against $15$ and $5$ for their hidden Markov counterparts.

\subsection{Simulation II: non-geometric durations}\label{sec:sim2}
The regime path is generated segment by segment as in Section~\ref{subsec:hsm}, with $K=3$ regimes, sequences of length $L=70$, $\bm{d}=(2,4,6)$, $\bm{a}=(6,4,2)$ and $\bm{b}=(0.5,0.35,0.2)$. The first segment starts at time $1$ in a uniformly drawn regime, the embedded chain has $p^{\ast}_{hk}=1/2$ for $h\neq k$, and the durations follow the shifted Poisson law $V_{ij}\mid B_{ij}=k\sim1+\mathrm{Pois}(\lambda_k)$. The window of $L=70$ curves right-censors the final segment. In Scenarios~A--C, $\mu=4$ and $\bm{\lambda}=(6,10,17)$, giving mean durations $7$, $11$ and $18$ with coefficients of variation $0.35$, $0.29$ and $0.23$, against $0.93$--$0.97$ for geometric durations with the same means. At $\mu=4$ most clean curves can be classified from their own shape. Scenario~D, the most informative test of explicit-duration modeling, therefore combines clean overlapping emissions ($\mu=1.5$) with short durations, $\bm{\lambda}=(3,5,8)$, giving mean durations $4$, $6$ and $9$ and about eleven segments per sequence (Figure~\ref{fig:sim2curves} in Appendix~\ref{app:detail-sec:sim2} shows the curves). Each emission is fitted under the i.i.d.\ mixture, the HMM and the HSMM with three sojourn families: the correctly specified shifted Poisson (P), the shifted negative binomial (NB) with size $r_k\in[0.05,50]$, and a nonparametric law (NP) on $\{1,\ldots,V_{\max}\}$. We take $V_{\max}=32$, both as the support bound of the NP sojourn and as the computational cutoff of the P and NB sojourns (Section~\ref{subsubsec:hsm-estep}); it excludes a fraction $7.5\times10^{-4}$ of the durations of the longest regime of Scenario~A and satisfies $V_{\max}\le L$, as required by the finite-horizon assumption of Proposition~\ref{prop:hsmm-id} (Appendix~\ref{app:hsmm-proof}). Table~\ref{tab:sim2core} reports Scenarios~D and~C, and Table~\ref{tab:sim2} in Appendix~\ref{app:detail-sec:sim2} the full comparison.

\begin{table}
\centering
\caption{Simulation II: selected mean ARI (standard deviation) over 100 replicates. Rows give the temporal layer, with Mixture for the i.i.d.\ mixture, the $t$ (T) or contaminated-Gaussian (C) emission and the P sojourn.}\label{tab:sim2core}
\begin{tabular}{@{}lcccc@{}}\toprule
&\multicolumn{2}{c}{Scenario D (clean overlap)}&\multicolumn{2}{c}{Scenario C (severe contamination)}\\
Method & funHDDC & Global FPCA & funHDDC & Global FPCA\\\midrule
Mixture-T & 0.755 (0.019)&0.609 (0.030)&0.770 (0.022)&0.522 (0.045)\\
HMM-T & 0.925 (0.012)&0.858 (0.020)&0.975 (0.015)&0.958 (0.053)\\
HSMM-T (P)&0.941 (0.010)&0.894 (0.016)&0.983 (0.006)&0.977 (0.007)\\
Mixture-C&0.762 (0.018)&0.621 (0.032)&0.913 (0.013)&0.869 (0.016)\\
HMM-C&0.926 (0.012)&0.865 (0.020)&0.991 (0.004)&0.984 (0.005)\\
HSMM-C (P)&0.942 (0.009)&0.897 (0.016)&0.993 (0.004)&0.989 (0.005)\\\bottomrule
\end{tabular}\end{table}

In Scenario D the HSMM with the correctly specified Poisson sojourn improves ARI over the HMM by 0.016--0.017, consistently across emissions. This is smaller than the gain from introducing temporal dependence, but is non-negligible when individual curves are ambiguous. Figure~\ref{fig:sim2labels} shows the MAP labels on the first two sequences of replicate~1 in Scenario~D. Over the whole replicate, T-funHDDC mislabels $181$ of the $1750$ curves, HMM-T-funHDDC $46$, and HSMM-T-funHDDC with P, NB and NP sojourns $33$, $34$ and $38$. With robust emissions under contamination the additional gain is at most 0.009 for $t$ and a negligible 0.002 for contaminated Gaussian, whose HMM already classifies almost perfectly. In Scenario D the advantage of state-specific subspaces over global FPCA persists under HSMM inference (roughly 0.04--0.05 ARI), and is larger than the incremental duration gain.
The correctly specified shifted-Poisson family recovers the duration distributions more accurately than the nonparametric family, which pays a substantial variance cost for its flexibility: in Scenario D, the fitted mean durations of HSMM-T-funHDDC have an absolute error of 0.208 curves and its fitted pmfs a total-variation distance of 0.036 from the true ones with the P sojourn, both averaged over the regimes and replicates, against 0.222 and 0.168 with the NP sojourn (Figure~\ref{fig:sim2sojourn}). All duration-recovery estimates (Table~\ref{tab:sim2soj}) and the negative-binomial comparisons are in Appendix~\ref{app:detail-sec:sim2}.

BIC recovers the generating structure whenever the emission accommodates the atypical curves. Among the funHDDC-family models with the same emission, it selects the HSMM with P sojourn in every replicate of the clean Scenarios~A and~D with every emission, and of the contaminated Scenarios~B and~C with the $t$ and contaminated-Gaussian emissions. It never selects the i.i.d.\ mixture or the NP sojourn, and in Simulation~I it selects the HMM over the i.i.d.\ mixture in every replicate. With the misspecified Gaussian emission under contamination, however, BIC prefers the HMM to every HSMM in every replicate. Because the cutoff $V_{\max}=32$ can place an HSMM likelihood below that of the nested HMM (Appendix~\ref{app:detail-sec:sim2}), this preference is not evidence about the sojourn law. Among the HSMMs, BIC prefers the NB to the P sojourn, although the true durations are P. The additional dispersion parameter absorbs part of the misfit caused by the atypical curves, so an overdispersed sojourn need not indicate non-geometric durations, a caution that recurs in the air-quality data of Section~\ref{sec:milano}. The marginal-entropy integrated completed likelihood (ICL, Appendix~\ref{subsec:hyperparameter-selection}) makes the same choices, except in three of these $200$ Gaussian replicates, where it prefers the HSMM (NB) to the HMM at BIC differences of one or two units.

\section{Applications}\label{sec:applications}
We analyze daily pollution profiles from the Milan area and continuous smartphone activity recordings, two ordered functional datasets with different persistence mechanisms, on held-out sequences with $K$ fixed for interpretability. Exchangeable mixtures, HMMs and HSMMs with robust emissions and their global-FPCA benchmarks, with $q$ chosen by the $90\%$ rule of Section~\ref{sec:sim-design}, are compared by BIC on the training sequences and by the held-out log-likelihood, that is, the log-likelihood of the held-out sequences under the fitted parameters. The intrinsic dimensions are selected by Cattell's scree test \citep{cattell1966scree} with threshold $0.2$ during the first ten iterations of every fit (Appendix~\ref{subsec:hyperparameter-selection}). External labels serve only as descriptive diagnostics. As a benchmark for the temporal layers, the MAP labels of the i.i.d.\ mixture are also smoothed by a running mode, which replaces each label by the most frequent label in a centered window of $15$ days (Milan) or $5$ windows (HAR); these widths were fixed in advance and not tuned on the held-out data. Preprocessing details and station-, subject- and basis-level results are in Appendix~\ref{app:application-details}.

\subsection{Air-quality regimes in the Milan area}\label{sec:milano}
\begin{figure}
\centering
\includegraphics[width=0.7\textwidth]{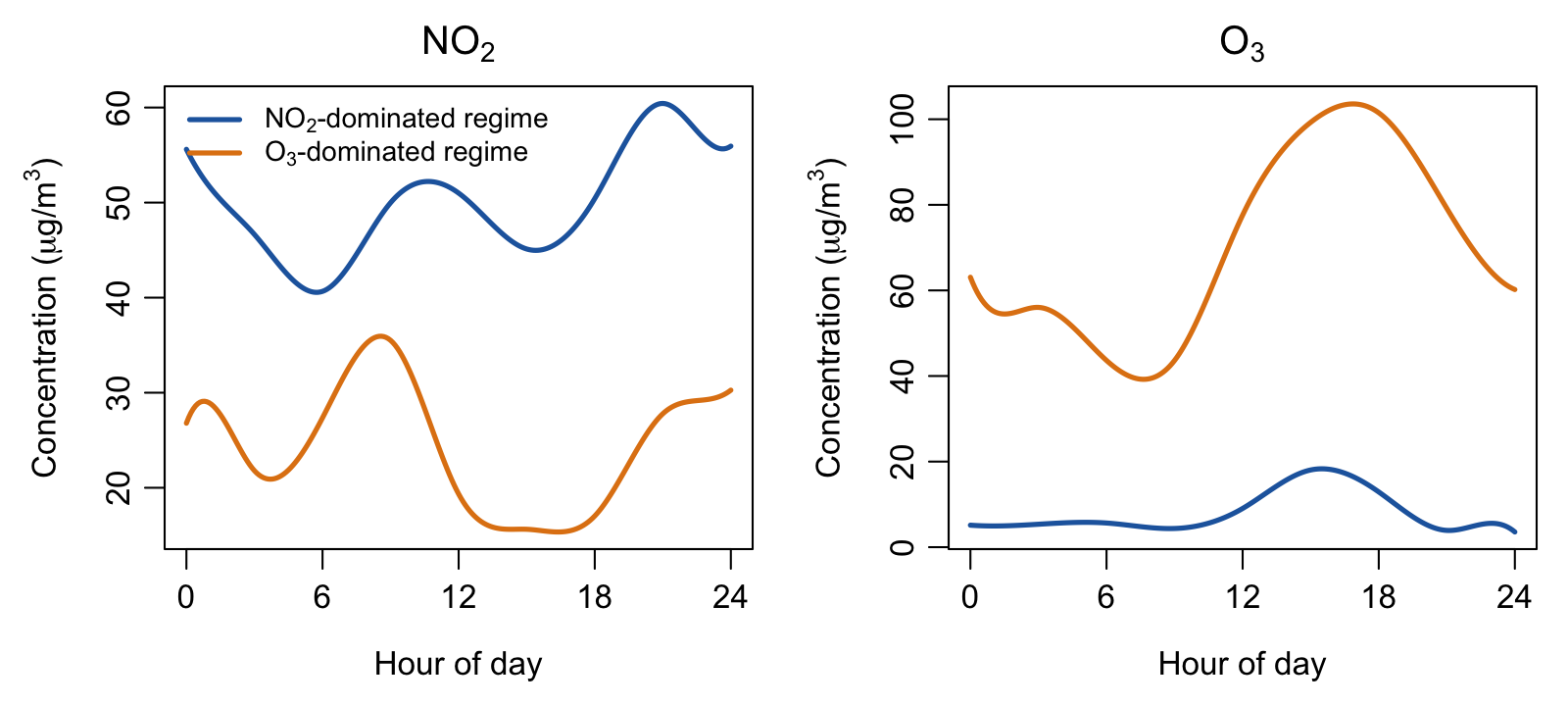}
\caption{Milan: regime mean curves of $\mathrm{NO_2}$ and $\mathrm{O_3}$ ($\mu\mathrm{g/m^3}$) under HSMM-T-funHDDC (NB). The HMM curves are indistinguishable.}
\label{fig:milanoprofiles}
\end{figure}

Daily $\mathrm{NO_2}$ and $\mathrm{O_3}$ profiles from four monitoring stations are represented by cubic B-splines ($R_s=11$ per pollutant). The training period is 2016--2022, and 2023--2026 is held out. Days with fewer than 18 valid hours for either pollutant remain in the calendar sequence as missing curves, which contribute no emission (Appendix~\ref{subsubsec:hsm-leftcens}). The first run of each station's sequence is left-censored and follows the residual-duration law of Appendix~\ref{subsubsec:hsm-leftcens}. The P and NB sojourns are evaluated with the cutoff $V_{\max}=366$ days (Section~\ref{subsubsec:hsm-estep}), above the longest completed MAP run of $227$ days (Appendix~\ref{app:detail-sec:milano}); re-evaluating the Milan HSMM fits of Tables~\ref{tab:milanoresults} and~\ref{tab:appsens} with $V_{\max}$ equal to the sequence length leaves their training and held-out log-likelihoods unchanged. Two regimes ($K=2$) describe an $\mathrm{NO_2}$-dominated pattern with morning/evening peaks and an $\mathrm{O_3}$-dominated pattern with an afternoon peak (Figure~\ref{fig:milanoprofiles}).

\begin{table}
\centering
\small
\caption{Milan air quality ($K=2$): BIC on 2016--2022 and log-likelihood of the held-out years 2023--2026. Cattell's test selects $d_k=(2,2)$, and global FPCA uses $q=5$ scores. BIC is reported as $\Lambda(\widehat{\bm\Theta})-(\tau/2)\log n_\bullet$ (larger is better). Best values in bold.}
\label{tab:milanoresults}
\begin{tabular}{@{}lcccc@{}}
\toprule
 & \multicolumn{2}{c}{BIC} & \multicolumn{2}{c}{Held-out log-lik.} \\
\cmidrule(lr){2-3}\cmidrule(l){4-5}
Temporal layer & $t$ & Contaminated & $t$ & Contaminated \\
\midrule
i.i.d.\ mixture & $-238845$ & $-247900$ & $-123873$ & $-129443$ \\
HMM & $-235451$ & $-245099$ & $-121895$ & $-127813$ \\
HSMM (P) & $-240841$ & $-247820$ & $-125061$ & $-130114$ \\
HSMM (NB) & $-235069$ & $-244604$ & $-121656$ & $-127511$ \\
HMM, heating-season transitions & $-235234$ & $-244805$ & $-121807$ & $-127696$ \\
HMM, monthly transitions & $\mathbf{-234876}$ & $\mathbf{-244376}$ & $\mathbf{-121570}$ & $\mathbf{-127495}$ \\
\midrule
\multicolumn{5}{@{}l}{\emph{Held-out agreement with the heating-season calendar (ARI)}} \\
 & \multicolumn{2}{c}{funHDDC} & \multicolumn{2}{c}{Global FPCA} \\
\cmidrule(lr){2-3}\cmidrule(l){4-5}
 & $t$ & Contaminated & $t$ & Contaminated \\
\midrule
i.i.d.\ mixture & 0.304 & 0.269 & 0.268 & 0.256 \\
i.i.d.\ mixture, running mode (15 days) & \textbf{0.386} & 0.353 & 0.340 & 0.318 \\
HMM & 0.344 & 0.341 & 0.345 & 0.329 \\
HSMM (P) & 0.370 & \textbf{0.401} & \textbf{0.410} & \textbf{0.412} \\
HSMM (NB) & 0.340 & 0.343 & 0.340 & 0.327 \\
\bottomrule
\end{tabular}
\end{table}

Among homogeneous temporal models, HSMM-T-funHDDC (NB) improves the held-out log-likelihood of HMM-T-funHDDC from $-121{,}895$ to $-121{,}656$, while the HSMM with P sojourn performs worse (Table~\ref{tab:milanoresults}); the NP sojourn is not fitted, because its left-censored version is not implemented (Appendix~\ref{subsubsec:hsm-leftcens}). The fitted NB sizes, $r_k=0.31$ and $0.15$, are far below the size $r_k=1$ of the geometric law and indicate substantial overdispersion. Yet this is not decisive evidence for intrinsic non-geometric persistence: a negative-binomial sojourn with size below one is a mixture of geometric sojourns and has a decreasing hazard (Remark~\ref{rem:nb-mixture} in Appendix~\ref{app:detail-sec:milano}), so these sizes are equally consistent with exit probabilities that vary across visits, as produced by seasonal transitions or by sub-regimes merged under the choice $K=2$. Table~\ref{tab:milanoresults} also reports HMMs whose transition matrix depends on a known calendar group, the heating season (15 October to 15 April) or the month, and the month-dependent HMM fits still better ($-121{,}570$ held-out log-likelihood), indicating that the homogeneous HSMM partly absorbs seasonal nonstationarity. Long warm-season runs coexist with many short episodes, which neither simple parametric sojourn family represents well (Appendix~\ref{app:detail-sec:milano}). Agreement with the heating-season calendar is modest.

\subsection{Human-activity regimes}\label{sec:har}
We use the continuous recordings of the Human Activities and Postural Transitions (HAPT) data \citep{ReyesOrtiz2016} rather than the activity-grouped, 50\%-overlapping windows of the original Human Activity Recognition (HAR) release \citep{Anguita2013}, whose file ordering is unsuitable for estimating state durations. The 61 continuous sessions from 30 volunteers are divided into nonoverlapping 2.56-second windows and represented as six-component functional observations (three acceleration and three angular-velocity channels). Each session is trimmed to its labeled protocol, so every sequence begins at the onset of a bout and the boundary-start convention of Assumption~\ref{id:A}\textup{(A5)} applies. All three sojourns use $V_{\max}=161$ windows, the length of the longest session, so for the P and NB sojourns each sequence uses $\min(V_{\max},L_i)=L_i$ and their likelihoods involve no truncation. The NP support $\{1,\ldots,161\}$ exceeds the longest training session, of $155$ windows, so the condition of Lemma~\ref{id:lemma-duration} that $V_{\max}$ not exceed the longest sequence fails on the training data, and the masses above $155$ enter the training likelihood only through their sum. No fitted run comes near this bound: the longest runs of the initial and fitted label sequences last $99$ and $89$ windows (Appendix~\ref{subsec:initialization}), so these masses receive weight only from censored final runs. The standard split by subject yields 42 training sessions from 21 subjects and 19 held-out sessions from 9 subjects. We impose three interpretable states: locomotion, upright static posture and lying. Windows spanning transitions or unlabeled periods remain in the sequences but are excluded from classification scoring.

\begin{table}
\centering
\small
\setlength{\tabcolsep}{4pt}
\caption{Human activity ($K=3$): BIC on the 42 training sessions and log-likelihood of the 19 held-out sessions. Cattell's test selects $d_k=(6,2,2)$ ($t$) and $(8,2,2)$ (contaminated) for locomotion, upright and lying, and global FPCA uses $q=25$ scores. BIC is reported as $\Lambda(\widehat{\bm\Theta})-(\tau/2)\log n_\bullet$ (larger is better). Best values in bold.}
\label{tab:harresults}
\begin{tabular}{@{}lcccc@{}}
\toprule
 & \multicolumn{2}{c}{BIC} & \multicolumn{2}{c}{Held-out log-lik.} \\
\cmidrule(lr){2-3}\cmidrule(l){4-5}
Temporal layer & $t$ & Contaminated & $t$ & Contaminated \\
\midrule
i.i.d.\ mixture & $-366503$ & $-451985$ & $-159925$ & $-193116$ \\
HMM & $-362960$ & $-450283$ & $-158609$ & $-192362$ \\
HSMM (P) & $-364897$ & $-451288$ & $-159388$ & $-192865$ \\
HSMM (NB) & $\mathbf{-362639}$ & $\mathbf{-450126}$ & $-158474$ & $-192297$ \\
HSMM (NP) & $-364421$ & $-451887$ & $\mathbf{-158452}$ & $\mathbf{-192257}$ \\
\midrule
\multicolumn{5}{@{}l}{\emph{Held-out agreement with the activity classes (ARI)}} \\
 & \multicolumn{2}{c}{funHDDC} & \multicolumn{2}{c}{Global FPCA} \\
\cmidrule(lr){2-3}\cmidrule(l){4-5}
 & $t$ & Contaminated & $t$ & Contaminated \\
\midrule
i.i.d.\ mixture & 0.842 & 0.891 & 0.470 & 0.654 \\
i.i.d.\ mixture, running mode (5 windows) & \textbf{0.889} & 0.902 & 0.489 & 0.686 \\
HMM & 0.858 & 0.906 & 0.669 & 0.682 \\
HSMM (P) & 0.839 & 0.871 & \textbf{0.685} & \textbf{0.694} \\
HSMM (NB) & 0.855 & 0.906 & 0.683 & 0.690 \\
HSMM (NP) & 0.846 & \textbf{0.909} & \textbf{0.685} & 0.692 \\
\bottomrule
\end{tabular}
\end{table}

The state-specific subspace representation provides the clearest empirical advantage: global-FPCA models have held-out ARI approximately 0.47--0.69, whereas robust funHDDC models attain 0.84--0.91 (Table~\ref{tab:harresults}). Temporal dependence improves predictive fit, but classification gains are modest. With $t$ emissions, the HMM increases the held-out log-likelihood by 1,316 relative to the i.i.d.\ mixture, and the HSMMs with NB and NP sojourns add 135 and 157, yet the running mode of the i.i.d.\ labels over five windows classifies better than every temporal model (ARI 0.889 against at most 0.858). The fitted durations characterize the experimental protocol, not natural activity behavior. Classification varies across subjects, and the preferred emission family depends on the basis dimension (Appendices~\ref{app:detail-sec:har} and~\ref{app:detail-sec:app-sens}).

\subsection{Robustness and interpretation}\label{sec:app-sens}
The sensitivity analyses (Appendix~\ref{app:detail-sec:app-sens}, Tables~\ref{tab:appsens} and~\ref{tab:milanstations}) distinguish robust conclusions from preprocessing-dependent classification. In Milan, each station separately favors the HSMM (NB) over the homogeneous HMM on held-out likelihood, although a calendar-dependent HMM is preferable overall. In HAR, overlapping windows exaggerate the apparent Markov contribution, and larger basis dimensions can lead the contaminated-Gaussian likelihood to favor classifications poorly aligned with the activity labels, while across all bases considered the HSMM (NB) retains a smaller positive gain in held-out log-likelihood over the HMM.

\section{Conclusion}
\label{sec:conclusion}

We have developed a family of hidden semi-Markov models
for clustering longitudinal multivariate functional
observations that combines state-specific functional
principal subspaces, Gaussian and robust emissions, and
explicit sojourn distributions within a likelihood-based
approach, with hidden Markov models as the
geometric-duration special case. Estimation exploits
low-rank functional covariance representations and
segmental forward--backward recursions, and the
identifiability results (Appendix~\ref{app:hsmm-proof})
clarify when the functional and semi-Markov parameters
can be recovered.
Accounting for latent temporal dependence generally
provides the largest improvement over exchangeable
functional clustering, especially when individual
curves are insufficiently informative. Robust
emissions protect the analysis against atypical
observations, and with robust emissions the
state-specific functional subspaces retain
discriminating variation that may be lost under
global dimension reduction; with a Gaussian
emission under contamination, the global FPCA
representation can classify better.
Explicit duration distributions provide additional
benefits when regimes overlap and their persistence
departs from geometricity, but the improvement is
not uniform. The applications confirm the value
of temporal modeling for predictive fit and
illustrate an important qualification: apparent
duration dependence, such as the overdispersed
sojourns of the air-quality regimes, may reflect
unmodeled time variation in the transition
mechanism or regimes merged by the choice of $K$.

The number of
states and the intrinsic dimensions were fixed in the simulations, and the
number of states was chosen for interpretability in the applications. The
global-FPCA comparison, on data generated from the funHDDC structure, does
not isolate the effect of state-specific subspaces alone
(Section~\ref{sec:sim-design}). Hidden semi-Markov models were not
fitted in Simulation~I, so the rate at which explicit-duration models are
selected under geometric sojourns was not assessed, and the sojourn
comparisons of Simulation~II used the computational cutoff $V_{\max}=32$.
Contaminated-Gaussian classification can be sensitive to the basis dimension
and atypical-observation weighting, and on clean data many
contaminated-Gaussian fits reached the iteration cap, where the
contamination parameters are weakly identified. The finite-support
nonparametric sojourn was not used with a left-censored initialization, and
the identifiability result does not cover left-censored windows. The
reported ICL uses marginal-state entropies, which make it conservative for
the temporal models relative to the i.i.d.\ mixture. Finally, the competitor
set does not encompass every existing functional hidden-state formulation.

Several extensions remain relevant. Calendar-dependent
semi-Markov dynamics would help distinguish
non-geometric durations from seasonal
nonstationarity. More flexible but parsimonious
duration distributions, heteroscedastic residual
covariance structures, automatic selection among
functional subspace parameterizations, dependence
between observational units, uncertainty in the
functional reconstruction, and scalable inference
for large longitudinal datasets also warrant
further work.

\section*{Disclosure Statement}
The authors report there are no competing interests to declare.


\bibliography{markov_hddc}
\newpage
\appendix
\counterwithin{equation}{section}

\section{Emission distributions and parameter updates}\label{app:emission-updates}
\subsection{Latent representations and E-step quantities}\label{app:emission-latent}
The three emissions of Section~\ref{subsec:emission-families} share the funHDDC scale matrix~\eqref{eq:covariance-funhddc}. The following derivations give the latent-variable representations of the robust emissions, the posterior expectations computed in the E-step and the M-step updates, which are common to the HMM and the HSMM.
The contaminated-Gaussian emission is obtained by introducing, for each functional observation, a latent variable $\Upsilon_{i\ell}\mid S_{i\ell}=k \sim \operatorname{Bernoulli}(\alpha_k)$, where $\Upsilon_{i\ell} = 1$ denotes an ordinary observation and $\Upsilon_{i\ell} = 0$ a contaminated one, with
\[
    \bm{c}_{i\ell}\mid \Upsilon_{i\ell}=1, S_{i\ell}=k \sim N_R(\bm{\mu}_k,\Sigma_k),\qquad \bm{c}_{i\ell}\mid \Upsilon_{i\ell}=0, S_{i\ell}=k \sim N_R(\bm{\mu}_k,\eta_k\Sigma_k).
\]
Here $\Sigma_k$ is the covariance matrix of the ordinary component, and the full regime covariance is $\operatorname{Cov}(\bm{c}_{i\ell}\mid S_{i\ell}=k) = \left(\alpha_k+(1-\alpha_k)\eta_k\right)\Sigma_k$. A value of $\alpha_k$ close to $1$ yields a nearly pure Gaussian component, while the restriction $\eta_k>1$ marks the inflated kernel.

For the multivariate-$t$ emission, the location $\bm{\mu}_k$ equals the mean when $\nu_k>1$, and the covariance exists when $\nu_k>2$ and equals $\nu_k\Sigma_k/(\nu_k-2)$. Thus $\Sigma_k$ is not itself the covariance matrix for finite $\nu_k$, although the eigenspaces of the two whitened matrices coincide. Equivalently, the $t$ emission is a scale mixture with latent $H_{i\ell}\mid S_{i\ell}=k \sim \text{Gamma}(\nu_k/2,\nu_k/2)$, in the shape-rate parameterization, and $\bm{c}_{i\ell}\mid H_{i\ell}=h_{i\ell},S_{i\ell}=k \sim N_R(\bm{\mu}_k,\Sigma_k/h_{i\ell})$.
For the contaminated Gaussian emission, the E-step computes, at the current parameter value $\bm{\Theta}^{(m)}$ of either temporal model,
\[
    \upsilon_{i\ell k}^{(m)} = \mathbb E\left[\Upsilon_{i\ell}\mid S_{i\ell}=k,\bm{c}_{i\ell};\bm{\Theta}^{(m)}\right]
\]
as
\[
    \frac{\alpha_k^{(m)} \phi_R\left(\bm{c}_{i\ell};\bm{\mu}_k^{(m)},\Sigma_k^{(m)}\right)}{\alpha_k^{(m)} \phi_R\left(\bm{c}_{i\ell};\bm{\mu}_k^{(m)},\Sigma_k^{(m)}\right) + \left(1-\alpha_k^{(m)}\right) \phi_R\left(\bm{c}_{i\ell};\bm{\mu}_k^{(m)},\eta_k^{(m)}\Sigma_k^{(m)}\right)}.
\]
For the $t$ emission, it computes
\[
h_{i\ell k}^{(m)} = \mathbb{E}\left[H_{i\ell}\mid S_{i\ell}=k,\bm{c}_{i\ell};\bm{\Theta}^{(m)}\right] = \frac{\nu_k^{(m)}+R}{\nu_k^{(m)}+\Delta_k^{(m)}(\bm{c}_{i\ell})},
\]
and
\[
    e_{i\ell k}^{(m)} = \mathbb{E}\left[\log H_{i\ell}\mid    S_{i\ell}=k,\bm{c}_{i\ell};\bm{\Theta}^{(m)}\right]
\]
as
\[
    \log h_{i\ell k}^{(m)} + \psi\left(\frac{\nu_k^{(m)}+R}{2}\right) - \log\left(\frac{\nu_k^{(m)}+R}{2}\right),
\]
where $\psi(\cdot)$ is the digamma function.

\subsection{M-step}\label{subsec:emission-updates}
Given the posterior state probabilities $\gamma_{i\ell k}^{(m)}$ from the HSMM E-step of Section~\ref{subsubsec:hsm-estep} or the HMM E-step of Appendix~\ref{subsubsec:hm-estep}, define $n_k^{(m)}=\sum_i\sum_{\ell=1}^{L_i}\gamma_{i\ell k}^{(m)}$. The updates below are the same for both temporal models, and only the E-step producing the $\gamma_{i\ell k}^{(m)}$ differs.
For Gaussian funHDDC emissions, $\bm{c}_{i\ell}\mid S_{i\ell}=k \sim N_R(\bm{\mu}_k,\Sigma_k)$, with the mean updated by
\[
    \bm{\mu}_k^{(m+1)} = \frac{\displaystyle\sum_{i=1}^{N} \displaystyle\sum_{\ell=1}^{L_i} \gamma_{i\ell k}^{(m)} \bm{c}_{i\ell}}{n_k^{(m)}}.
\] 
We then define the weighted scatter matrix as $\Omega_k^{(m+1)}$, and its whitened form $\tilde{\Omega}_k^{(m+1)}$, where
\[
    \Omega_k^{(m+1)} = \frac{1}{n_k^{(m)}} \displaystyle\sum_{i=1}^{N} \displaystyle\sum_{\ell=1}^{L_i} \gamma_{i\ell k}^{(m)} \left(\bm{c}_{i\ell}-\bm{\mu}_k^{(m+1)}\right) \left(\bm{c}_{i\ell}-\bm{\mu}_k^{(m+1)}\right)^\top.
\]
and $\tilde{\Omega}_k^{(m+1)} = W^{1/2}\Omega_k^{(m+1)}W^{1/2}$. The columns $\bm{u}_{k1}^{(m+1)}, \bm{u}_{k2}^{(m+1)}, \ldots,\bm{u}_{kd_k}^{(m+1)}$ are orthonormal eigenvectors of the symmetric matrix $\tilde{\Omega}_k^{(m+1)}$ associated with its $d_k$ largest eigenvalues $\lambda_{k1}^{(m+1)}\ge\lambda_{k2}^{(m+1)}\ge\cdots\ge\lambda_{kd_k}^{(m+1)}$. The variance parameters are updated by $a_{kj}^{(m+1)} = \lambda_{kj}^{(m+1)}$ for $j=1,\ldots,d_k$, and
\[
    b_k^{(m+1)} = \frac{\operatorname{tr}\left(\tilde{\Omega}_k^{(m+1)}\right) - \displaystyle\displaystyle\sum_{j=1}^{d_k} \lambda_{kj}^{(m+1)}}{R-d_k}.
\]
Only the leading $d_k$ eigenpairs and the trace of $\tilde{\Omega}_k^{(m+1)}$ are needed. The $O(Rd_k)$ cost of Section~\ref{subsec:emission-evaluation} concerns emission evaluation and excludes scatter construction and eigenpair computation, and $\Sigma_k^{(m+1)}$ follows from \eqref{eq:covariance-funhddc}.

For C-funHDDC, the precision weight is defined as
\[
    r_{i\ell k}^{C,(m)} = \upsilon_{i\ell k}^{(m)} + \frac{1-\upsilon_{i\ell k}^{(m)}}{\eta_k^{(m)}}.
\]
Then,
\[
    \alpha_k^{(m+1)} = \frac{\displaystyle\displaystyle\sum_{i=1}^{N} \displaystyle\displaystyle\sum_{\ell=1}^{L_i}\gamma_{i\ell k}^{(m)} \upsilon_{i\ell k}^{(m)}}{n_k^{(m)}},
\]
which in the reported numerical analysis are constrained by $\alpha_k\ge\alpha^\ast=0.75$ (and $\alpha_k<1$). The mean is updated by
\[
    \bm{\mu}_k^{C,(m+1)} = \frac{\displaystyle\sum_{i=1}^{N} \displaystyle\sum_{\ell=1}^{L_i} \gamma_{i\ell k}^{(m)} r_{i\ell k}^{C,(m)}\bm{c}_{i\ell}}{\displaystyle\sum_{i=1}^{N} \displaystyle\sum_{\ell=1}^{L_i} \gamma_{i\ell k}^{(m)} r_{i\ell k}^{C,(m)}}.
\]
The scatter matrix then becomes
\[
    \Omega_k^{C,(m+1)} = \frac{1}{n_k^{(m)}} \displaystyle\sum_{i=1}^{N} \displaystyle\sum_{\ell=1}^{L_i} \gamma_{i\ell k}^{(m)} r_{i\ell k}^{C,(m)} \left(\bm{c}_{i\ell} - \bm{\mu}_k^{C,(m+1)}\right)\left(\bm{c}_{i\ell} - \bm{\mu}_k^{C,(m+1)}\right)^\top.
\]
The eigenvectors, $a_{kj}$, and $b_k$ are obtained by applying the same eigendecomposition step to $\tilde{\Omega}_k^{C,(m+1)}$. With $\bm{\mu}_k^{C,(m+1)}$ and $\Sigma_k^{(m+1)}$ fixed, the inflation parameter is updated by
\[
    \eta_k^{(m+1)} = \max\left\{1+\varepsilon_\eta, \frac{\displaystyle\sum_{i=1}^{N} \displaystyle\sum_{\ell=1}^{L_i} \gamma_{i\ell k}^{(m)} \left(1-\upsilon_{i\ell k}^{(m)}\right) \Delta_k^{(m+1)}(\bm{c}_{i\ell})}{R\displaystyle\sum_{i=1}^{N} \displaystyle\sum_{\ell=1}^{L_i} \gamma_{i\ell k}^{(m)} \left(1-\upsilon_{i\ell k}^{(m)}\right)}\right\},
\]
where $\varepsilon_\eta>0$ is a small constant. 

For T-funHDDC, the mean is updated by
\[
    \bm{\mu}_k^{t,(m+1)} = \frac{\displaystyle\sum_{i=1}^{N} \displaystyle\sum_{\ell=1}^{L_i} \gamma_{i\ell k}^{(m)} h_{i\ell k}^{(m)}\bm{c}_{i\ell}}{\displaystyle\sum_{i=1}^{N} \displaystyle\sum_{\ell=1}^{L_i} \gamma_{i\ell k}^{(m)}h_{i\ell k}^{(m)}}.
\]
The scatter matrix is
\[
    \Omega_k^{t,(m+1)} = \frac{1}{n_k^{(m)}}\displaystyle\sum_{i=1}^{N} \displaystyle\sum_{\ell=1}^{L_i} \gamma_{i\ell k}^{(m)} h_{i\ell k}^{(m)} \left(\bm{c}_{i\ell}-\bm{\mu}_k^{t,(m+1)}\right) \left(\bm{c}_{i\ell}-\bm{\mu}_k^{t,(m+1)}\right)^\top.
\]
The eigenvectors, $a_{kj}$, and $b_k$ are obtained by applying the same eigendecomposition step to $\tilde{\Omega}_k^{t,(m+1)}$. The degrees of freedom are updated by solving
\[
    -\psi\left(\frac{\nu_k^{(m+1)}}{2}\right) + \log\left(\frac{\nu_k^{(m+1)}}{2}\right) + 1 + \frac{1}{n_k^{(m)}} \displaystyle\sum_{i=1}^{N} \displaystyle\sum_{\ell=1}^{L_i} \gamma_{i\ell k}^{(m)} \left(e_{i\ell k}^{(m)} - h_{i\ell k}^{(m)} \right) = 0.
\]
In practice, this equation is solved numerically over a bounded interval. If a common degrees-of-freedom parameter
$\nu_1=\cdots=\nu_K=\nu$ is imposed, then $\nu^{(m+1)}$ solves
\[
    -\psi\left(\frac{\nu^{(m+1)}}{2}\right) + \log\left(\frac{\nu^{(m+1)}}{2}\right) + 1 + \frac{1}{n_{\bullet}} \displaystyle\sum_{i=1}^{N} \displaystyle\sum_{\ell=1}^{L_i} \displaystyle\sum_{k=1}^{K} \gamma_{i\ell k}^{(m)}\left(e_{i\ell k}^{(m)} - h_{i\ell k}^{(m)}\right) = 0,
\]
where
\[
    n_{\bullet} = \displaystyle\sum_{i=1}^{N}L_i.
\] 

The parsimonious covariance variants are updated as in the funHDDC family, replacing ordinary mixture responsibilities by $\gamma_{i\ell k}^{(m)}$. For example, if $b$ is common across regimes, then
\[
    b^{(m+1)} = \frac{\displaystyle\sum_{k=1}^{K} n_k^{(m)}\left[\operatorname{tr}\left(\tilde{\Omega}_k^{(m+1)}\right) - \displaystyle\sum_{j=1}^{d_k} \lambda_{kj}^{(m+1)}\right]}{\displaystyle\sum_{k=1}^{K} n_k^{(m)}(R-d_k)}.
\]
If the first $d_k$ variance parameters are common within regime $k$, then
\[
    a_k^{(m+1)} = \frac{1}{d_k} \displaystyle\sum_{j=1}^{d_k} \lambda_{kj}^{(m+1)}.
\]
If the first $d_k$ variance parameters are common across all regimes, then
\[
    a^{(m+1)} = \frac{\displaystyle\sum_{k=1}^{K} n_k^{(m)} \displaystyle\sum_{j=1}^{d_k} \lambda_{kj}^{(m+1)}}{\displaystyle\sum_{k=1}^{K} n_k^{(m)}d_k}.
\] 

The preceding formulas give unconstrained candidate updates. When variance parameters are shared across regimes, these candidates need not preserve the signal-noise separation required by the model. For numerical estimation, we impose $b_k \ge \varepsilon_b$ and $a_{kj} \ge (1+\delta)b_k$, where $\varepsilon_b>0$ and $\delta>0$ are fixed throughout fitting. These restrictions preserve the theoretical condition $a_{kj}>b_k>0$. A candidate update is accepted when it satisfies these restrictions and the equalities defining the selected covariance variant. Otherwise, the variance parameters are updated jointly by constrained maximization of the covariance contribution to the expected complete-data log-likelihood.

\section{Hidden Markov benchmark: model and inference}
\label{app:hmm-model}

The hidden Markov model (HMM) constitutes the
geometric-duration special case of the functional
hidden semi-Markov model (HSMM) introduced in
Section~\ref{subsec:hsm}. We describe this
restriction and its likelihood-based inference
to document the benchmark models used in the
simulation studies and applications.
The HMM retains the longitudinal functional
representation of
Section~\ref{subsec:functional-representation},
the state-specific subspace structure of
Section~\ref{subsec:functional-subspaces},
and the Gaussian, contaminated-Gaussian and
multivariate-$t$ emission distributions of
Section~\ref{subsec:emission-families}.
The difference concerns exclusively the
stochastic process governing the latent
states. Consequently, the emission-parameter
updates are common to the two temporal models
and are collected in
Appendix~\ref{app:emission-updates}.

\subsection{Hidden Markov representation}
\label{app:hmm-definition}

Consider the $N$ longitudinal functional
sequences
\[
\mathcal D_i
=
\{\bm X_{i\ell}:\ell=1,\ldots,L_i\},
\qquad i=1,\ldots,N,
\]
and their coefficient representations
$\bm c_{i\ell}\in\mathbb R^R$, as defined
in Section~\ref{subsec:functional-representation}.
Let $S_{i\ell}\in\{1,\ldots,K\}$ denote the
latent state associated with the $\ell$-th
functional observation in sequence $i$.
Under the HMM specification, the latent
states form a homogeneous first-order
Markov chain,
\[
\Pr(S_{i\ell}=k\mid
S_{i1},\ldots,S_{i,\ell-1})
=
\Pr(S_{i\ell}=k\mid S_{i,\ell-1}),
\]
with initial distribution
\[
\rho_k=\Pr(S_{i1}=k),
\qquad
\rho_k\ge0,
\qquad
\sum_{k=1}^{K}\rho_k=1,
\]
and transition matrix
$P=(p_{hk})_{h,k=1}^{K}$, where
\begin{equation}
\label{eq:hmm-transition}
p_{hk}
=
\Pr(S_{i\ell}=k\mid S_{i,\ell-1}=h),
\qquad
p_{hk}\ge0,
\qquad
\sum_{k=1}^{K}p_{hk}=1.
\end{equation}
The sequences share the same model parameters
but are treated as independent, with no
transition between the end of one sequence
and the beginning of another.

The connection with the HSMM follows from
the distribution of consecutive visits to
the same state. For $p_{kk}<1$, the number
of successive observations spent in state
$k$ before a transition has the geometric
distribution
\begin{equation}
\label{eq:hmm-geometric}
\varphi_k(v)
=
(1-p_{kk})p_{kk}^{v-1},
\qquad v=1,2,\ldots,
\end{equation}
with survivor function
\[
\bar\varphi_k(v)
=
\Pr(V\ge v\mid B=k)
=
p_{kk}^{v-1}.
\]
The transition probabilities of the
embedded chain introduced in
Section~\ref{subsec:hsm} are
\begin{equation}
\label{eq:hmm-embedded}
p^\ast_{hk}
=
\frac{p_{hk}}{1-p_{hh}},
\qquad h\ne k,
\qquad
p^\ast_{hh}=0.
\end{equation}
Conversely, for an HSMM with geometric
sojourn distributions having state-specific
exit probabilities $q_k\in(0,1]$,
the corresponding HMM transition matrix is
\[
p_{kk}=1-q_k,
\qquad
p_{kh}=q_kp^\ast_{kh},
\quad h\ne k.
\]
The two representations therefore describe
the same latent-state process and induce
the same observed-data distribution when
they share the initial distribution and
state-dependent emission densities.

This equivalence also holds for finite
observation windows, provided the terminal
segment of the HSMM is treated as
right-censored. Specifically, an interior
segment contributes the geometric mass
\eqref{eq:hmm-geometric}, whereas a terminal
segment of observed length $v$ contributes
the survivor $\bar\varphi_k(v)$, without
requiring a transition beyond the final
observation. The corresponding likelihood
is therefore exactly the fixed-window HMM
likelihood.
The geometric distribution is memoryless:
the probability of leaving state $k$ is
$1-p_{kk}$, independently of the time
already spent in that state. By allowing
non-geometric sojourn distributions, the
HSMM relaxes precisely this restriction,
without modifying the functional emission
model.

\subsubsection{Relationship with exchangeable functional mixtures.}

The HMM also contains the ordinary
exchangeable functional mixture as a
special case. If
\[
p_{hk}=\pi_k,
\qquad h,k=1,\ldots,K,
\]
and $\rho_k=\pi_k$, where
$\pi_k\ge0$ and $\sum_k\pi_k=1$,
then the latent states are independent
and identically distributed, with
\[
\Pr(S_{i\ell}=k)=\pi_k.
\]
Consequently, the joint distribution
of the coefficient observations
factorizes as
\[
p(\bm c_{i1},\ldots,\bm c_{iL_i})
=
\prod_{\ell=1}^{L_i}
\left\{
\sum_{k=1}^{K}
\pi_k f_k(\bm c_{i\ell};\bm\theta_k)
\right\}.
\]
The ordinary funHDDC mixture and its
robust variants are therefore recovered
as exchangeable benchmarks when the
state-dependent emissions are retained
but longitudinal dependence is removed.

\subsubsection{Model complexity.}\label{app:hmm-complexity}

When the initial distribution and
transition matrix are freely estimated,
the HMM temporal layer contributes
\[
\tau_{\mathrm{HMM,temporal}}
=
(K-1)+K(K-1)
\]
free parameters, corresponding to the
initial probabilities and the $K$ rows
of the transition matrix.
If the initial distribution is constrained
to be the unique stationary distribution
of an irreducible transition matrix,
it is determined by $P$, and the
temporal contribution reduces to
$K(K-1)$ parameters.
The total number of parameters is obtained
by adding the relevant emission-parameter
count, including the covariance-sharing
restrictions, contamination parameters
or degrees of freedom when applicable.
These counts are described in
Appendix~\ref{subsec:hyperparameter-selection}.

\subsection{Likelihood and EM estimation}
\label{subsec:inference-hm}

Let
\[
\bm\Theta_{\mathrm H}
=
\{\bm\rho,P,\bm\theta_1,\ldots,\bm\theta_K\}
\]
denote the HMM parameters, conditional on
the selected number of states, intrinsic
dimensions and emission specification.
As in the HSMM, the functional observations
are conditionally independent given their
latent states, with
\[
\bm c_{i\ell}\mid S_{i\ell}=k
\sim f_k(\cdot;\bm\theta_k).
\]
For a latent state path
$\bm s_i=(s_{i1},\ldots,s_{iL_i})$,
the joint density of the observed
coefficients and the state path is
\begin{equation}
\label{eq:hmm-joint}
p(\bm c_i,\bm s_i;\bm\Theta_{\mathrm H})
=
\rho_{s_{i1}}
f_{s_{i1}}(\bm c_{i1};\bm\theta_{s_{i1}})
\prod_{\ell=2}^{L_i}
p_{s_{i,\ell-1},s_{i\ell}}
f_{s_{i\ell}}(\bm c_{i\ell};\bm\theta_{s_{i\ell}}),
\end{equation}
where
$\bm c_i=(\bm c_{i1},\ldots,\bm c_{iL_i})$.
Marginalizing over all possible state
paths gives the observed likelihood
contribution of sequence $i$,
\begin{equation}
\label{eq:hmm-likelihood}
\mathcal L_i(\bm\Theta_{\mathrm H})
=
\sum_{\bm s_i\in\{1,\ldots,K\}^{L_i}}
p(\bm c_i,\bm s_i;\bm\Theta_{\mathrm H}),
\end{equation}
and, by independence between sequences,
\[
\mathcal L(\bm\Theta_{\mathrm H})
=
\prod_{i=1}^{N}
\mathcal L_i(\bm\Theta_{\mathrm H}),
\qquad
\Lambda(\bm\Theta_{\mathrm H})
=
\sum_{i=1}^{N}
\log\mathcal L_i(\bm\Theta_{\mathrm H}).
\]

Let
$Z_{i\ell k}=\mathbf 1\{S_{i\ell}=k\}$,
as in the main text. The complete-data
log-likelihood, with the latent state
sequences treated as observed, is
\begin{align}
\label{eq:hmm-complete}
\Lambda_c(\bm\Theta_{\mathrm H})
={}&
\sum_{i=1}^{N}\sum_{k=1}^{K}
Z_{i1k}\log\rho_k
\\
&+
\sum_{i=1}^{N}
\sum_{\ell=2}^{L_i}
\sum_{h=1}^{K}\sum_{k=1}^{K}
Z_{i,\ell-1,h}Z_{i\ell k}
\log p_{hk}
\notag\\
&+
\sum_{i=1}^{N}
\sum_{\ell=1}^{L_i}
\sum_{k=1}^{K}
Z_{i\ell k}
\log f_k(\bm c_{i\ell};\bm\theta_k).
\notag
\end{align}
For Gaussian emissions this expression
provides the complete-data criterion
used by the EM algorithm. For
contaminated-Gaussian and multivariate-$t$
emissions, the latent contamination
indicators $\Upsilon_{i\ell}$ or scales
$H_{i\ell}$ are additionally introduced.
Their augmented conditional expectations
and emission updates are presented in
Appendix~\ref{app:emission-updates}.

\subsubsection{E-step: forward--backward recursions}
\label{subsubsec:hm-estep}

At iteration $m$, let
$\bm\Theta_{\mathrm H}^{(m)}$ denote
the current parameter estimates.
Define the forward probability
\[
F_{i\ell k}^{(m)}
=
p(\bm c_{i1},\ldots,\bm c_{i\ell},
S_{i\ell}=k;
\bm\Theta_{\mathrm H}^{(m)}).
\]
It is initialized by
\begin{equation}
\label{eq:hmm-forward-init}
F_{i1k}^{(m)}
=
\rho_k^{(m)}
f_k(\bm c_{i1};\bm\theta_k^{(m)}),
\end{equation}
and satisfies, for $\ell=2,\ldots,L_i$,
\begin{equation}
\label{eq:hmm-forward}
F_{i\ell k}^{(m)}
=
f_k(\bm c_{i\ell};\bm\theta_k^{(m)})
\sum_{h=1}^{K}
F_{i,\ell-1,h}^{(m)}
p_{hk}^{(m)}.
\end{equation}
Similarly, define the backward probability
\[
B_{i\ell k}^{(m)}
=
p(\bm c_{i,\ell+1},\ldots,\bm c_{iL_i}
\mid S_{i\ell}=k;
\bm\Theta_{\mathrm H}^{(m)}).
\]
With initialization
$B_{iL_i k}^{(m)}=1$,
the backward recursion is
\begin{equation}
\label{eq:hmm-backward}
B_{i\ell h}^{(m)}
=
\sum_{k=1}^{K}
p_{hk}^{(m)}
f_k(\bm c_{i,\ell+1};\bm\theta_k^{(m)})
B_{i,\ell+1,k}^{(m)},
\end{equation}
for $\ell=L_i-1,\ldots,1$.
The observed likelihood of sequence $i$
is obtained from the final forward
probabilities,
\[
\mathcal L_i(\bm\Theta_{\mathrm H}^{(m)})
=
\sum_{k=1}^{K}F_{iL_i k}^{(m)}.
\]
The forward and backward recursions are
implemented with scaling or on the
logarithmic scale to avoid numerical
underflow for long sequences.

The posterior probability that observation
$\ell$ belongs to state $k$ is
\begin{equation}
\label{eq:hmm-gamma}
\gamma_{i\ell k}^{(m)}
=
\Pr(S_{i\ell}=k
\mid\mathcal D_i;
\bm\Theta_{\mathrm H}^{(m)})
=
\frac{
F_{i\ell k}^{(m)}B_{i\ell k}^{(m)}
}{
\mathcal L_i(\bm\Theta_{\mathrm H}^{(m)})
}.
\end{equation}
For $\ell=2,\ldots,L_i$, the posterior
probability of a transition from
state $h$ to state $k$ is
\begin{equation}
\label{eq:hmm-omega}
\begin{aligned}
\omega_{i\ell hk}^{(m)}
&=
\Pr(S_{i,\ell-1}=h,S_{i\ell}=k
\mid\mathcal D_i;
\bm\Theta_{\mathrm H}^{(m)})\\
&=
\frac{
F_{i,\ell-1,h}^{(m)}
p_{hk}^{(m)}
f_k(\bm c_{i\ell};\bm\theta_k^{(m)})
B_{i\ell k}^{(m)}
}{
\mathcal L_i(\bm\Theta_{\mathrm H}^{(m)})
}.
\end{aligned}
\end{equation}
These posterior quantities satisfy
\[
\sum_{k=1}^{K}\gamma_{i\ell k}^{(m)}=1,
\qquad
\sum_{h=1}^{K}\sum_{k=1}^{K}
\omega_{i\ell hk}^{(m)}=1,
\]
and
\[
\sum_{k=1}^{K}\omega_{i\ell hk}^{(m)}
=
\gamma_{i,\ell-1,h}^{(m)}.
\]

The posterior state probabilities
$\gamma_{i\ell k}^{(m)}$ play the same role
as their segmental counterparts obtained
from the HSMM forward--backward recursions
in Section~\ref{subsubsec:hsm-estep}.
Once these probabilities are available,
both temporal models use the same
emission-parameter estimation formulas.
For contaminated-Gaussian and
multivariate-$t$ emissions, the additional
conditional expectations
$\upsilon_{i\ell k}^{(m)}$,
$h_{i\ell k}^{(m)}$, and
$e_{i\ell k}^{(m)}$
are computed from the current emission
parameters as in
Appendix~\ref{app:emission-updates}.

\subsubsection{M-step}
\label{subsubsec:hm-mstep}

The expected number of observations
assigned to state $k$ is
\[
n_k^{(m)}
=
\sum_{i=1}^{N}
\sum_{\ell=1}^{L_i}
\gamma_{i\ell k}^{(m)}.
\]
Maximizing the expected complete-data
log-likelihood with respect to the
initial distribution gives
\begin{equation}
\label{eq:hmm-rho-update}
\rho_k^{(m+1)}
=
\frac{1}{N}
\sum_{i=1}^{N}
\gamma_{i1k}^{(m)}.
\end{equation}
The transition probabilities are
updated by normalizing the posterior
expected transition counts:
\begin{equation}
\label{eq:hmm-P-update}
p_{hk}^{(m+1)}
=
\frac{
\displaystyle
\sum_{i=1}^{N}
\sum_{\ell=2}^{L_i}
\omega_{i\ell hk}^{(m)}
}{
\displaystyle
\sum_{i=1}^{N}
\sum_{\ell=1}^{L_i-1}
\gamma_{i\ell h}^{(m)}
}.
\end{equation}
The numerator is the expected number
of transitions from $h$ to $k$,
and the denominator is the expected
number of transitions originating
from state $h$. If the latter is
zero, the corresponding row is not
identified by the current E-step and
must be handled by the numerical
degeneracy rules of the fitting procedure.

For each emission family, the state-specific
parameters are updated using the posterior
weights $\gamma_{i\ell k}^{(m)}$ and,
where applicable, the auxiliary latent
expectations. The formulas are identical
to those used for the functional HSMM,
since the temporal model affects emission
estimation only through the posterior
state probabilities. The full updates
are provided in
Appendix~\ref{app:emission-updates}.
If the HMM is fitted subject to
stationarity of its initial distribution,
the unrestricted update
\eqref{eq:hmm-rho-update} is replaced by
an optimization respecting the
stationarity constraint. The
unrestricted updates above correspond
to the benchmark specification used
unless stated otherwise.

\subsection{Initialization and connection with HSMM fitting}
\label{app:hmm-initialization}

The HMM uses the same initial functional
partitions and emission-parameter
initialization procedures as the HSMM,
described in Appendix~\ref{subsec:initialization}.
The difference concerns the initialization
of the latent temporal parameters.
Let
\[
z_{i\ell k}^{(0)}
=
\mathbf{1}\!\left\{S_{i\ell}^{(0)}=k\right\},
\qquad
i=1,\ldots,N,\quad
\ell=1,\ldots,L_i,\quad
k=1,\ldots,K,
\]
With a positive smoothing constant
$\varepsilon_0$, the initial state
probabilities are
\begin{equation}
\label{eq:hmm-rho-init}
\rho_k^{(0)}
=
\frac{
\varepsilon_0+
\displaystyle\sum_{i=1}^{N}z_{i1k}^{(0)}
}{
K\varepsilon_0+N
}.
\end{equation}
The initial transition probabilities
are obtained from smoothed empirical
transition counts,
\begin{equation}
\label{eq:hmm-P-init}
p_{hk}^{(0)}
=
\frac{
\varepsilon_P+
\displaystyle\sum_{i=1}^{N}
\sum_{\ell=2}^{L_i}
z_{i,\ell-1,h}^{(0)}z_{i\ell k}^{(0)}
}{
K\varepsilon_P+
\displaystyle\sum_{i=1}^{N}
\sum_{\ell=2}^{L_i}
z_{i,\ell-1,h}^{(0)}
},
\end{equation}
where $\varepsilon_P>0$ is a small
smoothing constant.
The initial emission parameters are
computed from the corresponding
state assignments, with the robust
initialization steps for contaminated-
Gaussian and multivariate-$t$ emissions
applied as described in
Appendix~\ref{subsec:initialization}.

The fitted HMM also provides a natural
initialization for the HSMM with the
same emission family. Specifically,
its initial state distribution and
emission parameters are retained,
and its transition matrix is converted
into the embedded transition matrix
through \eqref{eq:hmm-embedded}.
The duration distribution in each state
is initialized to the geometric
distribution implied by
\eqref{eq:hmm-geometric}, or to an
appropriate approximation within the
selected sojourn family.
For the shifted negative-binomial
specification, setting the size
parameter to $r_k=1$ and the mean
parameter to
\[
m_k=\frac{p_{kk}}{1-p_{kk}}
\]
reproduces the geometric sojourn
distribution exactly.

This initialization is particularly
useful because the shifted
negative-binomial HSMM contains the
HMM as a special case. With the
same observation-window convention,
an unrestricted duration support,
and global maximizers of both
likelihoods, the maximized HSMM
log-likelihood cannot be smaller
than that of the corresponding HMM.
In numerical implementations,
finite duration truncation or
optimization restrictions may
affect this nesting relationship.
The use of fitted HMMs as starting
values for the HSMM therefore provides
both a computational strategy and
a useful diagnostic for comparing
the temporal specifications.

\section{Observation conventions, initialization and model selection}\label{app:estimation-details}\label{app:operational}
\subsection{Left-censored initial segments and missing curves}
\label{subsubsec:hsm-leftcens}

Two departures from the observation scheme of Section~\ref{subsec:hsm} arise in applications. First, a sequence defined by a calendar window starts inside an ongoing regime, so its first segment is left-censored. Conditionally on the state at the observation origin, we use the equilibrium forward-recurrence convention for the initial residual duration. This does not, by itself, impose the stationary state distribution of the full semi-Markov process. Under this convention, the observed length of the first segment follows the residual (forward-recurrence) law
\[
  \varphi^{0}_k(u;\bm{\beta}_k)=\frac{\bar{\varphi}_k(u;\bm{\beta}_k)}{E_k(\bm{\beta}_k)}, \qquad
  \bar{\varphi}^{0}_k(u;\bm{\beta}_k)=\sum_{v\ge u}\varphi^{0}_k(v;\bm{\beta}_k)=\frac{1}{E_k(\bm{\beta}_k)}\sum_{v\ge u}\bar{\varphi}_k(v;\bm{\beta}_k), \qquad u\ge1,
\]
where $E_k(\bm{\beta}_k)=\sum_{v\ge1}v\,\varphi_k(v;\bm{\beta}_k)$ is the mean duration. The left-censored version of the model replaces $\varphi^{\dagger}_k(u\mid u)$ by $\varphi^{0}_k(u)$ for a first segment that ends at $u<L_i$, and by $\bar{\varphi}^{0}_k(L_i)$ for a first segment that spans the whole window. Every other term, and $\rho_k$, now the probability of regime $k$ at the first observation time, are unchanged. Since $\sum_{u\ge1}\bar{\varphi}_k(u)=E_k$, the residual law sums to one, and $\mathcal{L}_i$ remains a density of the observations in the window. The forward, backward and posterior recursions carry over with this substitution, $\chi^{(m)}_{i0}(u,k)$ now carrying $\varphi^{0}_k(u)$ or $\bar{\varphi}^{0}_k(L_i)$. In the M-step, the first segments are removed from the counts of Section~\ref{subsubsec:hsm-mstep} and enter through counts of their own:
\[
\begin{aligned}
  \widehat{M}_k^{(m)}(u) &= \sum_{i=1}^{N}\sum_{r=u+1}^{L_i-1}\chi_i^{(m)}(r,u,k), &\qquad
  \widehat{M}_k^{\mathrm{cens},(m)}(u) &= \sum_{i:\,L_i>u}\chi_i^{(m)}(L_i,u,k),\\
  \widehat{M}_k^{0,(m)}(u) &= \sum_{i:\,L_i>u}\chi_{i0}^{(m)}(u,k), &\qquad
  \widehat{M}_k^{0,\mathrm{cens},(m)}(u) &= \sum_{i:\,L_i=u}\chi_{i0}^{(m)}(u,k).
\end{aligned}
\]
The duration parameter then maximizes
\[
\begin{aligned}
  \sum_{u}\Bigl[&\widehat{M}_k^{(m)}(u)\log\varphi_k(u;\bm{\beta}_k)+\widehat{M}_k^{\mathrm{cens},(m)}(u)\log\bar{\varphi}_k(u;\bm{\beta}_k)\\
  &+\widehat{M}_k^{0,(m)}(u)\log\varphi^{0}_k(u;\bm{\beta}_k)+\widehat{M}_k^{0,\mathrm{cens},(m)}(u)\log\bar{\varphi}^{0}_k(u;\bm{\beta}_k)\Bigr].
\end{aligned}
\]
A first segment thus contributes $\log\varphi^{0}_k(u)$ or $\log\bar{\varphi}^{0}_k(L_i)$ in place of its ordinary term, not in addition to it, and every later segment keeps its ordinary term, $\log\varphi_k(u)$ if it ends before $L_i$ and $\log\bar{\varphi}_k(u)$ if it is the terminal segment. The objective is maximized numerically for parametric duration families, and we do not use the left-censored version with the nonparametric family. For geometric durations $\varphi^{0}_k=\varphi_k$, by the memoryless property, so the hidden Markov likelihood is unchanged and the HMM remains nested in the negative-binomial HSMM. Second, a functional observation may be missing, for instance a day with too few valid hourly measurements. A missing curve contributes $f_k\equiv1$ in every state, so the latent path runs through it, and it is excluded from the emission updates and from $n_\bullet$. 

\subsection{Initialization}
\label{subsec:initialization}

The ECM algorithm requires initial values for the posterior state probabilities and the model parameters. Following the initialization strategy commonly used in the funHDDC family, we initialize the regime labels by applying a clustering method to the
pooled coefficient vectors
\[
    \mathcal{C} = \{\bm{c}_{i\ell}:i=1,\ldots,N,\ \ell=1,\ldots,L_i\}.
\]
A standard option is $k$-means, with trimmed $k$-means \citep{cuesta1997trimmed} as a more robust alternative when the data may contain outlying functional observations. Several random initializations are used, and the solution with the largest final BIC is retained. 

Let $S_{i\ell}^{(0)}$ be the initial state allocation and define $z_{i\ell k}^{(0)}=\mathbf 1\{S_{i\ell}^{(0)}=k\}$. The HMM initialization is given in Appendix~\ref{app:hmm-initialization}. For the hidden semi-Markov model, each initialized label sequence is collapsed into runs. Let $(\widehat b_{ij}^{(0)},\widehat V_{ij}^{(0)})$ for $ j=1,\ldots,\widehat M_i^{(0)}$ denote the initialized state and observed length of the $j$--th run of sequence $i$. Then,
\[
    \rho_k^{(0)} = \frac{\varepsilon_0 + \displaystyle\sum_{i=1}^{N} \bm{1}\!\left\{\widehat b_{i1}^{(0)}=k\right\}}{K\varepsilon_0+N},
\]
and, for $h\neq k$,
\[
    p_{hk}^{\ast(0)} = \frac{\varepsilon_P + \displaystyle\sum_{i=1}^{N} \displaystyle\sum_{j=2}^{\widehat{M}_i^{(0)}}\bm{1}\!\left\{\widehat{b}_{i,j-1}^{(0)}=h,\widehat{b}_{ij}^{(0)}=k\right\}}{(K-1)\varepsilon_P + \displaystyle\sum_{i=1}^{N} \displaystyle\sum_{j=2}^{\widehat{M}_i^{(0)}} \bm{1}\!\left\{\widehat{b}_{i,j-1}^{(0)}=h\right\}}.
\]
The duration distributions are initialized from the runs of the initial label sequences, treating the final run of each sequence as right-censored, that is, as a duration of at least its observed length. With left-censored sequences (Appendix~\ref{subsubsec:hsm-leftcens}), the first run instead follows the residual-duration law; its contribution is handled separately from the right-censored terminal run. For nonparametric durations, we start from a uniform distribution on $\{1,\ldots,V_{\max}\}$ and perform one update as described in Section~\ref{subsubsec:hsm-mstep}, using the run counts from the initial labels. We add $\varepsilon_\varphi$ to each resulting count before normalization. For parametric duration families, the initial parameters are obtained by maximizing the censored objective in the same subsection. If no finite maximizer exists, we retain a prespecified starting value. For a finitely supported NP sojourn, the support must accommodate all observed runs, including right-censored initial or terminal runs; a run longer than $V_{\max}$ is incompatible with that NP specification and cannot be repaired by administratively censoring it at $V_{\max}$.

The emission parameters are initialized by applying the Gaussian, contaminated Gaussian, or $t$-emission M-step formulas to the initial labels, with prespecified starting values $\nu_k^{(0)}$, $\alpha_k^{(0)}\in(\alpha^\ast,1)$ and $\eta_k^{(0)}>1$. Atypical curves inflate the scatter matrices of the initial clusters, and a contaminated-Gaussian fit started from such an inflated scatter may converge to a spurious maximum even from a good partition. For the $t$ and contaminated emissions we therefore also use a robust initial M-step. Within each initial cluster, the fraction $1-\alpha_k^{(0)}$ of curves with the largest squared Mahalanobis distances $\Delta_k(\bm{c}_{i\ell})$ is set aside, and the M-step is recomputed from the remaining curves. For C-funHDDC, $\eta_k^{(0)}$ is then set to the ratio of the mean squared distances of the set-aside and the retained curves, which estimates the inflation of the atypical curves.

\subsection{Model selection}
\label{subsec:hyperparameter-selection}

The number of states $K$, the intrinsic dimensions $d_k$, the parsimonious covariance structure, the use of state-specific or common degrees of freedom in the $t$-model, and the duration specification in the hidden semi-Markov model are treated as hyperparameters. The dimensions $d_k$ may be selected using the Cattell scree-test by comparing successive eigenvalue differences with a threshold $\varepsilon_{\mathrm{Cat}}$. Alternatively, one may perform a grid search over admissible values of $d_k$ and select the combination that maximizes the BIC. 

Let
\[
    n_{\bullet} = \displaystyle\sum_{i=1}^{N} L_i
\]
be the total number of functional observations. We use the BIC in the form
\[
    \mathrm{BIC} = \Lambda(\widehat{\bm{\Theta}}) - \frac{\tau}{2}\log n_{\bullet},
\]
where larger BIC values indicate the preferred model (equivalently, minimizing the conventional $-2\Lambda+\tau\log n_\bullet$ criterion), and $\Lambda(\widehat{\bm{\Theta}})$ is the maximized observed log-likelihood and $\tau$ is the number of free parameters. Let $\tau_U$ denote the number of identifiable orientation parameters with
\[
    \tau_U = \sum_{k=1}^{K} \left[Rd_k-\frac{d_k(d_k+1)}{2} - \sum_g\frac{m_{kg}(m_{kg}-1)}{2}\right],
\]
where $m_{kg}$ denote the sizes of the leading-eigenvalue blocks constrained to be equal by the selected covariance model and $\sum_g m_{kg}=d_k$. With unrestricted leading eigenvalues, the contribution is $Rd_k-d_k(d_k+1)/2$. When all leading eigenvalues within regime $k$ are equal, it reduces to $d_k(R-d_k)$. The corresponding HMM parameter count is given in Appendix~\ref{app:hmm-complexity}. For the HSMM-funHDDC model,
\[
    \tau_{\mathrm{HSMM}} = KR+\tau_U+\tau_D+(K-1)+K(K-2) + \tau_{\mathrm{dur}},
\]
where $\tau_D$ is the number of free eigenvalue parameters of the selected covariance structure, $K(K-2)$ is the number of free parameters in the embedded transition matrix $P^\ast$, and $\tau_{\mathrm{dur}}$ is the number of free duration parameters. For a nonparametric duration distribution on $\{1,\ldots,V_{\max}\}$, $\tau_{\mathrm{dur}} = K(V_{\max}-1)$, while a one-parameter family gives $\tau_{\mathrm{dur}}=K$ and a two-parameter family $\tau_{\mathrm{dur}}=2K$. For the most general Gaussian model $\text{HSMM-funHDDC}[a_{kj},b_k,U_k,d_k;\bm{\beta}_k]$, $\tau_D=\sum_{k=1}^{K}d_k+K$ and
\[
    \tau_{\mathrm{HSMM}} = KR + \displaystyle\sum_{k=1}^{K} d_k\left[R-\frac{d_k+1}{2}\right] + \left(\displaystyle\sum_{k=1}^{K}d_k+K\right) + (K-1)+K(K-2)+\tau_{\mathrm{dur}}.
\]
The contaminated-Gaussian version adds $2K$ parameters, $\alpha_k$ and $\eta_k$, and the $t$ version adds $K$ degrees of freedom, or one when they are common across regimes. 

In addition to BIC, we report a marginal-entropy approximation to the integrated completed likelihood (ICL) of \citet{Biernacki2000},
\[
  \mathrm{ICL} = \mathrm{BIC} + \sum_{i=1}^{N}\sum_{\ell=1}^{L_i}\sum_{k=1}^{K} \gamma_{i\ell k}\log(\gamma_{i\ell k}),
\]
where the posterior probabilities are evaluated at the fitted parameters and $0\log0=0$. This criterion subtracts the sum of marginal posterior state entropies from BIC, thereby penalizing uncertain state assignments. The sum of marginal entropies is used here in place of the joint posterior entropy of the state path. By subadditivity of entropy it is at least the joint entropy, with equality for the i.i.d.\ mixture, so this ICL is conservative for the temporal models relative to the i.i.d.\ mixture. Larger values of the resulting criterion are preferred.

\subsection{Convergence and classification}
\label{subsec:convergence-classification}

After initialization, the E-step and M-step are alternated until convergence. For the contaminated Gaussian models, the algorithm is an ECM algorithm because the inflation parameters $\eta_k$ are updated in a conditional maximization step. Let $\Lambda^{(m)} = \Lambda(\bm{\Theta}^{(m)})$ denote the observed log-likelihood at iteration $m$. We use the Aitken acceleration \citep{bohning1994distribution,mcnicholas2010serial}
\[
    a^{(m+1)} = \frac{\Lambda^{(m+2)}-\Lambda^{(m+1)}}{\Lambda^{(m+1)}-\Lambda^{(m)}},
\]
and the asymptotic estimate
\[
    \Lambda_{\infty}^{(m+2)} = \Lambda^{(m+1)} + \frac{\Lambda^{(m+2)}-\Lambda^{(m+1)}}{1-a^{(m+1)}}.
\]
The algorithm is terminated upon meeting the convergence condition below or reaching a maximum number of iterations $m_{\max}$. The Aitken criterion is used only when its quantities are finite, its denominators are numerically separated from zero, and $0\le a^{(m+1)}<1$. Convergence is declared when
\[
  0\le\Lambda_{\infty}^{(m+2)}-\Lambda^{(m+1)}\le\varepsilon.
\]
Otherwise, the stopping criterion is
\[
  0\le \Lambda^{(m+2)}-\Lambda^{(m+1)}\le \varepsilon.
\]
These checks apply only to successfully completed iterations. A likelihood decrease beyond numerical tolerance triggers a numerical check and recomputation, where an unresolved failure terminates the run without declaring convergence. Reaching $m_{\max}$ is likewise recorded as termination without convergence.

For pointwise regime classification, the maximum a posteriori (MAP) rule assigns
\[
    \widehat S_{i\ell} = \arg\max_{k=1,\ldots,K} \gamma_{i\ell k}.
\]
Because the observed likelihood is invariant to relabeling of the regimes, each fitted model is post-processed to a canonical ordering of the states before its parameters are reported or aggregated across restarts, resolving in practice the label-switching ambiguity of Proposition~\ref{prop:hsmm-id}. For sequence-level decoding, the segmental Viterbi algorithm gives the most likely sequence of regimes and durations, and the ordinary Viterbi algorithm the most likely state path of the HMM. For C-funHDDC emissions, the posterior probability that the functional observation
$\bm{c}_{i\ell}$ is contaminated is
\[
    \Pr(\Upsilon_{i\ell}=0\mid\mathcal D_i) = \displaystyle\sum_{k=1}^{K} \gamma_{i\ell k}(1-\upsilon_{i\ell k}).
\]

\section{Identifiability of the functional hidden-state models}\label{app:hsmm-proof}\label{sec:identifiability}

We establish sufficient conditions for identifying the
parameters of the proposed functional HSMM from the
distribution of the observed longitudinal sequences.
The results concern a known number of states $K$ and
a fixed, finite-dimensional functional representation,
with positive-definite Gram matrix $W$.
Identifiability must account for three distinct
components: the state-dependent emission distributions,
the functional principal subspaces, and the semi-Markov
dynamics governing state transitions and durations.
The conditions and principal result are stated here for completeness; the main article summarizes their statistical interpretation in Section~\ref{subsec:identifiability-summary}.

\subsection{Assumptions and principal result}\label{app:hsmm-assumptions}
\begin{assumption}[HSMM identifiability]
\label{id:A}
\leavevmode
\begin{enumerate}

\item[\textup{(A1)}]
The initial state distribution satisfies
$\rho_k>0$ for every $k=1,\ldots,K$.
The embedded transition matrix
$P^\ast=(p^\ast_{hk})$ is row-stochastic,
with $p^\ast_{hh}=0$.
Conditional on the current segment state,
the segment duration and the destination
state are independent, as specified in
Section~\ref{subsec:hsm}.

\item[\textup{(A2)}]
The state-dependent emission densities
$f_1,\ldots,f_K$ are pairwise distinct.
For contaminated-Gaussian emissions, the
$2K$ Gaussian kernels
\[
\left\{
(\bm\mu_k,\Sigma_k),
(\bm\mu_k,\eta_k\Sigma_k)
\right\}_{k=1}^K
\]
are pairwise distinct, with
$0<\alpha_k<1$ and $\eta_k>1$.
In addition, the state-specific means
$\bm\mu_1,\ldots,\bm\mu_K$ are pairwise
distinct in the contaminated-Gaussian case.

\item[\textup{(A3)}]
For every state $k$, the leading eigenvalues
of the whitened scale matrix satisfy
\[
a_{k1}\ge\cdots\ge a_{kd_k}>b_k>0,
\qquad 1\le d_k\le R-1.
\]
Repeated leading eigenvalues are permitted,
with the associated eigenspaces identified
rather than individual eigenvectors.

\item[\textup{(A4)}]
Let $L_{\max}=\max_i L_i$. For each state
$k$, the sojourn distribution assigns
positive probability to at least one
duration in $\{1,\ldots,L_{\max}-1\}$,
and the mapping
\[
\bm\beta_k\longmapsto
\bigl(
\varphi_k(1;\bm\beta_k),\ldots,
\varphi_k(L_{\max}-1;\bm\beta_k)
\bigr)
\]
is injective over the specified duration
parameter space. For a nonparametric
distribution supported on
$\{1,\ldots,V_{\max}\}$, this condition
holds when $V_{\max}\le L_{\max}$.

\item[\textup{(A5)}]
Each observed sequence begins at a state-entry
time, so that its first segment has the
state-specific full-duration distribution
$\varphi_k$. The final segment may be
right-censored by the observation window,
as specified in Section~\ref{subsec:hsm}.

\end{enumerate}
\end{assumption}

\begin{proposition}[Identifiability of the functional HSMM]
\label{prop:hsmm-id}

Let $K\ge2$ be known and suppose that
Assumption~\ref{id:A} holds.
Then the initial distribution $\bm\rho$,
the embedded transition matrix $P^\ast$,
the state-specific duration parameters
$\{\bm\beta_k\}_{k=1}^K$, and the
state-dependent emission distributions
$\{f_k\}_{k=1}^K$ are identifiable
from the joint distribution of the observed
coefficient sequences, up to a common
permutation of the state labels.
For the funHDDC emission parameterization,
the state-specific locations $\bm\mu_k$,
intrinsic dimensions $d_k$, leading scale
eigenvalues $a_{kj}$, residual scale
eigenvalues $b_k$, and spectral projectors
onto the leading functional subspaces are
also identifiable.

When the leading eigenvalues are simple,
the columns of $U_k$ are identifiable
up to sign changes. With repeated leading
eigenvalues, the corresponding eigenspaces
are identifiable, but their orthonormal
bases remain indeterminate up to rotations
within the equal-eigenvalue blocks.
The contamination parameters
$(\alpha_k,\eta_k)$ or the degrees of
freedom $\nu_k$ are identifiable for the
corresponding emission families.
\end{proposition}

The result follows by first identifying the
state-dependent emission distributions from
the marginal mixture distribution. The
Gaussian and multivariate-$t$ families admit
identifiable finite mixtures, while the
additional condition on the means of the
contaminated-Gaussian emissions ensures
that their component kernels can be paired
unambiguously within states. Linear
independence of the identified emission
densities then determines the finite-dimensional
distributions of the latent state process
from those of the observed coefficients.

Because the observation window begins at a
state-entry time, the probability of an initial
run of length $u$ in state $k$, followed by
state $h\ne k$, satisfies
\[
\Pr(S_{i1}=\cdots=S_{iu}=k,
    S_{i,u+1}=h)
=
\rho_k\varphi_k(u;\bm\beta_k)p^\ast_{kh}.
\]
Consequently, the observable state-path
distributions identify the duration masses
within the observation horizon and the
embedded transition probabilities. The
injectivity condition in
Assumption~\ref{id:A}\textup{(A4)} then identifies
the duration parameters. Finally, the
eigendecomposition of
$W^{1/2}\Sigma_kW^{1/2}$
identifies the intrinsic dimensions,
scale eigenvalues, and corresponding
spectral subspaces.
The following subsections give the auxiliary emission and spectral
lemmas, the identification proof, and explicit checks for the duration
families used in the experiments.

\subsection{Emission and spectral identification lemmas}\label{app:hsmm-lemmas}
\begin{lemma}[Linear independence of the emission densities]\label{id:lemma-li}
Under the emission conditions in Assumption~\ref{id:A}\textup{(A2)}, the densities $f_1,\ldots,f_K$ are linearly independent for each of the Gaussian, contaminated-Gaussian and multivariate-$t$ families.
\end{lemma}

\begin{proof}
Identifiability of finite mixtures of multivariate Gaussian distributions \citep{Yakowitz1968} and of multivariate-$t$ distributions with varying location, scale and degrees of freedom \citep{Holzmann2006} implies linear independence of their respective kernel families through the characterization of finite-mixture identifiability in \citet{Yakowitz1968}. Thus distinct Gaussian or multivariate-$t$ state densities are linearly independent.
For contaminated-Gaussian emissions, write
\[
 f_k=\alpha_k\phi_R(\cdot;\bm\mu_k,\Sigma_k)
       +(1-\alpha_k)\phi_R(\cdot;\bm\mu_k,\eta_k\Sigma_k).
\]
Assumption~\ref{id:A}\textup{(A2)} makes all $2K$ Gaussian kernels pairwise distinct, hence linearly independent. Suppose $\sum_{k=1}^K t_k f_k=0$ almost everywhere. In the expansion over the $2K$ distinct Gaussian kernels, the coefficients of the two kernels associated with state $k$ are $t_k\alpha_k$ and $t_k(1-\alpha_k)$. Their vanishing, with $0<\alpha_k<1$, implies $t_k=0$ for every $k$. The $K$ contaminated state densities are therefore linearly independent.
\end{proof}

\begin{lemma}[Spectral uniqueness of the funHDDC scale structure]\label{id:lemma}
Let $\Sigma_k$ have the form in \eqref{eq:covariance-funhddc}, and suppose Assumption~\ref{id:A}\textup{(A3)} holds. Then $\Sigma_k$ determines $b_k$, $d_k$, the ordered leading scale eigenvalues $a_{k1},\ldots,a_{kd_k}$, and the leading spectral projector $U_kU_k^\top$. For each distinct leading eigenvalue, its spectral projector is identified. A simple leading eigenvalue determines its unit eigenvector up to sign; a repeated leading eigenvalue determines only its eigenspace, with orthonormal bases equivalent up to rotation within that eigenspace.
\end{lemma}

\begin{proof}
Set $\widetilde\Sigma_k=W^{1/2}\Sigma_kW^{1/2}$. By construction,
\[
 \widetilde\Sigma_k
 =U_k A_k U_k^\top+b_k(I_R-U_kU_k^\top),
\]
with $U_k^\top U_k=I_{d_k}$ and $a_{kj}>b_k$. Therefore $b_k$ is the smallest eigenvalue of $\widetilde\Sigma_k$ and
\[
 d_k=\operatorname{rank}(\widetilde\Sigma_k-b_kI_R),
 \qquad
 U_kU_k^\top=\mathbf 1_{(b_k,\infty)}(\widetilde\Sigma_k).
\]
Here $\mathbf 1_{(b_k,\infty)}(\widetilde\Sigma_k)$ denotes its spectral projector onto the direct sum of eigenspaces with eigenvalues larger than $b_k$. These identities hold even if $b_k$ has multiplicity one, as happens when $d_k=R-1$, and even if some of the leading eigenvalues are tied. The multiset of eigenvalues strictly above $b_k$ recovers $a_{k1},\ldots,a_{kd_k}$ with multiplicities. Uniqueness of each eigenspace projector follows from the spectral theorem. An orthonormal basis of a one-dimensional eigenspace is unique up to sign, while two orthonormal bases of a higher-dimensional eigenspace differ by an orthogonal rotation.
\end{proof}


\subsection{Proof of the HSMM identifiability proposition}\label{app:hsmm-main-proof}
\begin{proof}[Proof of Proposition~\ref{prop:hsmm-id}]
We separate recovery of the emission distributions, recovery of the latent state-path probabilities, identification of the semi-Markov dynamics, and recovery of the functional parameters.
\emph{Step 1: emissions and the initial state distribution.}
Under the boundary-start convention of Assumption~\ref{id:A}\textup{(A5)}, the first coefficient vector has marginal density
\begin{equation}\label{id:hsmm-marginal}
 q(\bm c)=\sum_{k=1}^K\rho_k f_k(\bm c),
 \qquad \rho_k>0.
\end{equation}
In the Gaussian and multivariate-$t$ cases, finite-mixture identifiability \citep{Yakowitz1968,Holzmann2006} determines the unordered collection $\{(\rho_k,f_k)\}_{k=1}^K$ from $q$. In the contaminated-Gaussian case,
\[
 q(\bm c)=\sum_{k=1}^K\Bigl[
 \rho_k\alpha_k\phi_R(\bm c;\bm\mu_k,\Sigma_k)
 +\rho_k(1-\alpha_k)\phi_R(\bm c;\bm\mu_k,\eta_k\Sigma_k)
 \Bigr].
\]
The $2K$ kernels are pairwise distinct by Assumption~\ref{id:A}\textup{(A2)}, so their weights and parameters are identified up to permutation. The same assumption requires distinct state means in the contaminated case: consequently, exactly the two kernels from a given state share a mean, identifying their pairing. The condition $\eta_k>1$ distinguishes the ordinary kernel (the smaller scale matrix) from the inflated kernel. Summing their weights recovers $\rho_k$, and their weight ratio recovers $\alpha_k$. The identified pair then determines $f_k$. Hence, for all three emission families, $\bm\rho$ and the $f_k$ are determined up to the same state-label permutation. Fix one such labeling for the remainder of the proof.

\emph{Step 2: latent state-path probabilities.}
For an observed sequence of length at least $n$, conditional independence given the state sequence yields
\begin{equation}\label{eq:hsm-tensor}
 p^{(n)}(\bm c_1,\ldots,\bm c_n)
 =\sum_{\bm s\in\{1,\ldots,K\}^n}
 w_n(\bm s)\prod_{\ell=1}^n f_{s_\ell}(\bm c_\ell),
 \qquad
 w_n(\bm s)=\Pr(S_{i1}=s_1,\ldots,S_{in}=s_n).
\end{equation}
By Lemma~\ref{id:lemma-li}, $f_1,\ldots,f_K$ are linearly independent. Their $n$-fold tensor products are likewise linearly independent: expanding a putative linear relation and successively applying linear independence in each coordinate forces every coefficient to vanish. Thus the coefficients $w_n(\bm s)$ in \eqref{eq:hsm-tensor} are uniquely determined by the observed joint density, for every $1\le n\le L_{\max}$ (using the marginal of a sequence with length at least $n$). This identifies the finite-horizon law of the latent state process. No HMM assumption is used.

\emph{Step 3: durations and embedded transitions.}
Because each sequence begins at a segment boundary, the first segment satisfies $B_{i1}\sim\bm\rho$ and $V_{i1}\mid B_{i1}=k\sim\varphi_k(\cdot;\bm\beta_k)$. Under the semi-Markov factorization, its duration and the destination of the next segment are conditionally independent given $B_{i1}$. The embedded chain has no self-transitions. Consequently, for $h\ne k$ and $1\le u<L_{\max}$,
\begin{equation}\label{id:hsm-run}
 \Pr(S_{i1}=\cdots=S_{iu}=k,\ S_{i,u+1}=h)
 =\rho_k\varphi_k(u;\bm\beta_k)p^\ast_{kh}.
\end{equation}
The left-hand side is determined by Step 2. Summing over all destinations gives
\begin{equation}\label{id:hsm-duration-recovery}
 \varphi_k(u;\bm\beta_k)
 =\frac{1}{\rho_k}\sum_{h\ne k}
 \Pr(S_{i1}=\cdots=S_{iu}=k,\ S_{i,u+1}=h),
 \qquad 1\le u<L_{\max}.
\end{equation}
Assumption~\ref{id:A}\textup{(A4)} ensures that at least one such $u$ has positive mass for each $k$. For any such $u$, the ratio
\begin{equation}\label{id:hsm-transition-recovery}
 p^\ast_{kh}=
 \frac{\Pr(S_{i1}=\cdots=S_{iu}=k,\ S_{i,u+1}=h)}
 {\sum_{g\ne k}\Pr(S_{i1}=\cdots=S_{iu}=k,\ S_{i,u+1}=g)}
\end{equation}
identifies every entry of row $k$ of $P^\ast$. The probabilities $\varphi_k(1),\ldots,\varphi_k(L_{\max}-1)$ are identified from \eqref{id:hsm-duration-recovery}; their complement identifies the survivor $\Pr(V_{i1}\ge L_{\max}\mid B_{i1}=k)$. Injectivity in Assumption~\ref{id:A}\textup{(A4)} now identifies $\bm\beta_k$. No full-rank or irreducibility assumption on $P^\ast$ is needed: positivity of all $\rho_k$ makes every initial-state experiment observable in the population law.

\emph{Step 4: functional parameters.}
Once $f_k$ is known, its location $\bm\mu_k$ and scale matrix $\Sigma_k$ are identified. For Gaussian emissions these are the mean and covariance. For multivariate-$t$ emissions they are the location and \emph{scale}, together with the degrees of freedom $\nu_k$; no existence of the $t$ covariance is required. For contaminated-Gaussian emissions the kernel decomposition in Step 1 identifies $(\bm\mu_k,\Sigma_k,\alpha_k,\eta_k)$. Lemma~\ref{id:lemma} identifies the intrinsic dimension, leading and residual scale eigenvalues and leading spectral projector from $\Sigma_k$, with the prescribed eigenvector/sign or within-eigenspace rotational indeterminacies. This completes the proof.
\end{proof}

\subsection{Identifiability of the duration families}\label{app:duration-identification}
\begin{lemma}[Shifted Poisson, negative-binomial and finite-support durations]
\label{id:lemma-duration}
If $L_{\max}\ge2$, the shifted Poisson parameter $\lambda_k$ is identified by the first duration mass. If $L_{\max}\ge3$, the shifted negative-binomial parameters $(m_k,r_k)$ are identified by the first two duration masses for $m_k,r_k>0$. For a nonparametric duration distribution supported on $\{1,\ldots,V_{\max}\}$, the finite-horizon mass vector identifies the entire distribution if and only if $V_{\max}\le L_{\max}$ (apart from restrictions on the model class that might otherwise enforce identification).
\end{lemma}

\begin{proof}
Condition~\textup{(A4)} is about identifiability from the \emph{finite observation horizon}; it does not require observing complete realizations of every possible sojourn. We verify it for the duration families used in the manuscript, assuming parameters lie in the interior of their stated parameter spaces.
For the shifted Poisson law, $\varphi_k(1)=\exp(-\lambda_k)$, so $\lambda_k=-\log\varphi_k(1)$.

For the shifted negative-binomial law, suppress the state subscript and put
\[
 a=\varphi(1)=\left(\frac{r}{r+m}\right)^r,
 \qquad
 h=\frac{\varphi(2)}{\varphi(1)}=\frac{rm}{r+m}.
\]
Both $a$ and $h$ are determined by the first two masses, with $0<h<r$. Eliminating $m$ gives
\[
 a=\left(1-\frac{h}{r}\right)^r,
 \qquad r>h,
 \qquad m=\frac{hr}{r-h}.
\]
For fixed $h>0$, the logarithm of the right-hand side is strictly increasing in $r>h$, because
\[
 \frac{\mathrm d}{\mathrm dr}
 \left[r\log\left(1-\frac{h}{r}\right)\right]
 =\log(1-x)+\frac{x}{1-x}>0,
 \qquad x=\frac{h}{r}\in(0,1),
\]
where the inequality follows, for example, by differentiating the right-hand side in $x$ and noting it vanishes at $x=0$. Therefore $a$ and $h$ determine a unique finite $r$, and then $m$. The Poisson limit $r=\infty$ is a boundary point and is not part of this two-parameter interior model.
For the nonparametric model, if $V_{\max}\le L_{\max}$, the masses $\varphi(1),\ldots,\varphi(V_{\max}-1)$ are observed through the identified state law and the final mass follows by normalization. If $V_{\max}>L_{\max}$, distinct distributions can assign the same masses to durations below $L_{\max}$ and distribute the remaining mass differently over durations $L_{\max},\ldots,V_{\max}$; hence the unrestricted nonparametric distribution is not identifiable.
\end{proof}

\subsection{Scope and interpretation of the identification assumptions}\label{app:hsmm-remarks}
\begin{remark}[Boundary-start versus left-censored sequences]\label{id:rem-boundary}
The argument for \eqref{id:hsm-run} depends on an observed sequence starting at a segment-entry time. For an observation window beginning inside an ongoing segment, the first \emph{observed} run follows its residual-duration distribution. Its mass must replace $\varphi_k(u)$ in the first-run identity, and the preceding proof does not directly establish identification under that sampling convention. If $\rho_k$ is defined as the state probability at the observation origin, $\Pr(S_{i1}=k)=\rho_k$ still holds; it is the first-run duration law, not necessarily the initial state marginal, that changes. Interior completed runs may provide identification information, but an extension requires its own argument.
\end{remark}

\begin{remark}[Identification of contaminated-Gaussian emissions]
\label{id:rem-contam-pairing}
For contaminated-Gaussian emissions, identifiability of the
marginal Gaussian mixture determines its $2K$ component
distributions, but does not by itself establish their
pairing into $K$ state-dependent emissions. The
distinct-means condition in Assumption~\ref{id:A}\textup{(A2)} provides a sufficient way to resolve this
ambiguity: the two Gaussian components associated with
each state share the same mean, while different states
have distinct means. The constraint $\eta_k>1$ then
distinguishes the ordinary and inflated components
within each pair.

This condition is sufficient for the marginal-based
identification argument adopted here, but is not
claimed to be necessary for identifiability of the
full HSMM. In particular, joint distributions of
successive observations may provide additional
information about the pairing of components through
the latent-state dependence structure. Establishing
identifiability under weaker conditions is beyond
the scope of the present analysis.
\end{remark}

\subsection{Identifiability of the hidden Markov special case}
\label{app:hmm-identification}

\begin{proposition}[Identifiability of the functional HMM]\label{prop:hmm-id}
Let $K\ge2$ be known. Assume that the HMM transition matrix $P$ has full rank, its initial distribution satisfies $\rho_k>0$ for all $k$, and an observed sequence has length at least three. Assume the emission densities are linearly independent, with the Gaussian, contaminated-Gaussian or multivariate-$t$ conditions of Assumption~\ref{id:A}\textup{(A2)}, except that distinct state means are not required for contaminated-Gaussian emissions. Assume also the spectral condition \textup{(A3)}. Then $\bm\rho$, $P$, and the state emission parameters are identifiable from the observed law up to a common state permutation and, for the loading vectors, sign changes for simple leading eigenvalues or rotations within repeated-eigenvalue blocks.
\end{proposition}

\begin{proof}
The HMM identifiability can also be established directly by a three-observation decomposition. This independent result is included for completeness, but it is not used in the HSMM proof above.
The joint density of the first three coefficient vectors is
\begin{align}
 p^{(3)}(\bm c_1,\bm c_2,\bm c_3)
 &=\sum_{h,k,j=1}^{K}\rho_h P_{hk}P_{kj}
      f_h(\bm c_1)f_k(\bm c_2)f_j(\bm c_3)\\
 &=\sum_{k=1}^K\varrho_k\,g_k^{(-)}(\bm c_1)
      f_k(\bm c_2)g_k^{(+)}(\bm c_3),
 \label{id:hmm-threeview}
\end{align}
where
\[
 \varrho_k=\sum_h\rho_hP_{hk},\qquad
 g_k^{(-)}=\sum_h\frac{\rho_hP_{hk}}{\varrho_k}f_h,
 \qquad
 g_k^{(+)}=\sum_jP_{kj}f_j.
\]
Since $\bm\rho$ is strictly positive and $P$ is nonsingular and row-stochastic, $\varrho_k>0$ for each $k$. Relative to the linearly independent basis $(f_1,\ldots,f_K)$, the three factor systems in \eqref{id:hmm-threeview} have coefficient matrices $\operatorname{diag}(\bm\rho)P\operatorname{diag}(\bm\varrho)^{-1}$, $I_K$, and $P^\top$, up to orientation conventions. Each has full column rank and therefore Kruskal rank $K$. Since $3K\ge2K+2$ for $K\ge2$, uniqueness of the rank-$K$ three-way decomposition \citep[Theorem~9]{Allman2009} identifies these factor functions up to a common permutation and scaling. Normalization as probability densities removes the scaling ambiguity. In particular, it identifies $f_k$ and $g_k^{(+)}$ under the same state labels.

Linear independence of the $f_j$ gives a unique expansion $g_k^{(+)}=\sum_jP_{kj}f_j$, hence identifies every row of $P$. The one-observation marginal $\sum_k\rho_kf_k$ then identifies $\bm\rho$. For each identified $f_k$, the Gaussian, contaminated-Gaussian or multivariate-$t$ parameterization is identifiable as in Step 4 of the HSMM proof; in the contaminated case, its two kernels can be identified \emph{within the recovered state density}, so pairwise distinct state means are unnecessary. Lemma~\ref{id:lemma} recovers the funHDDC scale eigenvalues and spectral subspaces, with the stated indeterminacies.
\end{proof}

\begin{remark}[Parameter restrictions]\label{id:rem-constraints}
The restriction $\eta_k>1$ distinguishes the inflated Gaussian kernel from the ordinary one; the additional numerical constraint $\alpha_k>1/2$, when imposed, is not needed for the identification arguments. The full-rank restriction on $P$ belongs exclusively to Proposition~\ref{prop:hmm-id}; neither it nor stationarity of the HMM initial distribution is needed in Proposition~\ref{prop:hsmm-id}.
\end{remark}

\section{Computational performance}\label{app:computation}\label{app:detail-sec:sim-scaling}
The protocol is computationally modest. The median time per model and dataset, over all starts, screening and refinement passes, ranges from $5$ to $16$~s in Simulation~I and from $5$ to $44$~s in Simulation~II, the HSMM with contaminated emissions and NP sojourn being the most expensive. The Gaussian and $t$ fits meet the convergence criterion within $200$ iterations in $85$--$100\%$ of the cases. The contaminated-emission fits reach the iteration cap in about $80\%$ of the clean replicates, in which no curve is atypical and the contamination parameters $(\alpha_k,\eta_k)$ are weakly identified, but in at most $4\%$ of the contaminated ones. This does not affect classification: as $\alpha_k\to1$ or $\eta_k\to1$ the contaminated-Gaussian density tends to the Gaussian one, and in the clean scenarios the Gaussian and contaminated-Gaussian funHDDC-family models attain ARIs within $0.001$ of each other (Tables~\ref{tab:sim1} and~\ref{tab:sim2}). 

\begin{figure}
\centering
\includegraphics[width=\textwidth]{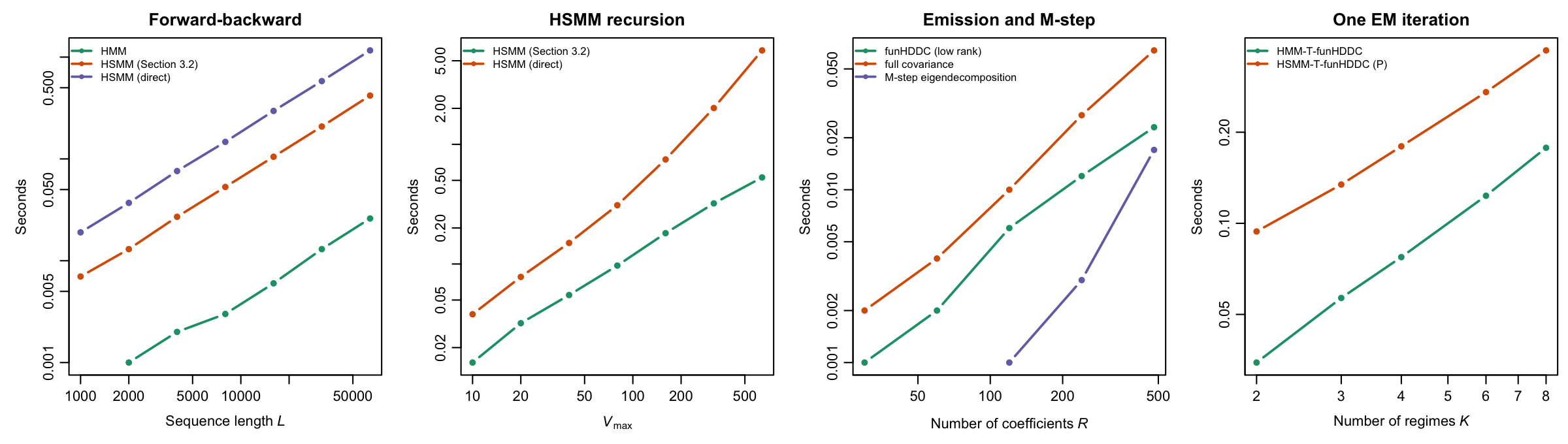}
\caption{Computational scaling (single core, median of five runs), from left: forward--backward time against $L$, semi-Markov and direct recursions against $V_{\max}$, emission evaluation and eigendecomposition against $R$, and one EM iteration against $K$.}
\label{fig:scaling}
\end{figure}

Figure~\ref{fig:scaling} reports single-core timings of the main computational steps (median of five runs on an Apple M4 Max processor). The hidden Markov and hidden semi-Markov forward--backward recursions are linear in the sequence length. The semi-Markov recursion of Section~\ref{subsubsec:hsm-estep}, of cost $O(L_iK^2+L_iV_{\max}K)$, is approximately $3$ times faster than the timed direct implementation at $V_{\max}=50$ and approximately $11$ times faster at $V_{\max}=640$, where it takes $0.5$~s for a sequence of $10{,}000$ curves. The direct implementation evaluates the same recursions at cost $O(L_iV_{\max}K^2)$ (Section~\ref{subsubsec:hsm-estep}). Since the ratio of the two costs, $V_{\max}K/(K+V_{\max})$, is below $K=3$, the larger measured speedups reflect block-emission and loop overheads of the direct implementation, not the asymptotic costs alone. The low-rank evaluation of the funHDDC emission on whitened coefficients grows linearly in $R$ and is about three times faster than a full-covariance density at $R=480$. The full eigendecomposition of the M-step grows about quadratically over this range and would dominate for larger $R$, where a partial eigendecomposition could replace it. One EM iteration grows about linearly in $K$ up to $K=8$, and a semi-Markov iteration costs two to three times a hidden Markov iteration at $V_{\max}=50$. The recursions against $L$ use $K=3$ and $V_{\max}=50$, and those against $V_{\max}$ a sequence of $L=10{,}000$ curves. The emission benchmark evaluates $20{,}000$ whitened curves for one state with $d_k=5$, and the EM iterations use $20$ sequences of $1{,}000$ curves with $R=60$ and $V_{\max}=50$.

\section{Simulation protocol and supplementary results}\label{app:simulation-results}
\subsection{Simulation protocol and numerical settings}\label{app:simulation-details}\label{app:detail-sec:sim-design}
The designs and the global-FPCA competitor are described in Section~\ref{sec:sim-design}. Contamination settings more extreme than Scenario~C are prone to spurious maxima in which a nearly empty state absorbs a few extreme curves, even when the fit is started from the true labels, and were not used. The number $q$ of global components is fixed by the $90\%$ rule before the study and not tuned, and the unrestricted covariance of each state of the competitor is the funHDDC structure with $d_k=q-1$ on the scores.

The numerical restrictions of Appendix~\ref{subsec:emission-updates} are $b_k\ge10^{-3}$ and $a_{kj}\ge(1+10^{-3})\,b_k$. Contaminated emissions are estimated with $\alpha_k\in(0.75,1)$ and $\eta_k\ge1+10^{-4}$, from $\alpha_k^{(0)}=0.9$ and, with the plain initial M-step, $\eta_k^{(0)}=2$. The robust initial M-step of Appendix~\ref{subsec:initialization} therefore sets aside $10\%$ of the curves of each initial cluster, and its $\eta_k^{(0)}$ is bounded below by $1.5$. The $t$ emissions have state-specific degrees of freedom, started at $\nu_k^{(0)}=50$ and updated on $[2,200]$, and the squared Mahalanobis distance is clamped at zero before it enters the $t$-density and the weights $h_{i\ell k}$, to avoid rounding artifacts. The size of the negative-binomial sojourn is restricted to $r_k\in[0.05,50]$ for numerical stability, so that the fitted law cannot reach its Poisson limit exactly. The smoothing constants of Appendix~\ref{subsec:initialization} are $\varepsilon_0=\varepsilon_P=\varepsilon_\varphi=1/2$, and the stopping rule of Appendix~\ref{subsec:convergence-classification} is applied with $\varepsilon=10^{-6}$ and $m_{\max}=200$. Because the likelihood has many local maxima, particularly under contamination, every model is estimated by the same multistart protocol:
\begin{enumerate}
    \item \emph{Shared starts.} Sixteen starting partitions of the pooled coefficients are generated once per dataset from a fixed seed and shared by all models: eight random labelings, six $k$-means partitions smoothed along each sequence by a running mode of width five, and two $k$-means partitions. With unequal $d_k$, which state carries which dimension is a discrete parameter that the EM iterations cannot change. Each starting partition is therefore relabeled so that the ranks of the intrinsic dimensions of its clusters, estimated by Cattell's scree test with threshold $0.2$, match the ranks of the $d_k$. This step and step~(iii) below are needed because the EM iterations cannot correct a start that assigns the dimensions to the wrong regimes.
    \item \emph{Screening.} Every start is run for ten iterations, the three with the largest BIC are continued to convergence, and the non-degenerate fit with the largest BIC is retained \citep{Biernacki2003}. For the $t$ and contaminated emissions, every start is screened twice, from the plain and from the robust initial M-step of Appendix~\ref{subsec:initialization}.
    \item \emph{Refinement.} Three passes follow, each of which replaces the current fit only if it increases the BIC. (i) Each model is restarted, with the same screening but two continued starts, from the final MAP partitions of all the other models fitted to the same dataset. (ii) Each HSMM is started from the fitted HMM with the same emission: its emission parameters and initial distribution, the embedded transitions $p^{\ast}_{hk}=p_{hk}/(1-p_{hh})$, and the sojourn law closest to the geometric sojourn of the HMM, namely the geometric law itself for the NB sojourn, since NB sojourn with size $r_k=1$ is geometric, its truncation to $\{1,\ldots,V_{\max}\}$ for the NP sojourn, and the shifted Poisson with the same mean for the P sojourn. The exact, untruncated HSMM (NB) likelihood contains the HMM likelihood as a special case. A fitted HSMM can nevertheless finish below the HMM because of duration truncation, parameter constraints or incomplete numerical optimization. (iii) For each funHDDC model, the dimensions of every pair of states are exchanged and the model is refitted from its final partition, until no exchange increases the BIC.
\end{enumerate}

A fit is treated as degenerate, and discarded, if its smallest expected state size $\sum_{i,\ell}\gamma_{i\ell k}$ is below $n_{\min}=\max\{0.01\,n_\bullet,\,2(\max_kd_k+1)\}$, since the contaminated-Gaussian and $t$ likelihoods admit spurious maxima in which a state collapses onto a few curves. As a check of the maximization, every model is also fitted once from the true labels. This fit is never used for selection. Across the funHDDC-family models, the protocol fit has at least the BIC of this fit, up to two units, in $7{,}736$ of the $7{,}749$ model--replicate pairs ($99.8\%$): all $5{,}998$ in Simulation~II and $1{,}738$ of $1{,}751$ in Simulation~I, where the lowest rate is $97\%$, for HMM-C-funHDDC in Scenario~C. The $51$ pairs whose fit from the true labels was itself degenerate are excluded.

The fitted duration distributions of the HSMM are compared with the true ones, after aligning the fitted states with the true regimes by the permutation that maximizes their agreement, through the absolute error of the mean duration and the total-variation (TV) distance between the fitted and true duration pmfs. Each (study, scenario, replicate) dataset is generated from its own fixed seed. We also record which candidate BIC and the marginal-entropy ICL of Appendix~\ref{subsec:hyperparameter-selection} select among the funHDDC-family models with the same emission.


\subsection{Simulation I: hidden Markov structure}
\label{app:detail-sec:sim1}

\begin{figure}
\centering
\includegraphics[width=\textwidth]{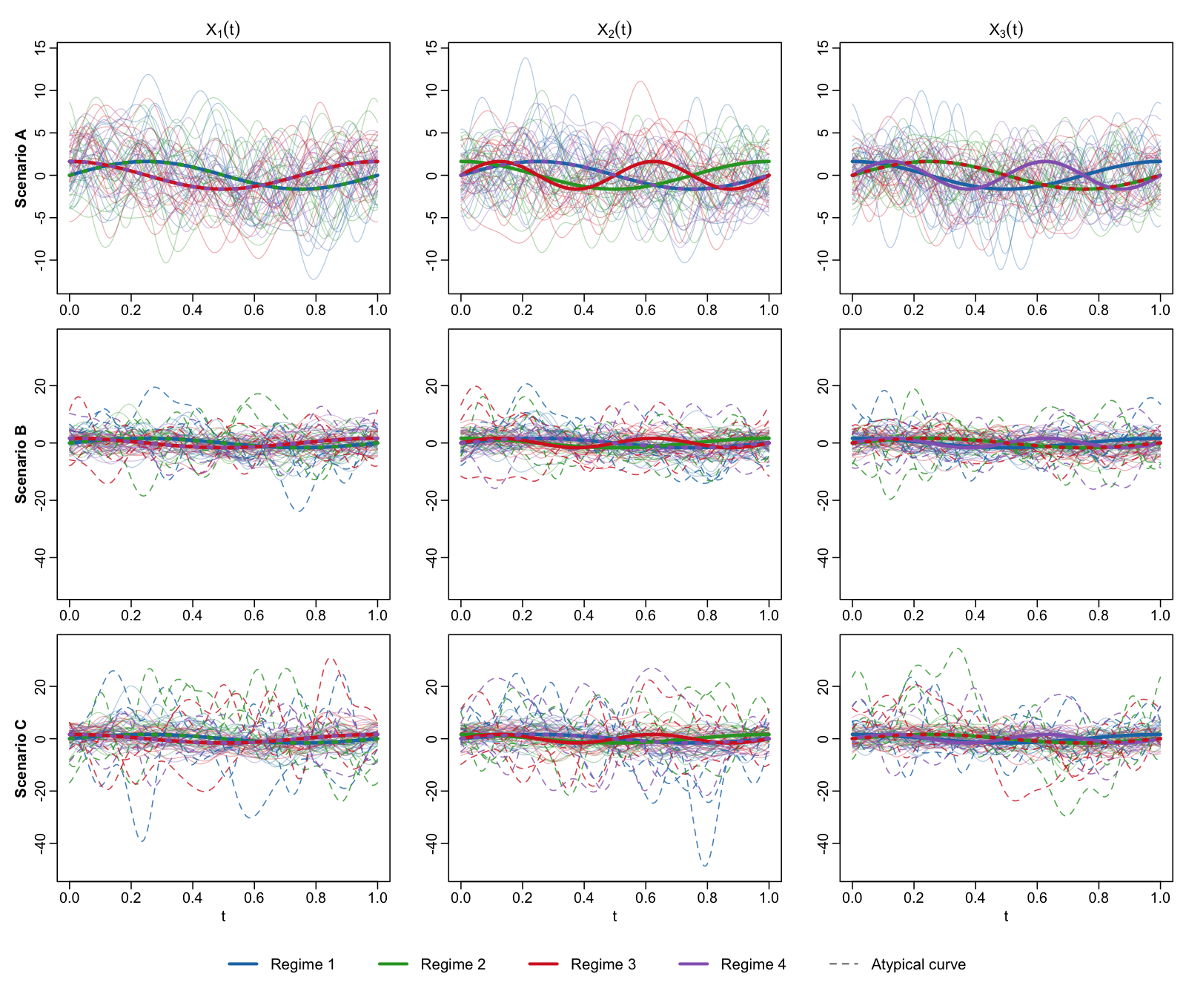}
\caption{Simulation I: simulated curves by regime in Scenarios~A--C, with the true regime means (dotted where two coincide). Dashed curves are atypical.}
\label{fig:sim1curves}
\end{figure}%

Figure~\ref{fig:sim1curves} shows the simulated curves by regime in Scenarios~A--C of the design of Section~\ref{sec:sim1}. 
In the clean scenario, every hidden Markov model reaches an ARI of $0.969$, against about $0.77$ for its i.i.d.\ counterpart (Table~\ref{tab:sim1}). Without atypical curves the three emissions are indistinguishable, so this gain of about $0.20$ is due entirely to the temporal layer. Under contamination the hidden Markov layer still improves on the i.i.d.\ mixture with every emission, but the accuracy it attains depends on the emission and on the severity of the contamination. Under mild contamination, HMM-C-funHDDC and HMM-T-funHDDC attain $0.970$ and $0.941$, close to their clean-data accuracy, whereas the misspecified Gaussian HMM falls to $0.676$. Under severe contamination the three emissions separate clearly. HMM-C-funHDDC, whose emission is correctly specified, keeps an ARI of $0.961$, HMM-T-funHDDC falls to $0.886$ with a larger spread, and HMM-funHDDC to $0.440$. With the robust emissions the gain of the temporal layer grows with the severity of the contamination, from $0.21$ to $0.24$ with contaminated emissions and from $0.25$ to $0.35$ with $t$ emissions, whereas with the Gaussian emission it shrinks, from $0.13$ to $0.09$. The fits from the true labels show that the lower accuracy of HMM-T-funHDDC under severe contamination is a property of the $t$ likelihood: its partitions of highest likelihood are not those closest to the truth.

The comparison with the global-FPCA competitors in Table~\ref{tab:sim1} separates the contribution of the state-specific subspaces from that of the temporal layer. For the i.i.d.\ mixtures, the funHDDC models are more accurate than their competitors in every scenario and with every emission, by $0.10$ to $0.12$ in the clean scenario and by $0.06$ to $0.24$ under contamination. Once the hidden Markov layer is added, the advantage shrinks, since the temporal layer recovers from the neighboring curves much of the regime information that the state-specific subspaces extract from a single curve. It nevertheless remains with both robust emissions in every scenario: $0.018$ to $0.026$ with contaminated emissions and $0.014$ to $0.030$ with $t$ emissions. With the misspecified Gaussian emission, the comparison is erratic. The competitor is far more accurate under mild contamination ($0.880$ against $0.676$) and less accurate under severe contamination ($0.402$ against $0.440$), and the Gaussian competitors have no non-degenerate fit in $14$ to $31$ of the contaminated replicates. Taking every model at the better of its protocol fit and its fit from the true labels does not change these comparisons.

\subsection{Simulation II: hidden semi-Markov structure}
\label{app:detail-sec:sim2}

\begin{figure}
\centering
\includegraphics[width=\textwidth]{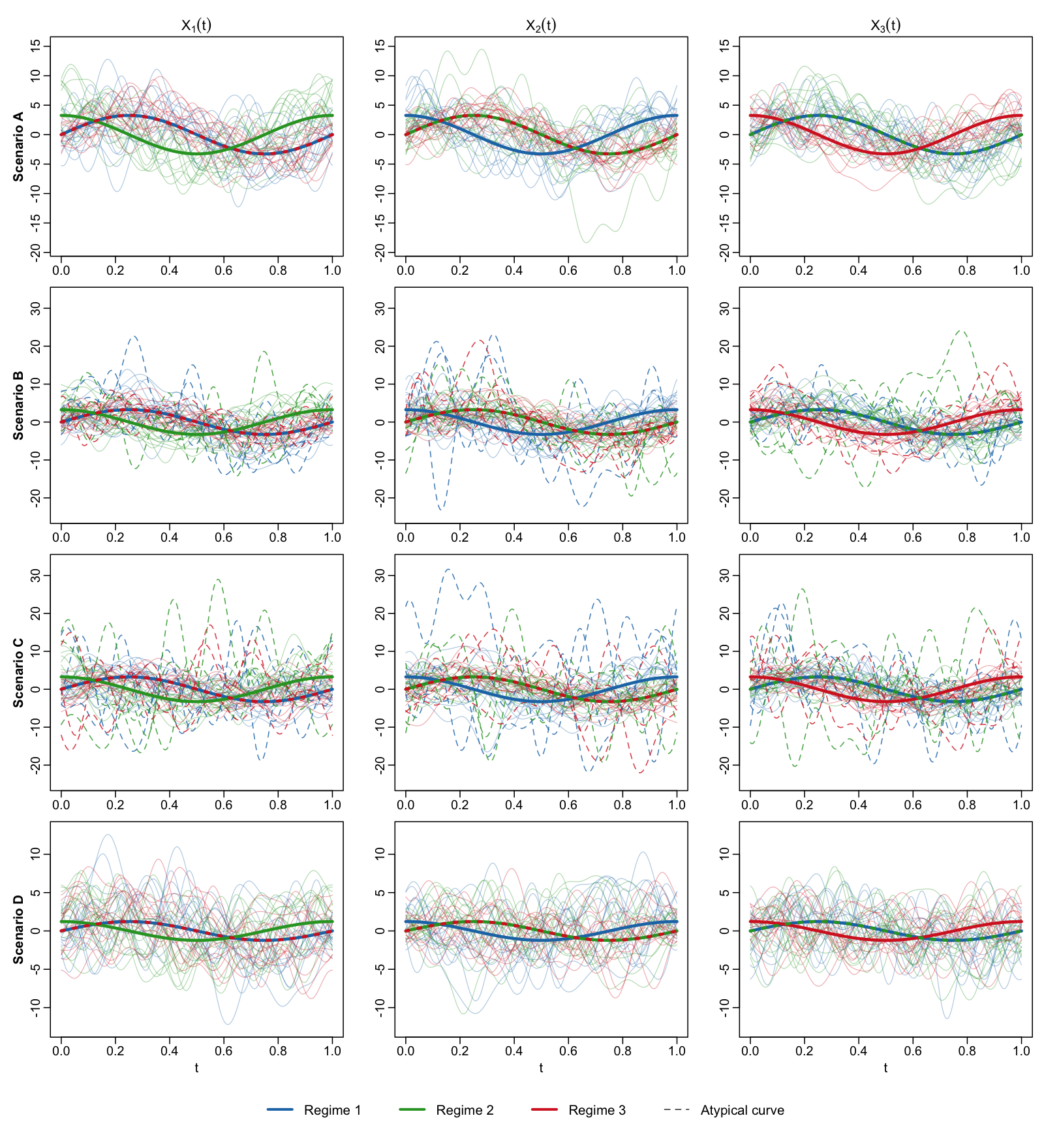}
\caption{Simulation II: simulated curves by regime in Scenarios~A--D, with the true regime means (dotted where two coincide). Dashed curves are atypical. Scenario~D has $\mu=1.5$.}
\label{fig:sim2curves}
\end{figure}

\begin{table}
\centering
\rotatebox{90}{%
\begin{minipage}{\textheight}
\centering
\footnotesize
\setlength{\tabcolsep}{4pt}
\caption{Simulation II: mean ARI (SD) over 100 replicates.}
\label{tab:sim2}
\begin{tabular}{@{}lcccccccc@{}}
\toprule
 & \multicolumn{4}{c}{funHDDC} & \multicolumn{4}{c}{FPCA} \\
\cmidrule(lr){2-5}\cmidrule(l){6-9}
Method & Scenario A & Scenario B & Scenario C & Scenario D
       & Scenario A & Scenario B & Scenario C & Scenario D \\
\midrule
Mixture     & 0.941 (0.009) & 0.625 (0.028) & 0.338 (0.032) & 0.763 (0.018) & 0.912 (0.012) & 0.626 (0.026) & 0.515 (0.024) & 0.633 (0.030) \\
Mixture-T   & 0.939 (0.010) & 0.862 (0.020) & 0.770 (0.022) & 0.755 (0.019) & 0.909 (0.012) & 0.820 (0.025) & 0.522 (0.045) & 0.609 (0.030) \\
Mixture-C   & 0.941 (0.009) & 0.935 (0.012) & 0.913 (0.013) & 0.762 (0.018) & 0.911 (0.012) & 0.901 (0.014) & 0.869 (0.016) & 0.621 (0.032) \\
HMM         & 0.994 (0.004) & 0.683 (0.024) & 0.378 (0.087) & 0.926 (0.013) & 0.990 (0.005) & 0.687 (0.023) & 0.566 (0.024) & 0.866 (0.020) \\
HMM-T       & 0.994 (0.003) & 0.982 (0.008) & 0.975 (0.015) & 0.925 (0.012) & 0.990 (0.005) & 0.977 (0.008) & 0.958 (0.053) & 0.858 (0.020) \\
HMM-C       & 0.994 (0.004) & 0.993 (0.004) & 0.991 (0.004) & 0.926 (0.012) & 0.990 (0.005) & 0.989 (0.005) & 0.984 (0.005) & 0.865 (0.020) \\
HSMM (P)    & 0.995 (0.003) & 0.684 (0.024) & 0.375 (0.086) & \textbf{0.943 (0.009)} & 0.993 (0.004) & 0.684 (0.024) & 0.560 (0.024) & 0.898 (0.016) \\
HSMM (NB)   & 0.995 (0.003) & 0.683 (0.025) & 0.378 (0.087) & 0.942 (0.009) & 0.993 (0.004) & 0.687 (0.024) & 0.566 (0.024) & 0.898 (0.017) \\
HSMM (NP)   & 0.995 (0.003) & 0.681 (0.025) & 0.385 (0.093) & 0.938 (0.010) & 0.991 (0.004) & 0.686 (0.024) & 0.566 (0.024) & 0.890 (0.020) \\
HSMM-T (P)  & 0.995 (0.003) & 0.990 (0.004) & 0.983 (0.006) & 0.941 (0.010) & 0.993 (0.004) & 0.988 (0.005) & 0.977 (0.007) & 0.894 (0.016) \\
HSMM-T (NB) & 0.995 (0.003) & 0.991 (0.005) & 0.983 (0.006) & 0.940 (0.010) & 0.993 (0.004) & 0.987 (0.005) & 0.973 (0.036) & 0.893 (0.017) \\
HSMM-T (NP) & 0.995 (0.003) & 0.987 (0.005) & 0.977 (0.016) & 0.934 (0.011) & 0.991 (0.004) & 0.984 (0.007) & 0.968 (0.038) & 0.885 (0.020) \\
HSMM-C (P)  & 0.995 (0.003) & \textbf{0.995 (0.003)} & \textbf{0.993 (0.004)} & 0.942 (0.009) & 0.993 (0.004) & 0.991 (0.004) & 0.989 (0.005) & 0.897 (0.016) \\
HSMM-C (NB) & 0.995 (0.003) & \textbf{0.995 (0.003)} & \textbf{0.993 (0.004)} & 0.942 (0.009) & 0.993 (0.004) & 0.991 (0.004) & 0.989 (0.005) & 0.896 (0.016) \\
HSMM-C (NP) & 0.995 (0.003) & 0.994 (0.004) & 0.991 (0.004) & 0.938 (0.010) & 0.991 (0.004) & 0.990 (0.004) & 0.987 (0.005) & 0.889 (0.019) \\
\bottomrule
\end{tabular}
\end{minipage}%
}
\end{table}

Figure~\ref{fig:sim2curves} shows the simulated curves in Scenarios~A--D of the design of Section~\ref{sec:sim2}.
Each of the three emissions is fitted under the i.i.d.\ mixture, the HMM and the HSMM with the P, NB and NP sojourns of Section~\ref{sec:sim2}, giving fifteen funHDDC-family models and fifteen global-FPCA competitors. All HSMM fits use the terminal survivor~\eqref{eq:hsm-effective-duration} throughout the recursions and the censored duration updates of Section~\ref{subsubsec:hsm-mstep}.

In Scenario~A every temporal model reaches an ARI of $0.994$--$0.995$, against $0.939$--$0.941$ for the i.i.d.\ mixtures (Table~\ref{tab:sim2}), and the semi-Markov layer adds at most $0.001$ to the hidden Markov layer. Scenario~D is the clean setting in which duration information matters. There the hidden Markov layer raises the ARI from about $0.76$ to $0.925$--$0.926$, and the hidden semi-Markov layer with P sojourn raises it further to $0.941$--$0.943$. This gain of $0.016$ to $0.017$ over the HMM is small relative to that of the hidden Markov layer, but it holds with every emission. The misspecified NB and NP sojourns retain most of it, with ARIs of $0.940$--$0.942$ and $0.934$--$0.938$.

Under contamination, the ordering depends on the emission. With robust emissions, the hidden Markov layer gives large gains over the i.i.d.\ mixture, from $0.06$ to $0.21$. The hidden semi-Markov layer adds a small gain with $t$ emissions ($0.008$ in both contaminated scenarios with the P sojourn, and at most $0.009$ with any sojourn) and a negligible one ($0.002$) with contaminated emissions, whose HMM already attains $0.993$ and $0.991$. HSMM-C-funHDDC with P or NB sojourn is the most accurate model in both contaminated scenarios ($0.995$ and $0.993$). With the misspecified Gaussian emission no temporal layer compensates for the atypical curves: the Gaussian models attain about $0.68$ under mild and $0.38$ under severe contamination. The semi-Markov layer therefore does not improve uniformly on the hidden Markov layer. Its gain in ARI is of the order of $0.01$ to $0.02$ when the regimes overlap or the emission is imperfect, and it vanishes when the emissions alone separate the regimes. Among the sojourn families, the P and NB sojourns are within $0.003$ of each other in every scenario, and in the clean scenarios and with robust emissions the NP sojourn is the least accurate. 

The global-FPCA competitors show where the state-specific subspaces matter (Table~\ref{tab:sim2}). In Scenario~D, funHDDC is more accurate than the competitor with every emission, temporal layer and sojourn family, in every replicate. The improvement is $0.13$ to $0.15$ for the i.i.d.\ mixtures, $0.06$ to $0.07$ for the HMM and $0.04$ to $0.05$ for the HSMM. The advantage of the state-specific subspaces therefore persists once temporal dependence is modeled by either layer, and it is about three times the gain from the duration layer. For instance, HSMM-T-funHDDC (P) attains $0.941$ against $0.894$ for FPCA-HSMM-$t$ (P), and the funHDDC HMMs ($0.925$--$0.926$) are more accurate than every FPCA HSMM ($0.885$--$0.898$). In Scenarios~A--C the robust funHDDC models are also more accurate than their competitors, by $0.03$ to $0.25$ for the i.i.d.\ mixtures and by $0.002$ to $0.017$ for the temporal models. With the misspecified Gaussian emission under severe contamination, the competitor is more accurate, by $0.18$ to $0.19$, and both approaches classify poorly. 
The selection of the temporal layer and the sojourn family by BIC and ICL is reported in Section~\ref{sec:sim2}.

\begin{table}
\centering
\small
\caption{Simulation II, clean scenarios: absolute error of the fitted mean duration (MAE, in curves) and total-variation distance to the true pmf (TV), averaged over regimes and 100 replicates.}
\label{tab:sim2soj}
\begin{tabular}{@{}lcccccccc@{}}
\toprule
 & \multicolumn{4}{c}{funHDDC} & \multicolumn{4}{c}{FPCA} \\
\cmidrule(lr){2-5}\cmidrule(l){6-9}
 & \multicolumn{2}{c}{Scenario A} & \multicolumn{2}{c}{Scenario D}
 & \multicolumn{2}{c}{Scenario A} & \multicolumn{2}{c}{Scenario D} \\
\cmidrule(lr){2-3}\cmidrule(lr){4-5}\cmidrule(lr){6-7}\cmidrule(l){8-9}
Model & MAE & TV & MAE & TV & MAE & TV & MAE & TV \\
\midrule
HSMM (P)    & 0.401 & 0.049 & 0.208 & 0.036 & 0.403 & 0.050 & 0.226 & 0.040 \\
HSMM (NB)   & 0.396 & 0.071 & 0.209 & 0.049 & 0.398 & 0.071 & 0.233 & 0.058 \\
HSMM (NP)   & 0.405 & 0.234 & 0.218 & 0.166 & 0.400 & 0.239 & 0.259 & 0.194 \\
HSMM-T (P)  & 0.401 & 0.049 & 0.208 & 0.036 & 0.403 & 0.050 & 0.226 & 0.040 \\
HSMM-T (NB) & 0.396 & 0.071 & 0.209 & 0.049 & 0.396 & 0.071 & 0.233 & 0.057 \\
HSMM-T (NP) & 0.402 & 0.235 & 0.222 & 0.168 & 0.400 & 0.239 & 0.256 & 0.199 \\
HSMM-C (P)  & 0.401 & 0.049 & 0.208 & 0.036 & 0.403 & 0.050 & 0.228 & 0.040 \\
HSMM-C (NB) & 0.396 & 0.071 & 0.209 & 0.049 & 0.398 & 0.071 & 0.230 & 0.057 \\
HSMM-C (NP) & 0.405 & 0.234 & 0.216 & 0.166 & 0.400 & 0.239 & 0.255 & 0.195 \\
\bottomrule
\end{tabular}
\end{table}

Every HSMM is also started from the fitted HMM with the same emission (step~(ii) of the protocol in Appendix~\ref{app:simulation-details}), and this warm start often gives the retained fit. Since the HSMM (NB) contains the HMM, the exact global maximum of its untruncated likelihood is at least that of the HMM. Finite duration cutoffs and incomplete numerical optimization may affect this comparison. With $t$ and contaminated emissions this holds in every replicate. With the Gaussian emission under contamination, the fitted HMM is so persistent that its geometric sojourn places a few percent of its mass beyond $V_{\max}=32$, and the HSMM (NB) can fall slightly below the HMM; with the duration support extended to the sequence length, it no longer does.

\begin{figure}[h!]
\centering
\includegraphics[width=0.72\textwidth]{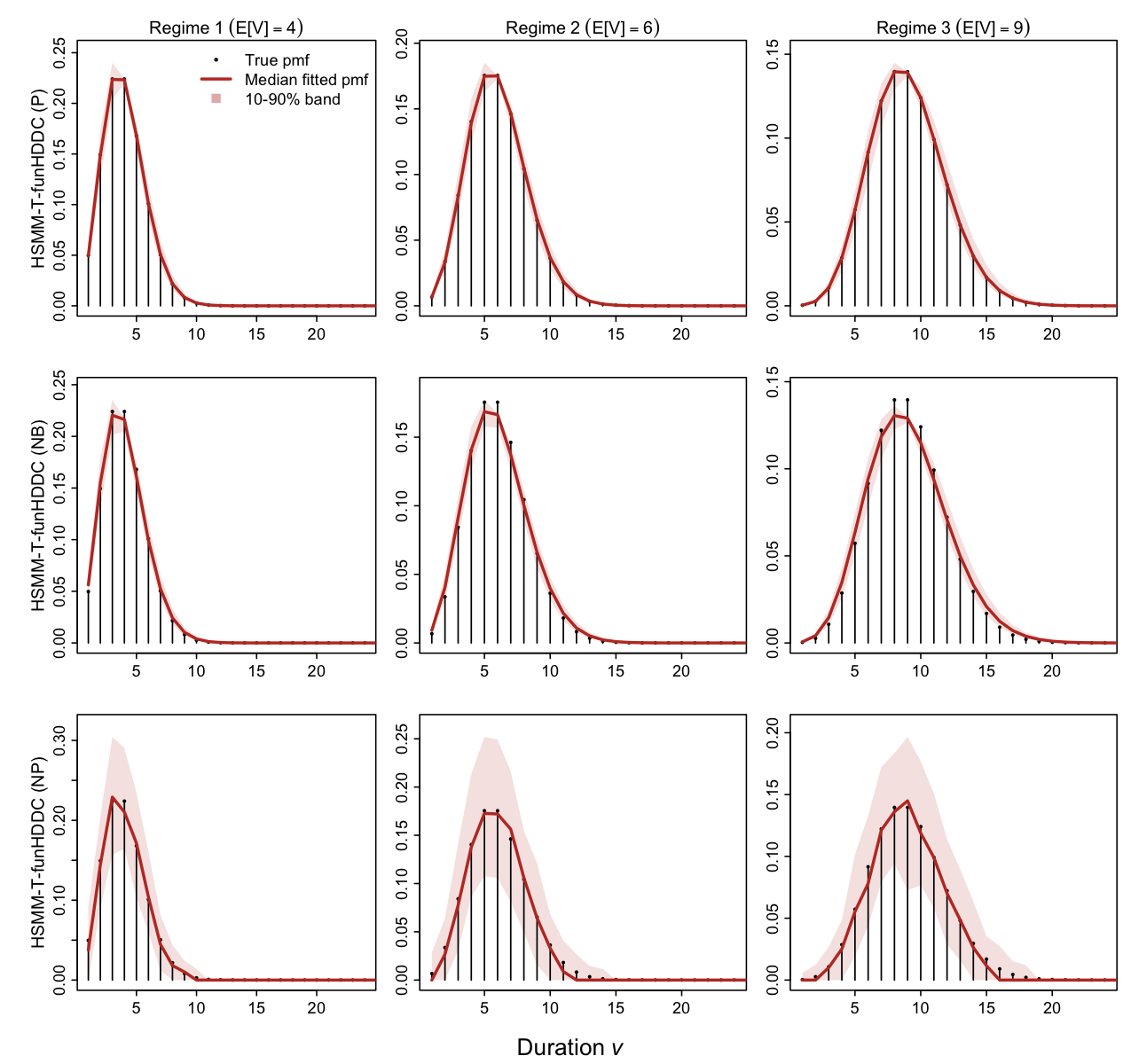}
\caption{Simulation II, Scenario~D: true duration pmfs (points) and median fitted pmfs of HSMM-T-funHDDC (lines, 10--90\% bands), by sojourn family (rows) and regime (columns).}
\label{fig:sim2sojourn}
\end{figure}

Table~\ref{tab:sim2soj} summarizes the recovery of the duration distributions in the clean scenarios. All three sojourn families recover the mean durations, with an absolute error of about $0.21$ curves in Scenario~D and $0.40$ curves in Scenario~A, whose durations are longer; they differ in how well they recover the whole distribution. The correctly specified P sojourn attains TV distances of $0.036$--$0.049$ and the NB sojourn $0.049$--$0.071$. The latter partly reflects the bound $r_k\le50$, which keeps the fitted law overdispersed relative to the Poisson: for the longest regime of Scenario~A the variance is at least $17+17^2/50\approx22.8$ instead of $17$. The NP sojourn, with $V_{\max}-1=31$ free masses per regime, attains $0.166$--$0.235$. The emission family has no visible effect. The FPCA-based HSMMs recover the durations as well as the funHDDC-based ones in Scenario~A and slightly less well in Scenario~D, where they also classify less accurately. Figure~\ref{fig:sim2sojourn} displays the median fitted pmfs in Scenario~D with their $10$--$90\%$ bands across the replicates. The parametric fits follow the true pmfs closely, whereas the NP fits are unbiased in the median but markedly more variable.

\begin{figure}[h!]
\centering
\includegraphics[width=0.72\textwidth]{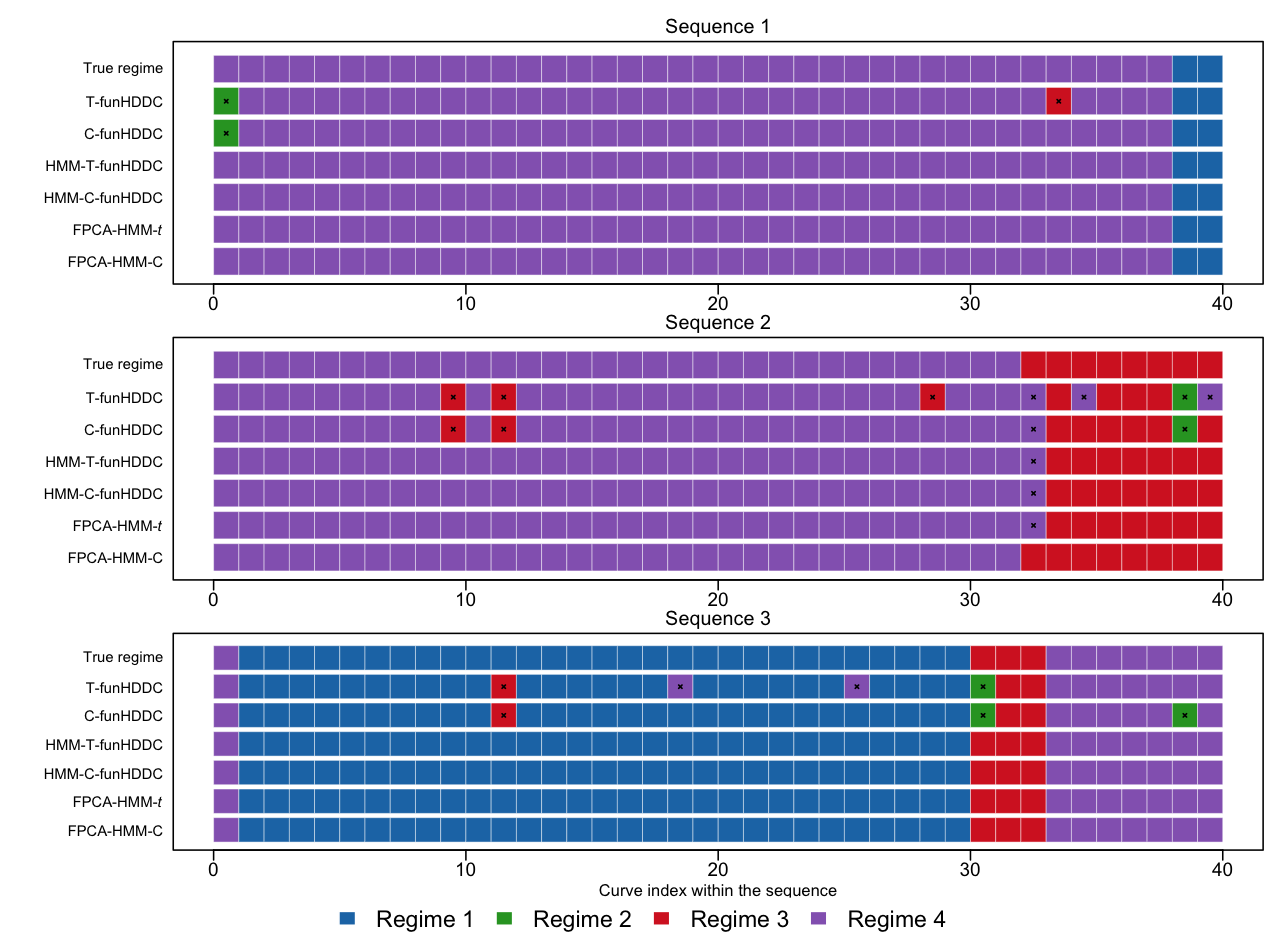}
\caption{Simulation I, Scenario~B: MAP labels of six models and the true regimes for the first three sequences of replicate~1. Crosses mark errors.}
\label{fig:sim1labels}
\end{figure}

\begin{figure}[h!]
\centering
\includegraphics[width=0.72\textwidth]{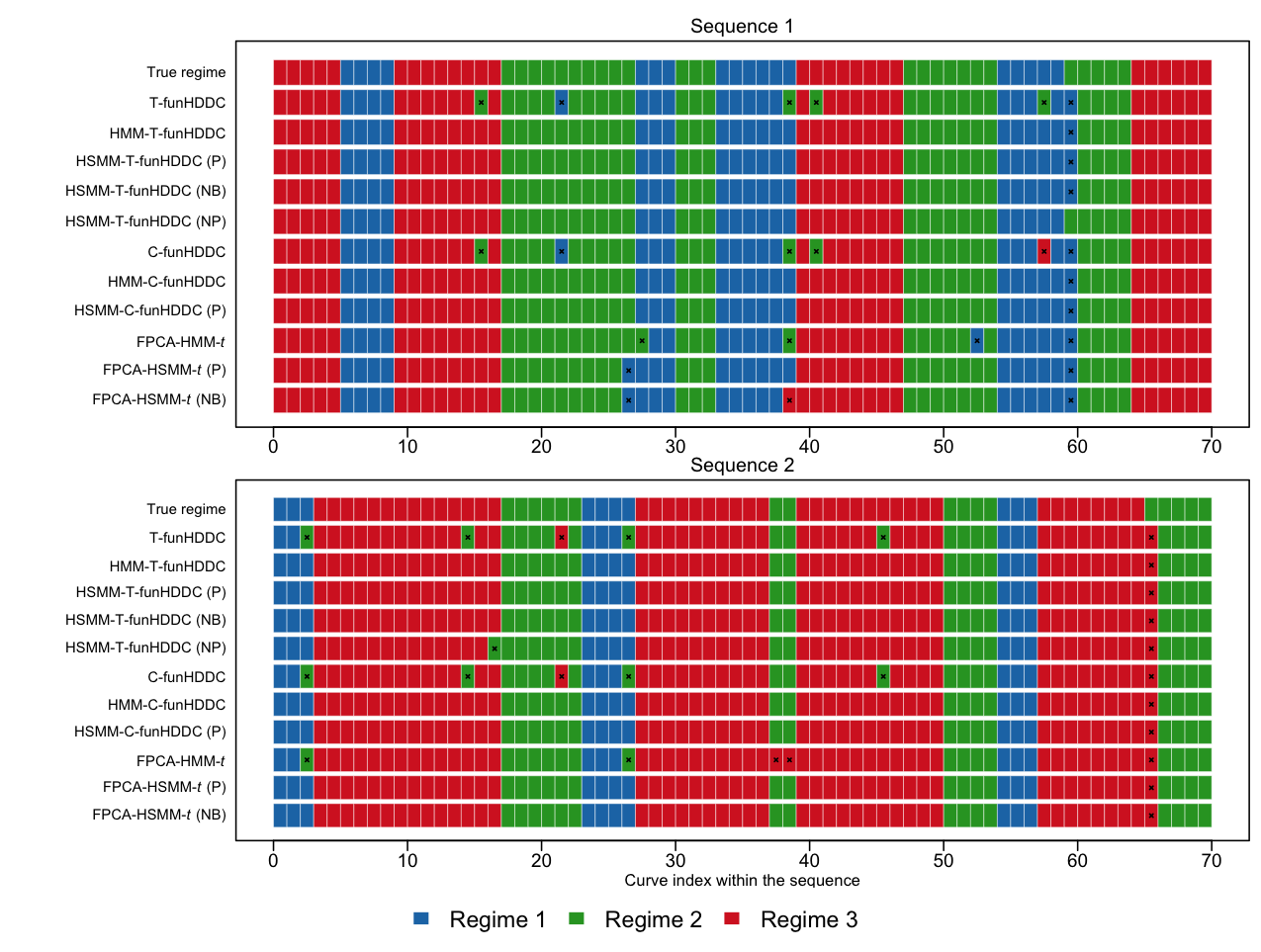}
\caption{Simulation II, Scenario~D: MAP labels of eleven models and the true regimes for the first two sequences of replicate~1. Crosses mark errors.}
\label{fig:sim2labels}
\end{figure}

\newpage
\section{Application methods and sensitivity analyses}\label{app:application-details}
\subsection{Air quality in the Milan area}
\label{app:detail-sec:milano}

We use hourly concentrations of nitrogen dioxide ($\mathrm{NO_2}$), a primary pollutant driven by road traffic and domestic heating, and of ozone ($\mathrm{O_3}$), a secondary photochemical pollutant, from four monitoring stations operated by ARPA Lombardia, retrieved from the Open Data Lombardia portal \citep{Maranzano2024}. Two stations are in the city of Milan (Pascal Citt\`a Studi, an urban background site, and Verziere, in the city center) and two in its metropolitan area (Pioltello Limito and Cormano). Each station and calendar day gives a bivariate curve on the daily cycle $[0,24]$ hours. Each pollutant is centered and scaled by the mean and standard deviation of its hourly values over the training period, and each component is projected on $R_s=11$ cubic B-splines ($R=22$; the condition number of $W_s$ is $27$). A day is kept when both pollutants have at least $18$ valid hours, and is otherwise a missing curve. Missing curves are not removed, since that would make non-consecutive days adjacent: they remain in the sequence with no emission (Appendix~\ref{subsubsec:hsm-leftcens}), which affects $7$--$11\%$ of the training days of each station and $3$--$6\%$ of the held-out days. The years 2016--2022 form the training sequences (four sequences of $2{,}557$ days, $9{,}247$ observed curves, so that $9.6\%$ of the training days are missing), and 1 January 2023 to 24 September 2026 the held-out sequences ($1{,}363$ days per station). Every sequence begins inside a season, so its first run is left-censored, and all HSMM fits use the residual duration law of Appendix~\ref{subsubsec:hsm-leftcens}. The four stations share the same meteorology, so their sequences are far from independent; we therefore also fit each station separately (Appendix~\ref{app:detail-sec:app-sens}).

\begin{figure}
\centering
\includegraphics[width=\textwidth]{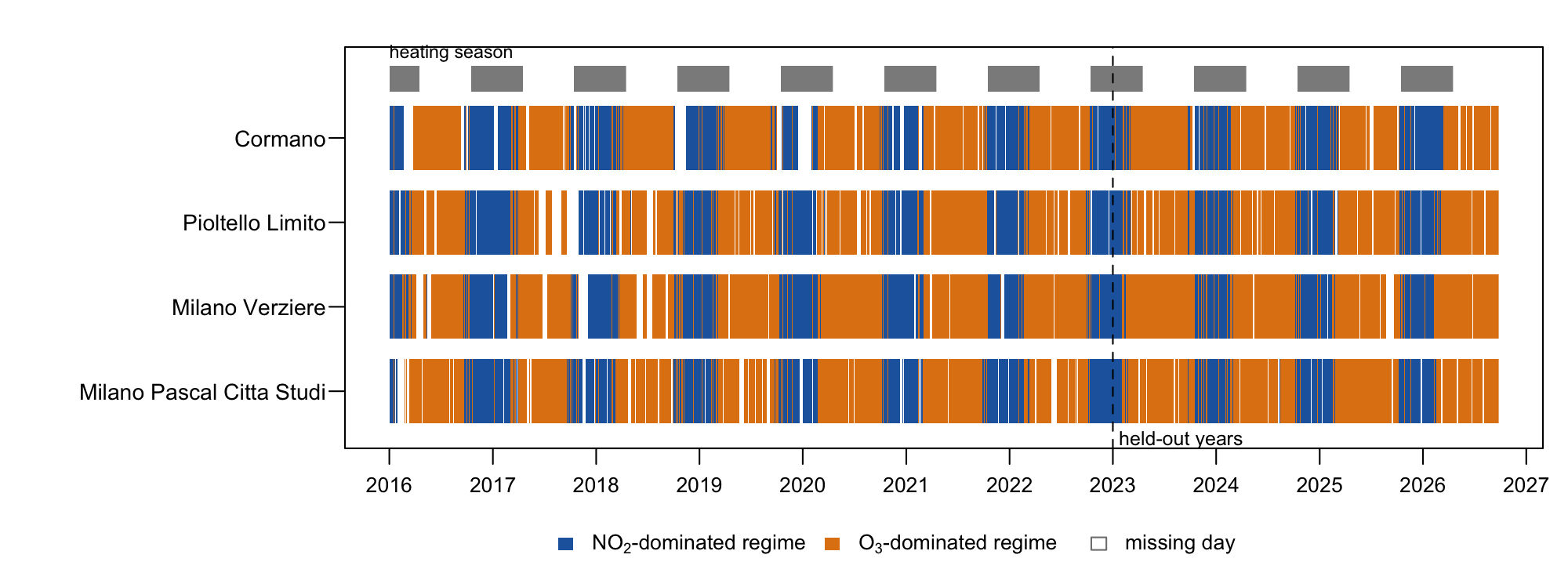}
\caption{Milan: MAP regime by station and day under HSMM-T-funHDDC (NB), fitted to 2016--2022 and decoded for 2023--2026. Gray bars mark the heating season (15 October--15 April), and white marks missing days.}
\label{fig:milanocalendar}
\end{figure}

We set $K=2$ as an interpretable choice rather than a statistically selected one, to contrast the two dominant daily pollution regimes of the Po Valley: a regime dominated by $\mathrm{NO_2}$, which accumulates under the stable, poorly ventilated conditions and heating emissions of the cold season, and a regime dominated by photochemical $\mathrm{O_3}$ \citep{Seinfeld2016, Bigi2014, Maranzano2022}. The fitted regimes have exactly this form as shown in Figure~\ref{fig:milanoprofiles}. The first has evening and morning $\mathrm{NO_2}$ peaks of $50$--$60\,\mu\mathrm{g/m^3}$ and almost no ozone. The second has low $\mathrm{NO_2}$ and an afternoon ozone peak of about $100\,\mu\mathrm{g/m^3}$. The regimes are not calendar seasons, however (Figure~\ref{fig:milanocalendar}). The $\mathrm{O_3}$-dominated regime occupies essentially every day from April to August (the $\mathrm{NO_2}$-dominated regime accounts for at most $1\%$ of those days), whereas from November to January the $\mathrm{NO_2}$-dominated regime prevails ($94$--$98\%$ of the days) but is interrupted by short $\mathrm{O_3}$-dominated episodes. Across the $4\times7$ training station-years, HSMM-T-funHDDC (NB) has $686$ completed runs. Exactly $28$ of them, one per station-year, are long $\mathrm{O_3}$-dominated runs of $110$ to $227$ days that start between February and June. Of the remaining $\mathrm{O_3}$-dominated runs, $87\%$ last less than ten days, and $74\%$ of the $\mathrm{NO_2}$-dominated runs are equally short. 

\begin{figure}
\centering
\includegraphics[width=\textwidth]{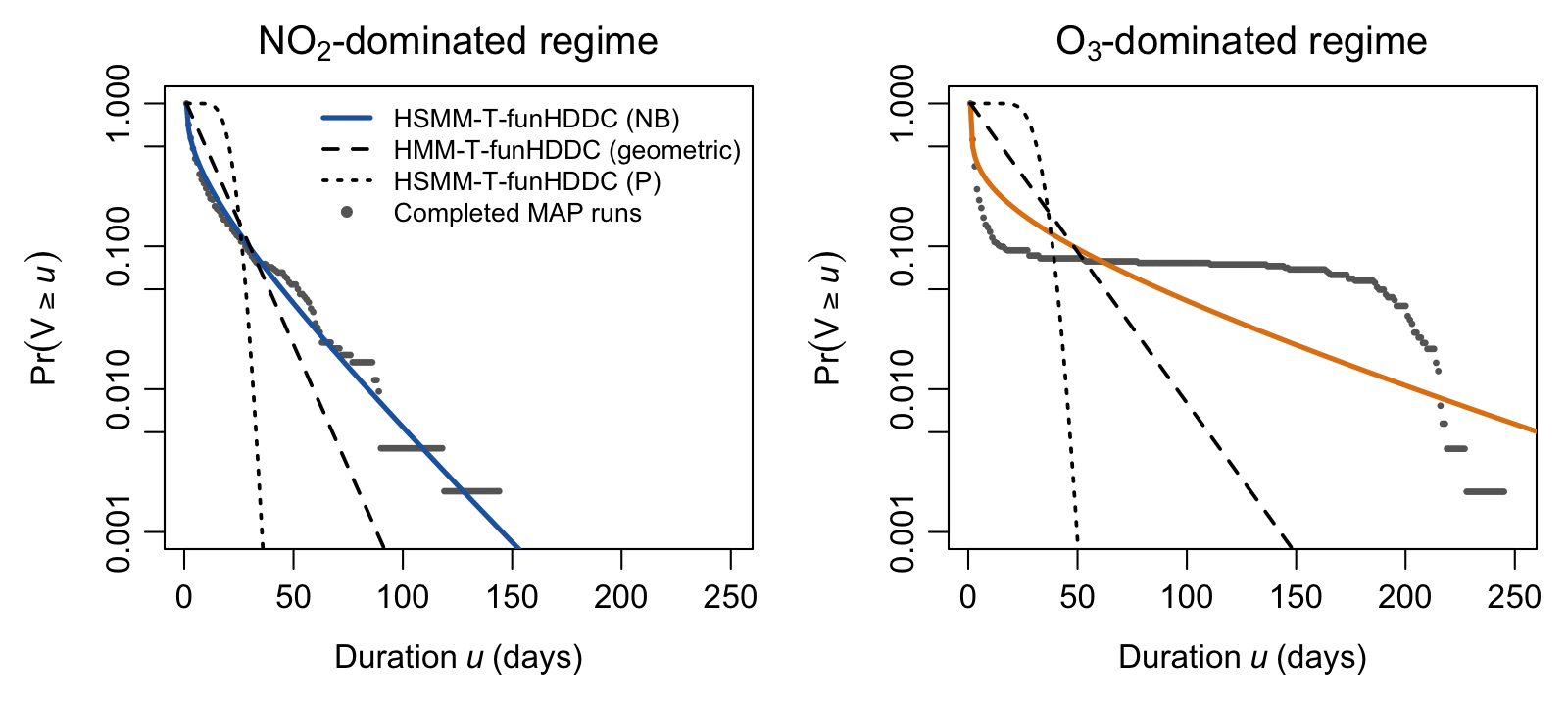}
\caption{Milan: survivor functions $\Pr(V\ge u)$ of the fitted durations under HSMM-T-funHDDC (NB), HSMM-T-funHDDC (P) and HMM-T-funHDDC, and the empirical survivor of the NB model's completed MAP runs (points), on a log scale.}
\label{fig:milanodurations}
\end{figure}

Table~\ref{tab:milanoresults} compares the models. Within each emission, BIC and the held-out likelihood agree that the HSMM (NB) is the best homogeneous model and the HMM the second, while the i.i.d.\ mixture and the HSMM (P) come last. With $t$ emissions, which are best by every criterion, BIC rises from $-238{,}845$ (i.i.d.) to $-235{,}451$ (HMM) and $-235{,}069$ (HSMM (NB)), and the held-out log-likelihood from $-123{,}873$ to $-121{,}895$ and $-121{,}656$. The gain of the HSMM over the HMM is thus about a tenth of that of the HMM over the i.i.d.\ mixture, but it holds on held-out years and at every station separately (Appendix~\ref{app:detail-sec:app-sens}). The fitted NB sizes are $0.31$ and $0.15$, well below the value $1$ that gives the geometric sojourn of the HMM, so the fitted durations are far more dispersed than geometric, and the underdispersed P sojourn fits worst of all (Figure~\ref{fig:milanodurations}). For the $\mathrm{NO_2}$-dominated regime, the NB sojourn follows the empirical run lengths closely. For the $\mathrm{O_3}$-dominated regime, it compromises between the many short runs and the long warm-season run, which appears as a separate mode near $200$ days that neither parametric family can reproduce.

\begin{remark}[Negative-binomial sojourns with size below one]\label{rem:nb-mixture}
Let $0<r_k<1$ and $\theta_k=r_k/(r_k+m_k)$. The shifted negative-binomial sojourn of Section~\ref{subsec:hsm} is then a mixture of the geometric sojourns of a hidden Markov model,
\[
\varphi_k(v;\bm\beta_k)=\int_{(0,\,1-\theta_k)}(1-s)\,s^{v-1}\,G_k(\mathrm ds),\qquad v\ge1,
\]
for a probability measure $G_k$ on $(0,1-\theta_k)$ over the persistence probability $s$, and its hazard $\varphi_k(v;\bm\beta_k)/\bar\varphi_k(v;\bm\beta_k)$ is strictly decreasing in $v$. A fitted size below one is therefore equivalent, in its sojourn law, to a regime whose persistence probability is drawn afresh at each visit. It cannot by itself separate intrinsic duration dependence from heterogeneity of the exit probabilities across visits, such as that induced by seasonal transitions or by sub-regimes merged under a small $K$.
\end{remark}

\begin{proof}
Write $r=r_k$ and $\theta=\theta_k$, so that $\widetilde V=V-1$ has $\Pr(\widetilde V=v)=\{\Gamma(v+r)/(\Gamma(r)\,v!)\}\,\theta^{r}(1-\theta)^{v}$ for $v=0,1,\ldots$. For $0<r<1$, the beta integral gives
\[
\frac{\Gamma(v+r)}{\Gamma(r)\,v!}=\frac{B(v+r,1-r)}{\Gamma(r)\,\Gamma(1-r)}=\frac{\sin(\pi r)}{\pi}\int_0^1t^{\,v+r-1}(1-t)^{-r}\,\mathrm dt .
\]
Multiplying by $\theta^{r}(1-\theta)^{v}$ and substituting $s=t(1-\theta)$, the powers of $1-\theta$ cancel and
\[
\Pr(\widetilde V=v)=\frac{\sin(\pi r)}{\pi}\,\theta^{r}\int_0^{1-\theta}s^{\,v+r-1}(1-\theta-s)^{-r}\,\mathrm ds=\int_0^{1-\theta}(1-s)\,s^{v}\,g(s)\,\mathrm ds,
\]
where $g(s)=\pi^{-1}\sin(\pi r)\,\theta^{r}s^{r-1}(1-\theta-s)^{-r}(1-s)^{-1}\ge0$. Summing over $v$ and using $\sum_{v\ge0}(1-s)s^{v}=1$ for $s\in(0,1)$ shows, by Tonelli's theorem, that $g$ integrates to one, so $G_k(\mathrm ds)=g(s)\,\mathrm ds$ is a probability measure, and $\varphi_k(v;\bm\beta_k)=\Pr(\widetilde V=v-1)$ has the stated form. Finally, $\Pr(\widetilde V=v+1)/\Pr(\widetilde V=v)=(1-\theta)(v+r)/(v+1)$ is strictly increasing in $v$ when $r<1$, so the probability mass function is strictly log-convex and its hazard is strictly decreasing.
\end{proof} 

These homogeneous models ignore that the regime process is seasonal, and the seasonality accounts for most of the advantage of the semi-Markov layer. We therefore also fit hidden Markov models whose transition matrix depends on a known calendar group, the heating season or the month, estimated with the same protocol and decoded on the held-out years with the calendar of those years. With two transition matrices (heating season and the rest of the year), the HMM improves on the homogeneous HMM but not on the HSMM (NB). With one transition matrix per month (22 transition parameters more than the homogeneous HMM), it overtakes the HSMM (NB). With $t$ emissions, its BIC is $-234{,}876$ and its held-out log-likelihood $-121{,}570$, against $-235{,}069$ and $-121{,}656$ (Table~\ref{tab:milanoresults}), and the same ordering holds with contaminated emissions. Its monthly persistence probabilities describe the seasonal cycle directly: the probability of remaining in the $\mathrm{O_3}$-dominated regime from one day to the next is at least $0.99$ from April to August and at most $0.36$ from November to January, and the reverse holds for the $\mathrm{NO_2}$-dominated regime. The overdispersed durations of the homogeneous semi-Markov model are thus largely a symptom of seasonal non-stationarity: a homogeneous model must mix short winter episodes with one long warm-season run per year, which an NB sojourn with a small size approximates better than a geometric law. We therefore do not read the Milan data as evidence for non-geometric durations as such. Whether durations remain non-geometric once seasonality is accounted for would require a semi-Markov model with season-dependent durations, which we leave to future work. 

The heating-season calendar describes, but does not validate, the segmentation. On the held-out years, the MAP regimes of HSMM-T-funHDDC (NB) agree with it with an ARI of $0.34$, against $0.30$ for the i.i.d.\ mixture (Table~\ref{tab:milanoresults}). A running mode over 15 days of the i.i.d.\ labels agrees with the calendar more closely (ARI $0.39$) than any temporal model with $t$ emissions. So do the HSMMs with P sojourn fitted with contaminated emissions or on the global-FPCA scores (ARI $0.40$--$0.41$), which segment the data into seasons with mean durations of about $140$ and $220$ days but have a lower likelihood than the corresponding HMMs. Agreement with a calendar thus measures how seasonal a segmentation is, not how well it fits the curves. The label-free criteria, by contrast, favor the finer regime structure found by the funHDDC models. Because the data span ten years and hundreds of completed runs, the comparison of temporal layers no longer rests on a single seasonal transition. What it shows is that the persistence of these regimes is governed mainly by the calendar.

\subsection{Human activity recognition}
\label{app:detail-sec:har}

\begin{figure}
\centering
\includegraphics[width=\textwidth]{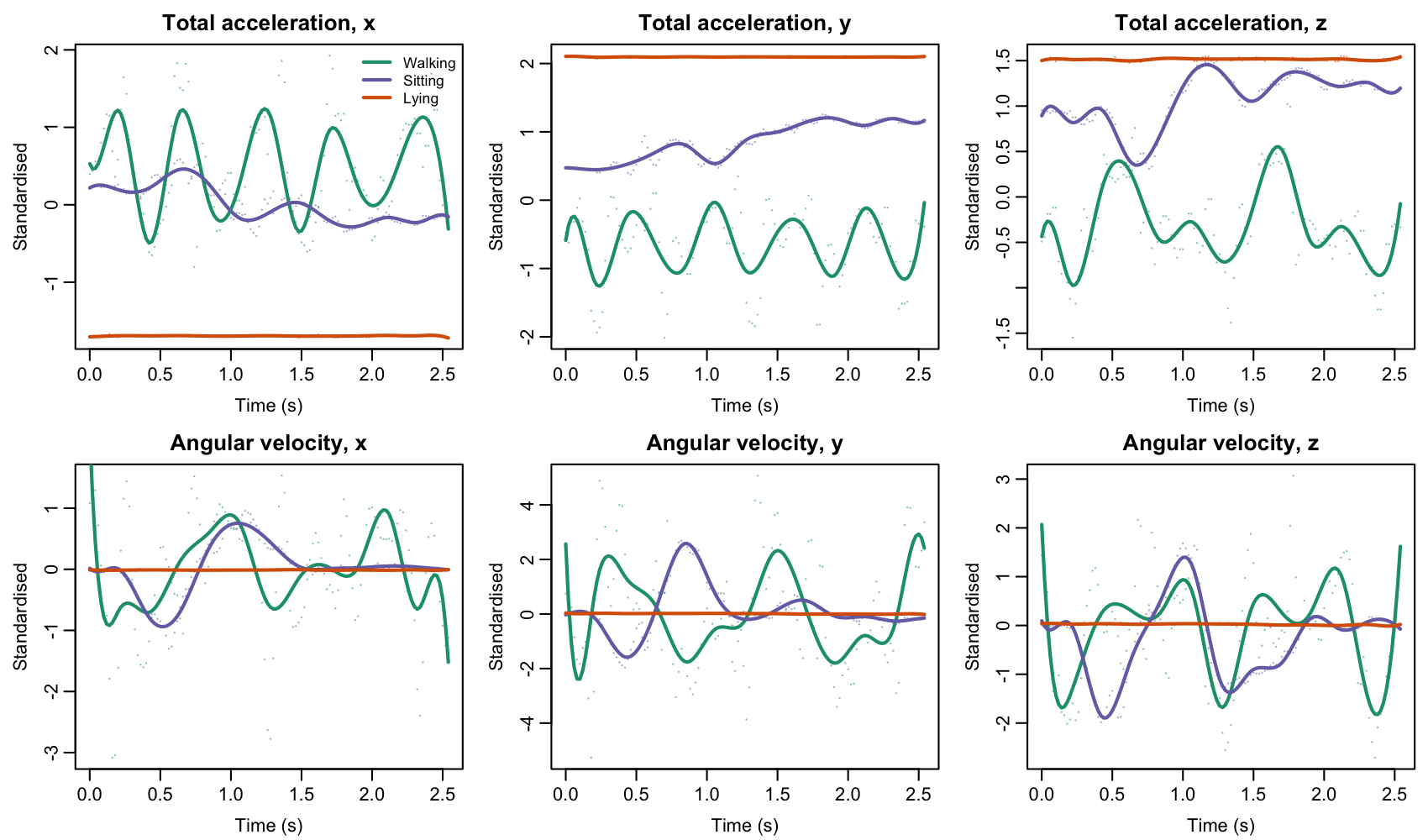}
\caption{Human activity: one window of walking, sitting and lying in the six channels (standardized samples and least-squares B-spline fit).}
\label{fig:harwindows}
\end{figure}

The Human Activity Recognition Using Smartphones data set \citep{Anguita2013} contains inertial signals of $30$ volunteers who performed a protocol of six activities (walking, walking upstairs and downstairs, sitting, standing and lying) with a smartphone on the waist, released as overlapping windows of $2.56$~s. Two features of that release preclude a temporal analysis. First, consecutive windows overlap by $50\%$, so adjacent observations share half their samples. Second, the windows of each subject are grouped by activity within each recording session rather than listed in time order. We checked the second point from the overlap itself: two windows are contiguous exactly when the second half of one equals the first half of the next, and in the training file every one of the $259$ changes of activity is a break in contiguity, while the blocks of each session appear grouped by activity (all standing bouts, then all sitting bouts, and so on), whereas the protocol interleaves them. Run lengths in the file order are therefore artifacts. We instead use the raw recordings of the same $30$ volunteers released with the HAPT data set \citep{ReyesOrtiz2016}, which provide each of the $61$ recording sessions as a continuous $50$~Hz stream with the start and end of every activity and postural transition. 

Each session, trimmed to its labeled protocol, is one sequence, tiled in time order by non-overlapping windows of $128$ samples ($2.56$~s). The median session has $128$ windows (range $52$--$161$). Each window is a curve with six components: the triaxial total acceleration, which contains the gravity direction, and the triaxial angular velocity. We do not add the body-acceleration channels of the original release because, within a window, they equal the total acceleration minus a nearly constant gravity term. Each channel is centered and scaled with the training subjects' samples and projected on $R_s=15$ cubic B-splines, so that $R=90$ and the condition number of $W_s$ is $27$. We keep the standard split into $21$ training subjects (42 sessions with 5,529 windows) and $9$ held-out subjects (19 sessions with 2,274 windows). Windows that overlap a postural transition, an unlabeled stretch (for instance turning at the end of a corridor) or two bouts stay in the sequences, so that the temporal layer sees the whole session, but they are not scored. The scored windows, $64\%$ of the total, lie entirely within one bout of one activity.

We set $K=3$ and aggregate the activities into locomotion (walking, upstairs and downstairs), upright static postures (sitting and standing) and lying. This is an interpretable modeling choice, not the only distinction the sensors support. Motion energy separates locomotion from the static postures, and the gravity direction in the total acceleration separates lying from upright as shown in Figure~\ref{fig:harwindows}, but other aggregations, or the six activities themselves, could also be studied. The durations in these data are set by the protocol, which prescribes how long each activity is performed. They therefore describe the experiment, not natural behavior, and we interpret them accordingly. 

\begin{figure}
\centering
\includegraphics[width=\textwidth]{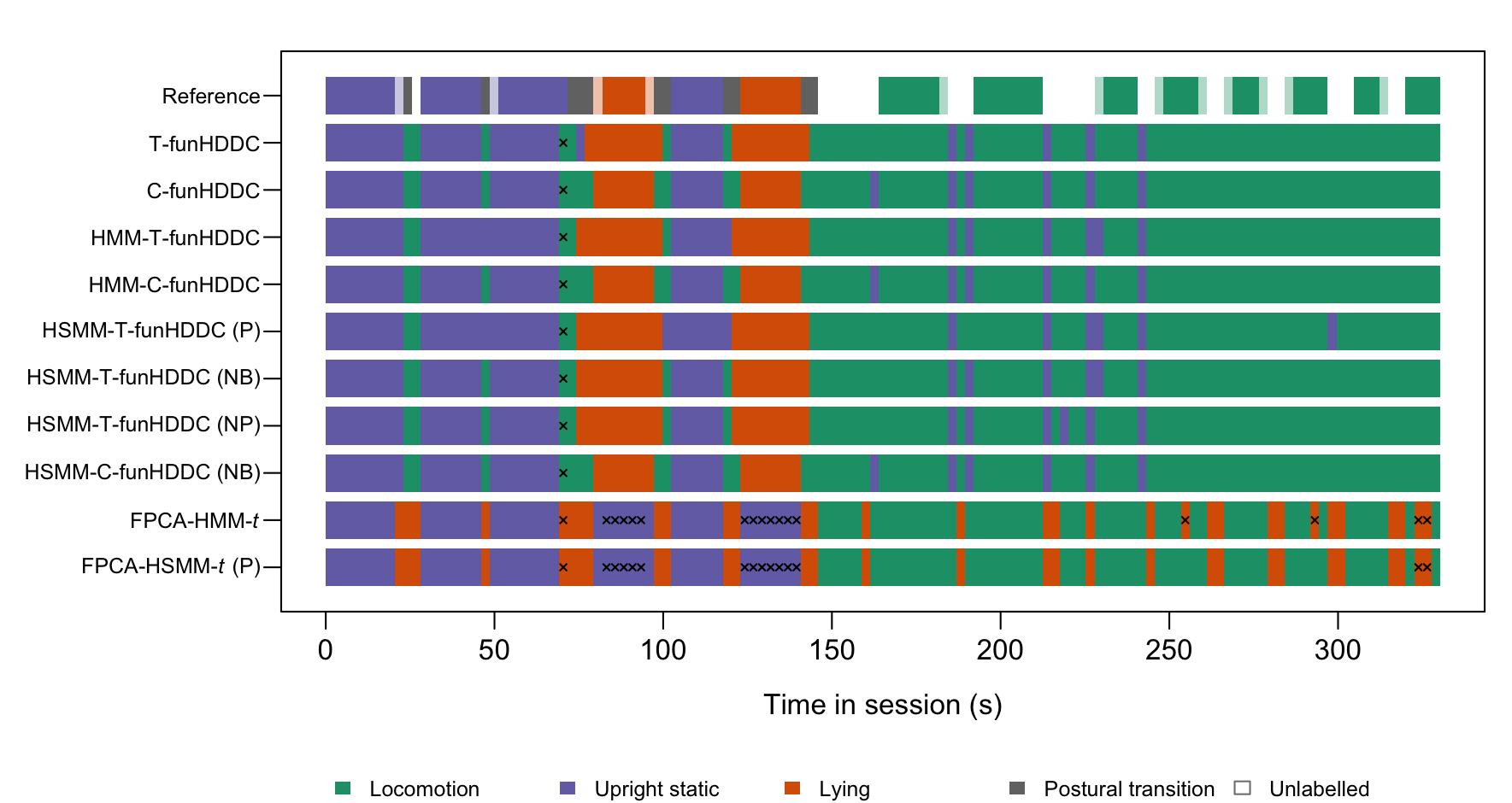}
\caption{Human activity: reference classes and held-out MAP labels of selected models in the first held-out session. Light shades mark unscored windows, gray marks postural transitions and white marks unlabeled stretches. Crosses mark errors on scored windows.}
\label{fig:harsession}
\end{figure}

Table~\ref{tab:harresults} reports the fits on the training sessions and their performance on the held-out subjects. With $R_s=15$, the emission family matters most for classification. The contaminated-Gaussian models classify the held-out windows best (ARI $0.87$--$0.91$) and the $t$ models next ($0.84$--$0.86$), although BIC and the held-out likelihood prefer the $t$ emission by a wide margin. As in Simulation~I under heavy contamination, fit and classification thus favor different emissions, and this ranking by classification is not stable to the number of basis functions, whereas the ranking of the temporal layers is (Appendix~\ref{app:detail-sec:app-sens}). The global-FPCA competitors classify far worse (ARI $0.47$--$0.69$) and in particular confuse lying with upright postures (Figure~\ref{fig:harsession}): with $q=25$ global components, the state-specific subspaces matter far more here than the temporal layer. 

The temporal layers improve the label-free criteria, but change the classification little. With $t$ emissions, the HMM raises the held-out log-likelihood by $1{,}316$ over the i.i.d.\ mixture ($754$ with contaminated emissions), and the HSMMs with NB and NP sojourns raise it by a further $135$ and $157$ ($65$ and $105$). The HSMM (P) lowers it, and BIC and ICL rank the HSMM (NB) first for both robust emissions. The held-out ARI, however, rises by only $0.015$--$0.016$ from the i.i.d.\ mixture to the HMM for both robust emissions and is essentially unchanged by the semi-Markov layer, and a running mode over five windows of the i.i.d.\ labels classifies better than every temporal model with $t$ emissions (ARI $0.889$ against at most $0.858$) and comparably with contaminated emissions ($0.902$ against $0.871$--$0.909$). Single windows of $2.56$~s are already informative about these three classes. The postural transitions, which are not scored, are mostly assigned to the locomotion regime (Figure~\ref{fig:harsession}), and the errors on scored windows are not concentrated at bout boundaries: $33\%$ of the errors of HMM-C-funHDDC lie next to an unscored window, against $32\%$ of all scored windows. The fitted mean durations of HSMM-T-funHDDC (NB) are $61$~s for locomotion, $16$~s for upright postures and $27$~s for lying. The protocol bouts have medians of $19$~s (walking), $12$--$13$~s (stairs) and $21$--$23$~s (static postures), and the MAP runs of locomotion extend across the unlabeled turns between walking bouts. Performance varies between subjects: the held-out ARI of HMM-C-funHDDC ranges from $0.69$ to $1.00$ over the nine held-out subjects, two subjects account for $39$ of its $48$ errors on scored windows, and $36$ of the $48$ are lying windows assigned to the locomotion regime. 

\begin{table}
\centering
\footnotesize
\setlength{\tabcolsep}{3.5pt}
\caption{Sensitivity analyses: held-out ARI, gain in held-out log-likelihood per 100 curves of the HMM over the i.i.d.\ mixture ($\Delta\ell$ HMM) and of the HSMM (NB) over the HMM ($\Delta\ell$ HSMM), and BIC gain of the HSMM (NB) over the HMM ($\Delta$BIC).}
\label{tab:appsens}
\begin{tabular}{@{}llcccc@{}}
\toprule
Analysis & Emission & ARI (i.i.d.\,/\,HMM\,/\,HSMM) & $\Delta\ell$ HMM & $\Delta\ell$ HSMM & $\Delta$BIC \\
\midrule
HAR, main analysis & $t$ & 0.842 / 0.858 / 0.855 & $+57.9$ & $+5.9$ & $+321$ \\
($R_s=15$) & Contaminated & 0.891 / 0.906 / 0.906 & $+33.2$ & $+2.9$ & $+157$ \\
\addlinespace
HAR, $50\%$-overlapping & $t$ & 0.842 / 0.846 / 0.844 & $+68.8$ & $+2.8$ & $+332$ \\
windows & Contaminated & 0.367 / 0.440 / 0.439 & $+55.7$ & $+1.1$ & $+93$ \\
\addlinespace
HAR, $R_s=11$ & $t$ & 0.861 / 0.865 / 0.858 & $+105.6$ & $+5.8$ & $+355$ \\
 & Contaminated & 0.821 / 0.817 / 0.819 & $+52.0$ & $+5.1$ & $+268$ \\
\addlinespace
HAR, $R_s=21$ & $t$ & 0.760 / 0.773 / 0.775 & $+56.8$ & $+5.8$ & $+313$ \\
 & Contaminated & 0.366 / 0.922 / 0.922 & $-506.3$ & $+4.5$ & $+243$ \\
\addlinespace
Milan, main analysis & $t$ & 0.304 / 0.344 / 0.340 & $+38.0$ & $+4.6$ & $+382$ \\
($R_s=11$) & Contaminated & 0.269 / 0.341 / 0.343 & $+31.3$ & $+5.8$ & $+495$ \\
\addlinespace
Milan, $R_s=7$ & $t$ & 0.310 / 0.373 / 0.363 & $+39.9$ & $+3.3$ & $+307$ \\
 & Contaminated & 0.295 / 0.370 / 0.352 & $+40.5$ & $+2.8$ & $+327$ \\
\addlinespace
Milan, $R_s=15$ & $t$ & 0.301 / 0.328 / 0.325 & $+36.5$ & $+4.5$ & $+440$ \\
 & Contaminated & 0.257 / 0.291 / 0.286 & $+34.8$ & $+4.4$ & $+519$ \\
\bottomrule
\end{tabular}
\end{table}

The six activities themselves are harder to recover. In a separate analysis with $K=6$ regimes, one per activity, every model confused sitting with standing and the three locomotion activities with one another, and neither the temporal layers nor larger intrinsic dimensions resolved these confusions. The three-class aggregation is therefore also the level at which the models can be assessed meaningfully with these channels and windows. 

\subsection{Sensitivity analyses}
\label{app:detail-sec:app-sens}

\begin{table}
\centering
\footnotesize
\caption{Milan, each station fitted separately ($t$ emissions): BIC and held-out log-likelihood of the HSMMs minus those of the HMM, and NB mean durations (days) and sizes for the $\mathrm{NO_2}$- and $\mathrm{O_3}$-dominated regimes (a size of $1$ gives the geometric sojourn).}
\label{tab:milanstations}
\begin{tabular}{@{}lcccccc@{}}
\toprule
 & \multicolumn{2}{c}{$\Delta$BIC (HSMM)} & \multicolumn{2}{c}{$\Delta$ held-out log-lik. (HSMM)} & \multicolumn{2}{c}{HSMM (NB)} \\
\cmidrule(lr){2-3}\cmidrule(lr){4-5}\cmidrule(l){6-7}
Station & P & NB & P & NB & Mean duration & Size \\
\midrule
Milano Pascal Citt\`a Studi & $-516$ & $+99$ & $-188$ & $+33$ & 10.0/16.4 & 0.36/0.13 \\
Milano Verziere & $-1404$ & $+95$ & $-897$ & $+34$ & 7.6/13.1 & 0.50/0.15 \\
Pioltello Limito & $-1362$ & $+111$ & $-737$ & $+29$ & 9.7/15.8 & 0.34/0.13 \\
Cormano & $-1309$ & $+89$ & $-746$ & $+56$ & 12.3/19.5 & 0.26/0.13 \\
\bottomrule
\end{tabular}
\end{table}

Table~\ref{tab:appsens} reports three sensitivity analyses for the $t$ and contaminated-Gaussian emissions, and Table~\ref{tab:milanstations} reports the station-by-station fits of the Milan data.
Repeating the HAR analysis with the $50\%$-overlapping windows of the original release inflates the apparent contribution of the hidden Markov layer, as expected when adjacent observations share half their samples. Per 100 held-out windows, its gain in log-likelihood over the i.i.d.\ mixture rises from $58$ to $69$ with $t$ emissions and from $33$ to $56$ with contaminated emissions, whereas the additional gain of the semi-Markov layer falls, from $5.9$ to $2.8$ and from $2.9$ to $1.1$. With overlapping windows, the contaminated-Gaussian fits of highest likelihood place about a quarter of the windows in the inflated component, with $\alpha_k$ at or near its lower bound $0.75$ in every regime, and classify poorly (held-out ARI $0.37$--$0.44$). The $t$ fits are unaffected ($0.84$--$0.85$). These contaminated fits have a higher BIC, by $800$ to $2{,}100$, than the fits reached from the plain initial M-step alone, which classify at $0.85$--$0.92$. The poor classification is therefore a property of the contaminated-Gaussian likelihood on these heavy-tailed data, not an optimization failure. In this analysis $V_{\max}=322$ windows, the length of the longest session; the longest training session has $310$ windows.

With $R_s=11$ or $21$ for HAR and $R_s=7$ or $15$ for Milan, the conclusions about the temporal layers hold with one exception. In seven of the eight analyses, the HMM gains $35$ to $106$ per 100 held-out curves over the i.i.d.\ mixture. In all eight, the HSMM (NB) gains a further $2.8$ to $5.8$ and is preferred to the HMM by BIC, by $243$ to $519$. The exception is the contaminated emission with $R_s=21$ for HAR. Its i.i.d.\ fit of highest likelihood again places a large share of the windows in the inflated component and classifies poorly (ARI $0.37$), and its held-out log-likelihood exceeds that of the HMM by $506$ per 100 windows. On the training sessions, however, the HMM fit has the higher log-likelihood ($-677{,}847$ against $-683{,}086$), so here the HMM fits the training data better but generalizes worse. The level of agreement with the activity classes is sensitive to the basis (held-out ARI of the HMM from $0.77$ to $0.92$), and so is the ranking of the emissions by that agreement. With $R_s=11$ the $t$ models classify best ($0.87$ for the HMM) and the contaminated models worst ($0.82$), and with $R_s=21$ the reverse holds ($0.77$ and $0.92$).
Fitted to each station separately, the four Milan stations give the same ranking: the HSMM (NB) is best, then the HMM, then the HSMM (P), by BIC and by the held-out log-likelihood (Table~\ref{tab:milanstations}). The fitted sizes are $0.26$--$0.50$ for the $\mathrm{NO_2}$-dominated regime and $0.13$--$0.15$ for the $\mathrm{O_3}$-dominated one, with mean durations of $8$--$12$ and $13$--$20$ days.

\end{document}